\documentclass[12pt]{article}
\usepackage[utf8]{inputenc}
\usepackage[english]{babel}
\usepackage{amsmath,amsthm,amssymb,amsfonts}
\usepackage{graphicx}
\usepackage{natbib}
\usepackage{url}
\usepackage[dvipsnames]{xcolor}
\usepackage{bm,mathrsfs}
\usepackage{enumitem}
\usepackage{adjustbox,booktabs,array,tabularx,multirow}
\usepackage{caption}
\usepackage{etoc}
\usepackage{microtype,needspace}
\usepackage[ruled,vlined]{algorithm2e}
\usepackage[colorlinks=true,linkcolor=red,citecolor=blue]{hyperref}

\DeclareCaptionLabelFormat{addS}{#1 S#2}
\newcommand{\spacingset}[1]{\renewcommand{\baselinestretch}{#1}\small\normalsize}

\newenvironment{revisionblue}{}{}
\DeclareRobustCommand{\revblue}[1]{#1}
\newcommand{\bv}{\mathbf{v}}
\newcommand{\calX}{\mathcal{X}}
\newcommand{\calA}{\mathcal{A}}
\newcommand{\calB}{\mathcal{B}}
\newcommand{\calQ}{\mathcal{Q}}
\newcommand{\calY}{\mathcal{Y}}
\newcommand{\calT}{\mathcal{T}}
\newcommand{\calZ}{\mathcal{Z}}
\newcommand{\calS}{\mathcal{S}}
\newcommand{\calE}{\mathcal{E}}
\newcommand{\frakG}{\mathfrak{G}}
\newcommand{\ba}{\boldsymbol{a}}
\newcommand{\bz}{\boldsymbol{z}}
\newcommand{\bZ}{\boldsymbol{Z}}
\newcommand{\bpi}{\boldsymbol{\pi}}
\newcommand{\bone}{\boldsymbol{1}}
\newcommand{\bzero}{\boldsymbol{0}}
\newcommand{\lb}{\left\{}
\newcommand{\rb}{\right\}}
\newcommand{\mH}{\mathbf{H}}
\newcommand{\calG}{\mathcal{G}}
\newcommand{\calV}{\mathcal{V}}
\newcommand{\calU}{\mathcal{U}}
\newtheorem{theorem}{Theorem}
\newtheorem{assumption}{Assumption}
\newtheorem{lemma}{Lemma}
\newtheorem{corollary}{Corollary}
\newtheorem{condition}{Condition}
\newtheorem{proposition}{Proposition}
\newtheorem{definition}{Definition}
\newtheorem{example}{Example}
\newtheorem{remark}{Remark}
\newtheorem{prop}{Proposition}
\newtheorem{inequality}{Inequality}
\newtheorem{regularitycondition}{Regularity condition}

\title{\bfseries On the permutation equivariance principle for causal estimands}
\author{Jiaqi Tong$^{1}$ and Fan Li$^{1,\ast}$\\[0.2cm]
$^{1}$Yale School of Public Health\\
$^{\ast}$fan.f.li@yale.edu}
\date{}
\hypersetup{pdftitle={On the permutation equivariance principle for causal estimands},pdfauthor={Jiaqi Tong and Fan Li}}

\begin{document}
\spacingset{1}
\maketitle
\bigskip
\begin{abstract}
In many causal inference problems, multiple action variables share a common causal role yet lack a natural ordering. \revblue{We consider $K\geq2$ action variables, each evaluated under treatment or control conditions,} and formalize permutation equivariance, the principle that permuting the variables permutes the corresponding estimands in a trackable manner, hence preserving their scientific meaning. We characterize this principle algebraically and present a complete class of weighted permutation equivariant estimands capturing main effects and interactions of all orders. We discuss the interpretation and choice of weights and characterize residual-free estimands, whose inclusion--exclusion sum recovers the endpoint contrast between the all-treated and all-control configurations. \revblue{Applying our general framework to network interference yields a new hierarchy of direct, indirect, and overall effects, whose first-order aggregates recover the average effects of \citet{hu2022average}. We also identify the overall effects as mixed derivatives of expected welfare under independent Bernoulli assignment.} We illustrate the framework through the contexts of factorial studies, causal mediation, and network interference.
\end{abstract}

\vfill
\newpage
\spacingset{1.8}

% Main manuscript
\section{Introduction}
\subsection{Background}
Causal inference concerns the study of potential outcome contrasts defined with respect to the same target population. Many modern causal inference problems involve multiple variables that share a common causal role, including treatment factors in factorial experiments or observational studies \citep{mukerjee2006modern,dasgupta2015causal,zhao2022regression,yu2026balancing}, unordered mediators in causal mediation \citep{lange2014assessing,xia2022decomposition,ohnishi2024bayesian}, and treatment assignments to units under network interference \citep{hudgens2008toward,manski2013identification,aronow2017estimating}. \revblue{These motivating causal inference settings can all be represented using $K\geq2$ \emph{action variables}, such as treatment factors, mediators, or unit treatment assignments, with the binary labels $0$ (control) and $1$ (treatment) indexing the two conditions under comparison.} In each case, one is often interested in causal contrasts that summarize the separate or interactive effects of these action variables.

When the action variables have no inherent ordering, a natural question is whether each main or interaction effect maintains a consistent scientific interpretation if the variables are relabeled. For example, if two mediators exchange labels, the main effect associated with each mediator should follow that mediator to its new label, without changing its scientific interpretation. On the other hand, an estimand definition that assigns roles according to label order can make the attributed effects depend on an arbitrary labeling, which may be undesirable. To ensure that the scientific meaning does not depend on such labeling, we require the vector of causal estimands to be \emph{permutation equivariant}; that is, permuting the variables permutes the corresponding estimands in a trackable manner. \revblue{Thus, each effect remains associated with the variable or group of variables it is defined to represent.} We give a concrete context below with three canonical examples.

\subsection{Related literature and canonical examples}\label{subsec:related-literature}

\begin{example}[\emph{Factorial experiment}]\label{eg:factorial-trial/os}
Factorial experiments are concerned with treatment effect contrasts among combinations of multiple treatment factors \citep{mukerjee2006modern}. Because the factor labels usually carry no inherent ordering, main effects and interactions should retain their scientific interpretations when these labels are permuted. Related symmetry considerations arise in the study of isomorphism for fractional factorial designs \citep{ma2001isomorphism,cheng2004geometric}. Isomorphism concerns statistical equivalence between designs, whereas permutation equivariance concerns the behavior of a vector of causal estimands under factor relabeling. This distinction motivates a formulation at the level of estimands that also applies to factorial observational studies \citep{yu2026balancing} and other causal inference problems.
\end{example}

\begin{example}[\emph{Multiple unordered mediators}]\label{eg:meidation-multiple}
Causal mediation analysis concerns the decomposition of a treatment effect into indirect pathways operating through mediators and a direct pathway that bypasses them. With multiple causally unordered mediators, one is interested in quantifying the contributions of individual mediators and their interactions \citep{lange2014assessing,xia2022decomposition}. \revblue{When mediator effects are defined by changing one mediator at a time, the order of these changes can affect how interactions are allocated and hence the effect attributed to each mediator. As the numerical example in Section~\ref{sec:whatispi} illustrates, exchanging the labels can inadvertently reverse the ranking of the effects attributed to the same mediators.} Permutation equivariance provides a criterion for preserving their intended interpretations and motivates a principled construction for an arbitrary number of unordered mediators.
\end{example}

\begin{example}[\emph{Network interference}]\label{eg:network-interference}
Causal inference under network interference concerns settings in which a unit's outcome can be affected by the treatments assigned to other units \citep{hudgens2008toward,aronow2017estimating}. \revblue{Unit numbers serve only as labels, and calling a person unit 1 rather than unit 2 should not change the causal question. For example, the average direct and indirect effects of \citet{hu2022average} distinguish the effect of a unit's treatment on its own outcome from its effects on other units' outcomes. This distinction depends on whose treatment and outcome are being compared, regardless of their numerical labels. For a renumbering to describe the same causal setting, each unit's treatment assignment and assignment probability, outcome, attributes, and network connections must follow that unit to its new label. The same principle applies to policy evaluation and targeting \citep{viviano2025policy}, where a treatment allocation policy should retain its meaning when units are renumbered. Permutation equivariance formalizes this requirement, such that effects indexed by units or groups of units follow those units when their labels change.}
\end{example}

\revblue{These examples motivate a unified treatment of permutation equivariance that connects the definition of complex estimands to their scientific interpretation. Estimand construction must specify the variables being contrasted and how the remaining variables are held fixed or averaged over. These decisions determine the causal question and how the resulting estimands behave under relabeling.}

\subsection{Our contributions}\label{subsec:contributions}
This article makes three main contributions. First, we provide a rigorous characterization of the permutation equivariance principle for vectors of causal estimands, drawing on group action theory and combinatorial algebra \citep{mccullagh2000invariance,o2014analysis}. We derive an algebraic criterion for determining whether relabeling the action variables induces the corresponding permutation of the estimand vector. We then construct a complete class of weighted estimands containing a main effect for each variable and an interaction for every group of two or more variables. \revblue{The weights specify how each contrast is averaged over configurations of the remaining variables. We distinguish permutable weights, which follow their associated variables under relabeling, from invariant weights, for which we establish conditions under which permutation equivariance holds.}

Second, we characterize the weights that yield a \emph{residual-free} decomposition, which is an inclusion--exclusion sum of main effects and interactions that equals the endpoint contrast between the configurations with all action variables in state $1$ and all action variables in state $0$. Within the class of nonnegative, normalized invariant weights, we show that this identity holds for every potential outcome response surface precisely when each component holds the remaining variables in state $1$. This result gives a common interpretation to decompositions of a natural indirect effect in mediation and of the all-treated versus all-control contrast under interference. \revblue{We also examine the causal assumptions and assignment support needed to identify and estimate the proposed estimands.}

\revblue{Third, we apply our general framework to derive a new hierarchy of direct, indirect, and overall effects under network interference, using weights induced by independent Bernoulli assignment. The first-order aggregates recover the average effects of \citet{hu2022average}, while higher-order components describe interactions generated by sets of treatment assignments. We show that the overall effects and interactions are derivatives of expected welfare with respect to the corresponding treatment probabilities. Their complete collection exactly decomposes welfare differences between Bernoulli assignment policies, with the residual-free endpoint decomposition obtained as a boundary case. We also discuss unbiased design-based estimators when all treatment probabilities lie strictly between zero and one.}

\revblue{In addition, our framework formulates equivariance relative to relabelings that preserve the scientific roles of the action variables. This accommodates partially ordered actions through permutations within unordered groups that preserve the ordering between groups. We also examine equivariance for estimands defined through exposure mappings and marginal structural models,} and extend the construction to ratio effect measures.

\revblue{The remainder of this article is organized as follows. Section~\ref{subsec:primitive-response} introduces the setup and applications. Section~\ref{sec:what-is-equivariance} characterizes permutation equivariance. Sections~\ref{ss:class-estimands} and~\ref{ss:w-IEP} develop weighted estimands and the residual-free decomposition. Section~\ref{sec:change-scale} presents ratio estimands, Section~\ref{sec:network-policy-effects} applies the framework to network interference, and Section~\ref{sec:conclusion} concludes. The Supplementary Material provides proofs, worked examples, a group theory primer, and further application and identification details.}

\section{\revblue{Setup and notation}}\label{subsec:primitive-response}

\revblue{We use \emph{action variables} to specify the choices entering a causal comparison. Let $\calX=\{X_1,\ldots,X_K\}$ denote $K\geq2$ such variables, each with two states labeled $0$ (control) and $1$ (treatment). In a factorial experiment, an action variable represents a treatment factor. In mediation, its state selects whether a mediator takes its potential value under treatment $0$ or $1$. In a network experiment, it represents the treatment assigned to a unit.}

\revblue{We index each configuration by the set of action variables in state $0$. For a subset $\calA\subseteq\calX$, define the corresponding state vector $\ba(\calA)\in\{0,1\}^K$ by}
\begin{revisionblue}\[
a_k(\calA)
=
\begin{cases}
0, & X_k\in\calA,\\
1, & X_k\notin\calA.
\end{cases}
\]\end{revisionblue}
\revblue{Thus, the variables in $\calA$ take state $0$, and all remaining variables take state $1$. For example, with three variables, $\calA=\{X_1,X_3\}$ gives $\ba(\calA)=(0,1,0)$. The empty set $\calA=\emptyset$ represents the configuration with all variables in state $1$, whereas $\calA=\calX$ represents the configuration with all variables in state $0$. The collection of all subsets, denoted by $2^{\calX}:=\{\calA:\calA\subseteq\calX\}$, is called the power set, and its elements correspond one-to-one to the $2^K$ binary configurations.}

\revblue{The \emph{primitive response surface} records the causal response of interest under each configuration. We denote this function by}
\begin{revisionblue}\begin{align*}
f:2^{\calX}\longrightarrow\mathbb R,
\qquad
\calA\longmapsto f(\calA).
\end{align*}\end{revisionblue}\revblue{Here, $f(\calA)$ is the response under the configuration $\ba(\calA)$. In the examples below, it is a mean potential outcome. Causal effects are constructed by comparing means indexed by different configurations. The binary action states describe the choices being compared; the potential outcome itself need not be binary.}

\begin{example}
\revblue{Consider causal mediation with a binary treatment and $K$ causally unordered mediators. Let $X_k(a)$ denote the value that mediator $k$ would take under treatment $a\in\{0,1\}$. For $(a_1,\ldots,a_K)\in\{0,1\}^K$, define}
\begin{revisionblue}\[
\mu(a_1,\ldots,a_K)
=
E\{Y(1,X_1(a_1),\ldots,X_K(a_K))\}.
\]\end{revisionblue}\revblue{This mean evaluates the outcome under treatment $1$, with mediator $k$ set to its potential value under treatment $a_k$. The corresponding response surface is}
\begin{revisionblue}\[
f(\calA)
=
\mu\{a_1(\calA),\ldots,a_K(\calA)\}
=
E\!\left[
Y\{1,X_1(a_1(\calA)),\ldots,X_K(a_K(\calA))\}
\right].
\]\end{revisionblue}\revblue{Thus, the mediators indexed by $\calA$ take their potential values under treatment $0$, and the remaining mediators take their potential values under treatment $1$. Each binary action state selects a treatment condition for a given mediator.}
\end{example}

\begin{example}\label{eg:factorial}
\revblue{Consider a factorial experiment with $K$ treatment factors and $N$ units (typically $K<N$). The action variables index the factors. Under a design-based formulation, units and potential outcomes are held fixed, and randomness arises solely from treatment assignment. The response surface is}
\begin{revisionblue}\[
f(\calA)
=
\frac{1}{N}\sum_{i=1}^N
Y_i\{\ba(\calA)\},
\qquad
\ba(\calA)\in\{0,1\}^K,
\]\end{revisionblue}\revblue{where $Y_i\{\ba(\calA)\}$ is the potential outcome that unit $i$ would have under the specified factor combination. Thus, $f(\calA)$ is the mean potential outcome across the study units under that combination. Of note, our development also applies to a superpopulation formulation, where the mean potential outcome is replaced by the average over a notional population expectation $f(\calA)=E[Y_i\{\ba(\calA)\}]$.}
\end{example}

\begin{example}
\revblue{In a network experiment, each action variable represents the treatment assigned to one unit. The number of action variables therefore equals the number of units, $K=N$, and $\ba(\calA)\in\{0,1\}^N$ specifies their joint treatment assignment. Here, the coordinates index individual units rather than the treatment factors in Example \ref{eg:factorial}. Interference means that a unit's outcome can depend on other units' treatments as well as its own. Accordingly, $Y_i\{\ba(\calA)\}$ denotes the potential outcome of unit $i$ under the full assignment. The design-based response surface averages these outcomes across the study units:}
\begin{revisionblue}\[
f(\calA)
=
\frac{1}{N}\sum_{i=1}^N
Y_i\{\ba(\calA)\},
\qquad
\ba(\calA)\in\{0,1\}^N.
\]\end{revisionblue}
\end{example}
\revblue{The finite-population response surfaces in the factorial and network examples can be viewed as special cases of the general design-based framework of \citet{harshaw2025general}.} \revblue{A response value summarizes one configuration, whereas an effect compares responses across configurations. For example, $f(\{X_1\})$ is the response when the first action variable is in state $0$ and all remaining variables are in state $1$. Changing the first variable to state $1$, while keeping the others unchanged, gives the causal contrast $f(\emptyset)-f(\{X_1\})$. Section \ref{sec:what-is-equivariance} illustrates these causal contrasts and Section~\ref{ss:class-estimands} presents their general weighted construction.
\revblue{For reference, Supplementary Table~S\ref{tab:application-comparison} compares the action variables, mean responses, weight choices, relabelings, and identification considerations across the three applications.}
}

\section{What is permutation equivariance?}\label{sec:what-is-equivariance}

\subsection{Introductory examples}\label{sec:whatispi}
We discuss two examples in the context of causal mediation analysis and one example in a factorial experiment to define permutation equivariance. \revblue{Supplementary Appendix~\ref{app:dimension-reduction} also examines permutation equivariance for exposure mappings and marginal structural models, providing compatibility conditions and worked examples that either satisfy or violate permutation equivariance.}

\begin{example}\label{eg1}
Consider the problem of causal mediation with two mediators that are not causally ordered. Let $X_k(a)$ denote the potential value of the $k$th ($k=1,2$) mediator under condition $a$ ($a=0,1$). Let $Y(1,X_1(a),X_2(a^*))$ be the potential outcome under treatment, with the potential mediators taking the values they would naturally have under treatment levels $a$ and $a^*$. \cite{lange2014assessing} defined the natural indirect main effects through $X_1$ and $X_2$ as:
\begin{align}\label{eg:delta}
    &\Delta_{\{X_1\}}=\mu(1,0)-\mu(0,0),~~\Delta_{\{X_2\}}=\mu(1,1)-\mu(1,0),
\end{align}
where $\mu(a,a^*)=E\{Y(1,X_1(a),X_2(a^*))\}$. Here, exchanging the labels for the two mediators leads to two new estimands: $\Delta_{\{X_1\}}^\dagger=\mu(0,1)-\mu(0,0)$ and $\Delta_{\{X_2\}}^\dagger=\mu(1,1)-\mu(0,1)$. Neither $\Delta_{\{X_1\}}^\dagger$ nor $\Delta_{\{X_2\}}^\dagger$ overlaps with the original estimands, $\Delta_{\{X_1\}}$ and $\Delta_{\{X_2\}}$. In particular, $\Delta_{\{X_1\}}^\dagger$ now represents the natural indirect main effect through $X_2$ while holding $X_1$ at its natural value under the control condition, whereas the originally defined natural indirect main effect $\Delta_{\{X_2\}}$ holds $X_1$ at its natural value under the treatment condition. That is, swapping the mediator labels alters the scientific interpretation of the estimands (also see an illustration after Example~\ref{eg2}). As a result, \eqref{eg:delta} is not permutation equivariant and may be undesirable in applied settings, especially when practitioners may not fully recognize how changes in labeling inadvertently affect causal interpretation.
\end{example}

\begin{example}\label{eg2}
Following Example \ref{eg1}, \cite{xia2022decomposition} defined the ``exit indirect effects'' to study causal mediation with unordered mediators:
\begin{align}\label{eg:tau}
    &\Delta_{\{X_1\}}=\mu(1,1)-\mu(0,1),~~\Delta_{\{X_2\}}=\mu(1,1)-\mu(1,0).
\end{align}
In contrast to the estimands in \eqref{eg:delta}, those in \eqref{eg:tau} are permutation equivariant, because exchanging the labels results in $\Delta_{\{X_1\}}^\dagger=\mu(1,1)-\mu(1,0)=\Delta_{\{X_2\}}$ and $\Delta_{\{X_2\}}^\dagger=\mu(1,1)-\mu(0,1)=\Delta_{\{X_1\}}$. Both $\Delta_{\{X_1\}}$ and $\Delta_{\{X_2\}}^\dagger$ track the same natural indirect mediation main effect. In other words, under any arbitrary relabeling, \eqref{eg:tau} consistently preserves the same scientific interpretation.

\Needspace{5\baselineskip}
\revblue{To further illustrate this distinction, consider the hypothetical response surface, defined by $\mu(0,0)=0$, $\mu(1,0)=2$, $\mu(0,1)=1$, and $\mu(1,1)=10$. Then, the natural indirect main effects in Example~\ref{eg1} are $\Delta_{\{X_1\}}=2$ and $\Delta_{\{X_2\}}=8$. Exchanging the labels and applying the same definitions gives $\Delta_{\{X_1\}}^\dagger=1$ and $\Delta_{\{X_2\}}^\dagger=9$. Thus, relabeling changes the magnitudes of the natural indirect main effects and may even change which mediator has the larger effect (see Table~\ref{tab:numerical-attribution} for a quick summary). By contrast, the natural indirect main effects in Example~\ref{eg2} satisfy $\Delta_{\{X_1\}}=\Delta_{\{X_2\}}^\dagger=9$ and $\Delta_{\{X_2\}}=\Delta_{\{X_1\}}^\dagger=8$, so the effect estimand associated with each original mediator remains identical after relabeling. The definitions in Example~\ref{eg1} may be appropriate when the ordering used to define the effects is scientifically justified. Otherwise, for unordered mediators, arbitrary labels should in principle not change these attributed effects. Permutation equivariance is a principle that characterizes this consistency. Supplementary Appendix~\ref{app:worked-example} further illustrates this point for weighted estimands.}
\end{example}

\begin{table}[htbp]
\centering

\setlength{\belowcaptionskip}{5pt}
\renewcommand{\baselinestretch}{1}\small
\renewcommand{\arraystretch}{1.15}
\caption{Natural indirect main effects before and after exchanging mediator labels for the same hypothetical response surface. The last two columns track the same original mediators. After exchanging the labels, $\Delta_{\{X_2\}}^\dagger$ denotes the main effect of the original $X_1$, whereas $\Delta_{\{X_1\}}^\dagger$ denotes the main effect of the original $X_2$.}
\label{tab:numerical-attribution}
\begin{tabular}{@{}llcc@{}}
\toprule
Effect definitions & Labeling & Original $X_1$ & Original $X_2$ \\
\midrule
Example~\ref{eg1} & Original & $\Delta_{\{X_1\}}=2$ & $\Delta_{\{X_2\}}=8$ \\
Example~\ref{eg1} & Exchanged & $\Delta_{\{X_2\}}^\dagger=9$ & $\Delta_{\{X_1\}}^\dagger=1$ \\
\addlinespace
Example~\ref{eg2} & Original & $\Delta_{\{X_1\}}=9$ & $\Delta_{\{X_2\}}=8$ \\
Example~\ref{eg2} & Exchanged & $\Delta_{\{X_2\}}^\dagger=9$ & $\Delta_{\{X_1\}}^\dagger=8$ \\
\bottomrule
\end{tabular}

\end{table}

\begin{example}\label{eg3}
Consider the problem of factorial experiments with two unordered factors. Let $X_k$ denote the $k$th factor ($k=1,2$). Let $Y(X_1=a, X_2=a^*)$ denote the potential outcome when the two factors are set to levels $a$ and $a^*$, respectively. With a slight abuse of notation, we also use $\mu(a, a^*)$ to denote the expected potential outcome $N^{-1}\sum_{i=1}^NY_i(X_1=a, X_2=a^*)$. Following \cite{dasgupta2015causal}, the main effects in a balanced full factorial trial can be defined as
\begin{align*}
    &\Delta_{\{X_1\}}=\frac{1}{2}\left\{\mu(1,1)-\mu(0,1)\right\}+\frac{1}{2}\left\{\mu(1,0)-\mu(0,0)\right\},\\
    &\Delta_{\{X_2\}}=\frac{1}{2}\left\{\mu(1,1)-\mu(1,0)\right\}+\frac{1}{2}\left\{\mu(0,1)-\mu(0,0)\right\}.
\end{align*}
The main effects defined above are permutation equivariant, because permuting the labels of $X_1$ and $X_2$ results in the same set of estimands.
\revblue{More generally, the behavior under relabeling depends on what the weights represent. Now consider the main effects $\Delta_{\{X_1\}}=\frac{1}{3}\{\mu(1,1)-\mu(0,1)\}+\frac{2}{3}\{\mu(1,0)-\mu(0,0)\}$ and $\Delta_{\{X_2\}}=\frac{1}{2}\{\mu(1,1)-\mu(1,0)\}+\frac{1}{2}\{\mu(0,1)-\mu(0,0)\}$}.
\revblue{If these numerical weights remain attached to their original positions when the factors are relabeled, the two main effects do not form a permutation equivariant system.} \revblue{Alternatively, following \cite{zhao2022regression}, suppose the weights describe the assignment probabilities of the corresponding factors. Then the main effects are defined as}
\begin{align*}
    &\Delta_{\{X_1\}}=\sum_{a=0}^1\Pr(X_2=a)\{\mu(1,a)-\mu(0,a)\},\\
    &\Delta_{\{X_2\}}=\sum_{a=0}^1\Pr(X_1=a)\{\mu(a,1)-\mu(a,0)\}.
\end{align*}

\revblue{Each main effect averages the effect of one factor over the assignment distribution of the other. In the preceding example, $\Pr(X_2=1)=1/3$ and $\Pr(X_1=1)=1/2$. When the factor labels are exchanged, each factor retains its original assignment probability. The effect associated with each original factor is therefore preserved.} \revblue{We call such weights that follow the action variables under relabeling ``permutable weights'', whereas the weights that are held fixed under relabeling are called ``invariant weights''.
Section~\ref{ss:class-estimands} further discusses the choice and interpretation of weights, and Appendix~\ref{sec:application-eg-factorial-os} considers factorial observational studies \citep{yu2026balancing}.}

\end{example}

\subsection{Definition and verification of permutation equivariance}
\label{subsec:setup}

\revblue{We present a general setup for defining permutation equivariance. For illustration, we first consider invariant weights; Supplementary Appendix~\ref{app:small-k} extends the formulation to permutable weights. We begin with a self-contained introduction to the relevant group concepts. Readers unfamiliar with group action notation may consult Supplementary Appendix~\ref{app:group-primer}, which summarizes the notation in Table~S\ref{tab:group-primer-notation} and further explains these concepts through worked examples. A permutation of $\calX$ is a bijection $\sigma:\calX\to\calX$ that relabels the action variables. There are \(K!\) possible ways to relabel the \(K\) action variables. The symmetric group \(S_K\) consists of all $K!$ permutations $\sigma$.
In certain scientific contexts, not all permutations are meaningful. We call a relabeling scientifically admissible if it preserves the scientific relationships among action variables. For example, suppose mediators $X_1$ and $X_2$ are measured contemporaneously before mediator $X_3$. Scientifically admissible permutations allow exchanging the labels of $X_1$ and $X_2$, but not exchanging either label with that of $X_3$. That is, $\{X_1,X_2,X_3\}$ are partially ordered.
Remark~\ref{remark:partially-ordered} presents a general treatment of partially ordered action variables.
 }

\revblue{Let \(\frakG\) be the subgroup of \(S_K\) consisting of these admissible relabelings. Each $\sigma\in\frakG$ acts on $2^{\calX}$ according to $\sigma(\calA)=\{\sigma(X):X\in\calA\}$. The \emph{orbit} of a subset $\calA\subseteq\calX$ consists of all subsets obtainable from $\calA$ through these admissible relabelings. Thus, an orbit groups together subsets that are equivalent up to admissible changes of labels. The orbits partition $2^{\calX}$, with each subset belonging to exactly one orbit. Under $\frakG=S_K$, permutations preserve cardinality and can map any subset to any other subset of the same cardinality. The orbits are therefore the cardinality classes $[q]=\{\calA\subseteq\calX:|\calA|=q\}, q=0,\ldots,K$. For example, $[1]$ contains all singleton subsets, since any action variable can be relabeled as any other. The corresponding \emph{quotient set}, $\calQ=\{[0],[1],\ldots,[K]\}$, is the collection of these orbits, with each orbit treated as a single element. Thus, $\calQ$ groups subsets by their cardinality rather than their specific labels. Example \ref{eg:orbits} illustrates these definitions with $K=3$.}

\begin{example}\label{eg:orbits}
\revblue{Let $\calX=\{X_1,X_2,X_3\}$ and $\frakG=S_3$. A permutation
that exchanges $X_1$ and $X_2$ maps $\{X_1,X_3\}$ to
$\{X_2,X_3\}$. Because any two subsets of the same size can be
related by a permutation, the four orbits are $[0]=\{\varnothing\}$, $[1]=\big\{\{X_1\},\{X_2\},\{X_3\}\big\}$, $[2]=\big\{\{X_1,X_2\},\{X_1,X_3\},\{X_2,X_3\}\big\}$, $[3]=\{\calX\}$. Thus, $\calQ=\{[0],[1],[2],[3]\}$ has four elements, one for each possible subset size.}
\end{example}

\revblue{We choose a reference ordering of the subsets of $\calX$ and collect the corresponding response values in $\bv=\{f(\calA):\calA\subseteq\calX\}^\top$. Relative to this ordering, a relabeling $\sigma$ is represented by a column permutation matrix $\mathbf{P}_c(\sigma)$ satisfying $\mathbf{P}_c(\sigma)\bv = \{f(\sigma^{-1}(\calA)):\calA\subseteq\calX\}^\top$, where the coordinates on the right follow the same reference ordering. For example, let $K=3$ and consider the permutation $\sigma(X_1)=X_2$, $\sigma(X_2)=X_3$, and $\sigma(X_3)=X_1$. Suppose the original response values at $\{X_1\},\{X_2\},\{X_3\}$ are $10,20,30$, respectively. Relabeling attaches these same values to $\{X_2\},\{X_3\},\{X_1\}$, respectively. Thus, in the fixed order $\{X_1\},\{X_2\},\{X_3\}$, the singleton coordinates of $\mathbf{P}_c(\sigma)\bv$ are $(30,10,20)^\top$. The inverse identifies where each coordinate's value originated: the new $\{X_1\}$ coordinate receives the original value at $\sigma^{-1}(\{X_1\})=\{X_3\}$.}

More generally, let $\bm\Delta$ denote a vector of $d$ causal contrasts:
\begin{align}\label{eq:def-contrast}
\bm\Delta=\mathbf{H}\bv,
\end{align}
where $\mathbf{H}\in\mathbb R^{d\times 2^K}$ is a linear contrast map whose columns follow this reference ordering. Its rows encode the scientific contrasts of interest. The following definition formalizes permutation equivariance in terms of matrix operations.

\begin{definition}
\label{def:permutation-invariance}
The contrast vector $\bm\Delta$ is \revblue{permutation equivariant with respect to $\frakG$ if, for every $\sigma\in\frakG$} and the associated column permutation matrix $\mathbf{P}_c(\sigma)$, there exists a row permutation matrix $\mathbf{P}_r(\sigma)$ such that
\begin{align}\label{eq:intertwining}
\mathbf{P}_r(\sigma)\mathbf{H}=\mathbf{H} \mathbf{P}_c(\sigma).
\end{align}
\end{definition}

\revblue{By \eqref{eq:def-contrast}, forming contrasts from the relabeled response vector gives $\mathbf{H}\{\mathbf{P}_c(\sigma)\bv\}$, whereas relabeling the original contrasts gives $\mathbf{P}_r(\sigma)\bm\Delta$. Permutation equivariance requires these operations to agree for every possible response vector $\bv$:}
\[
\mathbf{H}\{\mathbf{P}_c(\sigma)\bv\}
=
\mathbf{P}_r(\sigma)\{\mathbf{H}\bv\}
=
\mathbf{P}_r(\sigma)\bm\Delta.
\]
\revblue{This requirement is equivalent to the matrix identity \eqref{eq:intertwining}, known as an \emph{intertwining relation}. It expresses an intrinsic algebraic symmetry of $\mathbf{H}$; that is, admissible permutations of its columns induce corresponding permutations of its rows, thereby preserving the combinatorial structure of the contrast system independently of the response values. Individual contrast coordinates need not remain numerically unchanged; rather, their values are rearranged consistently with the relabeling of the associated action variables.} \revblue{To clarify the combinatorial algebra, Supplementary Appendix~\ref{app:group-primer} provides explicit examples of orbits and quotient sets, explains the conventions for row and column permutations, and verifies Equation~\eqref{eq:intertwining} for an estimand vector with $K=2$.}

\begin{remark}[\revblue{\emph{Partially ordered action variables}}]
\label{remark:partially-ordered}
\revblue{For partially ordered action variables, suppose $\calX$ is partitioned into ordered blocks $\calX_1,\ldots,\calX_T$, with $|\calX_t|=K_t$. Admissible relabelings are restricted to those that preserve block membership, so $\frakG$ is a subgroup of $S_{K_1}\times\cdots\times S_{K_T}$. Examples include mediator blocks measured at successive time points or intervention components delivered at successive visits, provided the variables within a block are scientifically unordered. The previous construction applies within a selected block, with the action configuration outside that block held fixed as part of the estimand definition. Equivariance then concerns only admissible permutations within the selected block, preserving the ordering between blocks. Supplementary Appendix~\ref{app:controlled-histories} develops this application.}
\end{remark}

Beyond Definition \ref{def:permutation-invariance}, permutation equivariance can be alternatively characterized in the following way. For a matrix $\mathbf{H}$, let $\mathcal R(\mathbf{H})$ denote the multiset of its rows, including multiplicities. A multiset is an unordered collection that retains repeated elements. Here, the multiplicity of a row is the number of times it appears in $\mathbf{H}$. Theorem \ref{thm:characterization} converts the existential condition in Definition \ref{def:permutation-invariance} into a direct counting criterion.

\begin{theorem}\label{thm:characterization}
The contrast vector $\bm\Delta=\mathbf{H}\bv$ is \revblue{equivariant with respect to $\frakG$} if and only if, for every $\sigma\in\frakG$,
\begin{align*}
\mathcal R\{\mathbf{H} \mathbf{P}_c(\sigma)\}=\mathcal R(\mathbf{H}).
\end{align*}
\end{theorem}
Thus, permutation equivariance holds when every admissible relabeling leaves the multiset of contrast rows unchanged. Supplementary Appendix~\ref{app:row-multiset-examples} verifies this criterion for the three introductory examples by displaying their contrast matrices and comparing the rows before and after relabeling. Theorem~\ref{thm:characterization} also motivates Algorithm~\ref{algo:compute} in Supplementary Appendix~\ref{app:algorithm}, which identifies the corresponding row permutation matrix $\mathbf{P}_r(\sigma)$ for a given relabeling $\sigma$. Finally, to accommodate permutable weights, we expand the response vector $\bv$ to include products of weights and primitive responses; see Supplementary Appendix~\ref{app:small-k}.

\section{Permutation equivariant and complete estimands}\label{ss:class-estimands}

\revblue{A vector of estimands is \emph{complete} if it includes one contrast for each nonempty subset of the action variables. The index $\calY$ identifies exactly the variables jointly contrasted by $\Delta_{\calY}$. For example, $\Delta_{\{X_1\}}$ is the first-order contrast indexed by the singleton $\{X_1\}$. In a factorial experiment with $K=2$ (Example \ref{eg3}), the first-order contrast for \(X_1\) is its main effect, or equivalently the average treatment–control contrast for \(X_1\) across the two states of \(X_2\). With equal weights, $\Delta_{\{X_1\}}
=\tfrac12\{\mu(1,1)-\mu(0,1)\}
+\tfrac12\{\mu(1,0)-\mu(0,0)\}$. The index \(\{X_1\}\) indicates that each comparison changes only the state of \(X_1\); the state of the other factor is held fixed within each comparison, and the resulting contrasts are then averaged over its states. More generally, the complete weighted system of estimands is defined as follows.}

\begin{definition}\label{def:pi-estimands-defi}
For $K$ action variables, let $\bm\Delta=\{\Delta_{\calY}:\emptyset\neq\calY\subseteq\calX\}$, where
\begin{align}\label{eq:estimandsclass-unweighted}
\Delta_{\calY}
={}&
\sum_{\calT\subseteq\calY^c}w(\calT,\calY)
\left\{
\sum_{\calZ\subseteq\calY}(-1)^{|\calZ|}f(\calZ\cup\calT)
\right\},
\end{align}
where the nonnegative weights are user-defined and should satisfy $\sum_{\calT\subseteq\calY^c}w(\calT,\calY)=1$ for every nonempty $\calY\subseteq\calX$.
The system contains $2^K-1$ contrasts. Conditional on $\calT$, the inner alternating sum is the mixed finite difference generated by the variables in $\calY$; the outer sum standardizes that contrast over reference states of the remaining variables outside $\calY$.
\end{definition}

\revblue{Definition~\ref{def:pi-estimands-defi} nests several existing estimands from the causal inference literature and encompasses all introductory examples as special cases. First, to recover the estimands of \citet{xia2022decomposition}, we may set \(w(\calT,\calY)=\mathbf{1}(\calT=\emptyset)\) for every nonempty \(\calY\), where \(\mathbf{1}(\cdot)\) denotes the indicator function. That is, each contrast indexed by a nonempty \(\calY\) is defined by holding all variables outside \(\calY\) at state \(1\). While \cite{xia2022decomposition} did not provide explicit definitions for $K\geq3$, $\{\Delta_{\calY}=\sum_{\calZ\subseteq \calY}(-1)^{|\calZ|}f(\calZ):\emptyset\neq\calY\in 2^{\calX}\}$ serve as a natural extension of their estimands for general $K$. When \(K=2\) and \(\calY=\{X_1\}\), this yields \(\Delta_{\{X_1\}}=f(\emptyset)-f(\{X_1\})=\mu(1,1)-\mu(0,1)\), as in Example~\ref{eg2}.} Second, to recover the factorial effects of \citet{dasgupta2015causal} in \(2^K\) factorial experiments, one can use the equal weight scheme \(w(\calT,\calY)=2^{-|\calY^c|}\). This choice recovers their first-order effects and yields \(2^{q-1}\) times their order-\(q\) factorial interaction effect. When \(K=2\), it yields the invariant weights used in Example~\ref{eg3}. \revblue{
Finally, to recover the estimands for factorial experiments under the coherent weighting framework of \citet{zhao2022regression}, one can define the weights using the assignment distribution of the factors (Supplementary Remark~\ref{remark:compatibility}). For example, setting \(w(\emptyset,\{X_1\})=\Pr(X_2=1)\) and \(w(\{X_2\},\{X_1\})=\Pr(X_2=0)\) averages the contrast for \(X_1\) over the two assignment states of \(X_2\), giving the factorial effect with permutable weights in Example~\ref{eg3}; the effect of \(X_2\) is defined analogously.
For further illustration, we also provide an example under independent Bernoulli experimental design below.}

\begin{remark}[\revblue{\emph{Assignment-induced weights}}]\label{remark:assignment-induced-weights}
\revblue{Suppose the action variables follow independent Bernoulli assignments with probability $\bpi=(\pi_1,\ldots,\pi_K)^\top$. If $\calT\subseteq\calY^c$ indexes the action variables outside $\calY$ assigned state $0$, then the weights follow}
\begin{revisionblue}\begin{align}\label{eq:bernoulli-weights}
w_{\bpi}(\calT,\calY)
=
\prod_{k:X_k\in\calT}(1-\pi_k)
\prod_{k:X_k\in\calY^c\setminus\calT}\pi_k.
\end{align}\end{revisionblue}\revblue{These are permutable weights, for which relabeling the action variables naturally relabels their treatment assignment probabilities.}
\end{remark}

\revblue{A few considerations on weight specification are worth noting. First, choosing weights for $\Delta_{\calY}$ requires specifying how the variables outside $\calY$ should be fixed or averaged over. The equal weight scheme, $w(\calT,\calY)=2^{-|\calY^c|}$, averages equally over all configurations of these variables. The point mass scheme, $w(\calT,\calY)=\mathbf{1}(\calT=\emptyset)$, holds all variables outside $\calY$ at state $1$. The assignment-induced weight scheme averages over the assignment distribution, which is either known by design or estimated from the data. Different weight schemes may yield different definitions of $\Delta_{\calY}$ when the contrast being averaged varies across configurations of the remaining variables. Second, the interpretation of the weights determines how they behave under relabeling. Assignment probabilities correspond to particular action variables and follow those variables when their labels change; they are hence permutable. Invariant weights, by contrast, remain fixed under relabeling. Third, in settings such as mediation analysis, it may be unclear how to associate weights with scientifically meaningful assignment probabilities, making invariant weights more appropriate. In factorial experiments and network interference, permutable weights can be defined using the assignment probabilities of the factors or network units and are more natural constructions. The choice between permutable and invariant weights hence depends on the scientific context. Most typically, permutable weights define effects averaged over the assignment distribution, whereas invariant weights define effects averaged under a fixed rule independent of the assignment mechanism.}

Theorem \ref{thm:estimand-pi} below proves that $\boldsymbol{\Delta}$ in Definition \ref{def:pi-estimands-defi} is permutation equivariant.

\begin{theorem}\label{thm:estimand-pi}
\revblue{Fix an admissible group $\frakG$.}
\begin{enumerate}[label=(\roman*)]
\item For each $q=1,\ldots,K$, the order-$q$ subvector $\{\Delta_{\calY}:|\calY|=q\}$ is equivariant with permutable weights. \revblue{With invariant weights, the same conclusion holds if $w(\calT,\calY)=w\{\sigma(\calT),\sigma(\calY)\}$ for all admissible permutations $\sigma\in\frakG$, where $\calY$ identifies the variables being contrasted and $\calT$ identifies remaining variables assigned state $0$.}
\item Relabeling by \revblue{$\sigma\in\frakG$} sends $\Delta_{\calY}$ to $\Delta_{\sigma(\calY)}$.
\item The full vector $\{\Delta_{\calY}:\emptyset\neq\calY\subseteq\calX\}$ is complete and equivariant \revblue{under the weight conditions in (i).}
\end{enumerate}
\end{theorem}

\revblue{Several remarks on Theorem~\ref{thm:estimand-pi} are in order. First, causal estimands with permutable weights are naturally permutation equivariant, whereas equivariance under invariant weights requires that each pair $(\calT,\calY)$ receive the same weight as any pair $(\calT',\calY')$ obtained from it by an admissible relabeling of the action variables. For example, the equal weight scheme $w(\calT,\calY)=2^{-|\calY^c|}$ and the point mass weight scheme $w(\calT,\calY)=\mathbf{1}(\calT=\emptyset)$ both satisfy condition~(i) of Theorem~\ref{thm:estimand-pi} for invariant weights. Example \ref{con:invariant-weights-pi} further illustrates this condition for $K=2$, and Supplementary Remark~\ref{remark:df-pi-inva} gives the degrees of freedom for general $K$. Second, the contrast $\Delta_{\calY}$ captures the interaction among the $q=|\calY|$ action variables in $\calY$ and is relabeled as $\Delta_{\sigma(\calY)}$ under an admissible permutation $\sigma$. For example, an admissible exchange of $X_3$ and $X_4$ relabels the third-order interaction $\Delta_{\{X_1,X_2,X_3\}}$ as $\Delta_{\{X_1,X_2,X_4\}}$.
}

\begin{example}\label{con:invariant-weights-pi}
\revblue{When $K=2$ and $\frakG=S_2$, the permutation equivariant vector of estimands with invariant weights can be parameterized by a single weight parameter $\lambda\in[0,1]$ as follows:
\begin{equation*}
    \mathbf{\Delta}=(\Delta_{\{X_1\}},\Delta_{\{X_2\}},\Delta_{\{X_1,X_2\}})^\top=\mathbf{H}\mathbf{v}=\begin{pmatrix}
    \lambda&-\lambda&(1-\lambda)&-(1-\lambda)\\
    \lambda&1-\lambda&-\lambda&-(1-\lambda)\\
    1&-1&-1&1
    \end{pmatrix}\mathbf{v},
\end{equation*}
where $\bv=(\mu(1,1),\mu(0,1),\mu(1,0),\mu(0,0))^\top$.
Here, $\lambda$ is the common weight assigned to state $1$ when averaging over the action variable not being contrasted. For the hypothetical response surface preceding Table~\ref{tab:numerical-attribution}, the estimands are $(2+7\lambda,1+7\lambda,7)$. Setting $\lambda=0$, $1/2$, or $1$ evaluates each action with the other in state $0$, averaged equally over both states, or in state $1$, giving $(2,1,7)$, $(5.5,4.5,7)$, or $(9,8,7)$, respectively. Each choice is permutation equivariant. In contrast, for assignment probabilities $(\pi_1,\pi_2)$, the permutable weight estimands are $(2+7\pi_2,1+7\pi_1,7)$, because each main effect averages over the other variable's assignment probability. Supplementary Appendix~\ref{app:worked-example} shows these calculations.}

\end{example}

By Theorem \ref{thm:estimand-pi}, the proposed estimands in \cite{xia2022decomposition} (with invariant weights) and in \cite{dasgupta2015causal} (with invariant weights) and \cite{zhao2022regression} (with permutable weights) are all permutation equivariant and complete. Below, we also provide an alternative characterization for $\Delta_{\calY}$.

\begin{proposition}\label{prop:alternative-repre}
For any $X_k\in \calY$, \eqref{eq:estimandsclass-unweighted} can alternatively be expressed as
\begin{align*}
    \Delta_{\calY}=&\sum_{\calT\subseteq \calY^c}w(\calT,\calY)\left[\sum_{\calZ\subseteq \calY\backslash\{X_k\}}(-1)^{|\calZ|}\{f(\calZ\cup \calT)-f(\calZ\cup\{X_k\}\cup \calT)\}\right].
\end{align*}
\end{proposition}
For example, when $K=2$, consider the interaction effect where $\mathcal{Y}=\{X_1,X_2\}$. Proposition \ref{prop:alternative-repre} defines this estimand as $\Delta_{\{X_1,X_2\}} = \{\mu(1,1)-\mu(0,1)\} - \{\mu(1,0)-\mu(0,0)\}$.
This estimand is interpreted as the extent to which the effect of variable $X_1$ is modified by variable $X_2$. The same difference in differences also describes how the effect of $X_2$ is modified by $X_1$. More generally, Proposition \ref{prop:alternative-repre} shows that $\Delta_{\calY}$ quantifies the extent to which the effect of one variable in $\calY$, $X_k$, is modified by the interaction of the remaining $|\calY|-1$ variables, with the choice of $X_k$ being arbitrary within $\calY$ because the expression is symmetric in the variables in $\calY$. On a technical level, Proposition \ref{prop:alternative-repre} suggests that an iterative characterization of the vector of estimands in Definition \ref{def:pi-estimands-defi} is possible; see, e.g., \cite{zhao2022regression} for related discussions in the context of factorial experiments.

\section{Residual-free estimands}\label{ss:w-IEP}

Within the complete class of estimands in Definition~\ref{def:pi-estimands-defi}, the weights specify how each effect is averaged over configurations of the action variables outside the contrast. We next characterize the weight schemes for which the inclusion–exclusion sum recovers the endpoint contrast for every response surface. This contrast is the response difference between the all-treated and all-control configurations, representing the total change to be decomposed into main and interaction effects. We call any estimand system with this property \emph{residual-free}.

\begin{definition}
\label{def:residual-free}
For a fixed weight scheme, the estimand system $\bm\Delta$ is residual-free if the following identity holds for all primitive response surfaces $f$:
\begin{equation}\label{eq:residual-free-identity}
f(\emptyset)-f(\calX)
=
\sum_{\emptyset\neq\calY\subseteq\calX}
(-1)^{|\calY|+1}\Delta_{\calY}.
\end{equation}

\end{definition}

\revblue{The quantity $f(\emptyset)-f(\calX)$ is the endpoint contrast, comparing the all-treated and all-control configurations.} The alternating sum on the right of \eqref{eq:residual-free-identity} has the M\"obius inclusion--exclusion form on the Boolean lattice \citep{rota1964foundations}. Departures from the identity are measured by the residual
\[
\operatorname{res}(f)
=
\sum_{\emptyset\neq\calY\subseteq\calX}
(-1)^{|\calY|+1}\Delta_{\calY}
-
\{f(\emptyset)-f(\calX)\}.
\]
Supplementary Lemma~\ref{lemma:alt-repre-cY-supp} gives its coordinate representation. The following result characterizes the unique nonnegative, normalized invariant weighting scheme that yields a residual-free decomposition of the endpoint contrast.

\begin{theorem}\label{prop:nie}
An estimand system in Definition~\ref{def:pi-estimands-defi} with nonnegative, normalized invariant weights is residual-free if and only if $w(\calT,\calY)=\mathbf 1(\calT=\emptyset)$ for all nonempty $\calY\subseteq\calX$ and all $\calT\subseteq\calY^c$. For each such $\calY$, the corresponding estimand is uniquely given by
 $\Delta_{\calY}
=
\sum_{\calZ\subseteq\calY}(-1)^{|\calZ|}f(\calZ)$.
\end{theorem}

Theorem~\ref{prop:nie} has the following implications for constructing and using the decomposition. First, the residual-free property uniquely fixes the reference configuration within the stated weight class, such that every effect is evaluated with all action variables outside the contrast held in the treatment state. Second, Equation~\eqref{eq:residual-free-identity} connects the decomposition to familiar causal contrasts. In causal mediation, it expresses the natural indirect effect as an inclusion-exclusion sum of main and interaction components. \revblue{In network interference, it gives the all-treated versus all-control boundary case of the expansion for finite policy changes developed in Subsection~\ref{subsec:network-finite-policy}.} \revblue{Third, the endpoint contrast and its components can have different identification requirements. In causal mediation, consistency, positivity, and sequential ignorability identify the natural indirect effect \citep{imai2010identification}, while identifying the exit indirect effects and their interaction may require additional restrictions on cross-world dependence \citep{xia2022decomposition}. Under network interference, the distinction depends on design support. A design restricted to the all-control and all-treated assignments identifies the endpoint contrast but generally leaves its finer main and interaction components unidentified. An independent Bernoulli experimental design with treatment probabilities strictly between zero and one assigns positive probability to every combination of control and treatment across units and permits unbiased design-based estimation of the finer main and interaction estimands (Section~\ref{subsec:network-design-estimation}). Supplementary Appendix~\ref{sec:identification} discusses these identification and estimation requirements in further detail.}

\section{Generalization to ratio estimands}\label{sec:change-scale}

The preceding results describe causal effects on an additive scale. For binary outcomes, scientific questions may instead concern relative changes in risk or odds. We can accommodate these comparisons by applying the weighted estimand construction on the log scale and then exponentiating. Specifically, let $h(x)=x$ for risk ratios and $h(x)=x/(1-x)$ for odds ratios.
Assuming $0<f(\calA)<1$, for each nonempty \(\calY\subseteq\calX\), with \(\calT\subseteq\calY^c\) and \(\calZ\subseteq\calY\) as in Definition~\ref{def:pi-estimands-defi}, we define the following set of estimands
\begin{equation*}
\Delta_{\calY}
=
\prod_{\calT\subseteq\calY^c}
\prod_{\calZ\subseteq\calY}
\left[h\{f(\calZ\cup\calT)\}\right]^{(-1)^{|\calZ|}w(\calT,\calY)}.
\end{equation*}
Here, the weights average contrasts on the log scale and appear as exponents on the ratio scale. Permutation equivariance and completeness follow from the additive construction on the transformed scale, while inclusion--exclusion sums become multiplicative or geometric inclusion--exclusion products. In particular, with the residual-free point mass weight scheme, the inclusion--exclusion product recovers the endpoint ratio contrast $h\{f(\emptyset)\}/h\{f(\calX)\}$. Supplementary Appendix~\ref{app:ratio-extensions} gives explicit risk-ratio and odds-ratio formulas in Corollaries~\ref{eg:rr} and~\ref{eg:or}, respectively, and illustrates the construction for causal mediation analysis and balanced full factorial designs with $K=2$.

\section{\revblue{An application to network interference with new results}}
\label{sec:network-policy-effects}

\revblue{Assuming independent Bernoulli experimental design, we apply our general theory to extend the estimands of \citet{hu2022average} under network interference \citep{hudgens2008toward,aronow2017estimating}. Throughout, we focus on the additive scale and assume that higher potential outcome values are preferable and interpret the expected population mean outcome under a policy as its welfare.} \revblue{Specifically, Section~\ref{subsec:network-source-effects} introduces a new hierarchy of source set direct, indirect, and overall effects. The overall effects specialize Definition~\ref{def:pi-estimands-defi}, so their full collection is permutation equivariant and complete by Theorem~\ref{thm:estimand-pi}. Section~\ref{subsec:network-cross-curvature} recovers the estimands of \citet{hu2022average} as first-order aggregates and characterizes the overall effects as mixed derivatives of expected welfare, extending their first-order policy slope interpretation.}
\revblue{Section~\ref{subsec:network-finite-policy} derives a decomposition of finite policy changes through the overall effects. This decomposition recovers the residual-free endpoint decomposition in Definition~\ref{def:residual-free} as a boundary case. Finally, Section~\ref{subsec:network-design-estimation} discusses unbiased design-based estimators when all treatment assignment combination probabilities lie strictly between zero and one, without additional stochastic exchangeability assumptions.}

\subsection{\revblue{Source set direct, indirect, and overall effects}}
\label{subsec:network-source-effects}

\revblue{We consider the network interference setting, and let $Y_i(\bz)$ denote the potential outcome of unit $i\in[N]:=\{1,\ldots,N\}$ under assignment $\bz\in\{0,1\}^N$. The finite-population average welfare is $W_N(\bz)=N^{-1}\sum_{i=1}^N Y_i(\bz)$. We treat the potential outcomes as fixed and let $\bZ$ follow an independent Bernoulli design with treatment probabilities $\bpi=(\pi_1,\ldots,\pi_N)^\top$.} \revblue{A nonempty source set $\calS\subseteq[N]$ consists of the units whose assignments are contrasted. Holding the background assignment outside $\calS$, $\bz_{-\calS}$, fixed, we define $\tau_{i,\calS}(\bz_{-\calS})=\sum_{\calB\subseteq\calS}(-1)^{|\calB|}Y_i(\bone_{\calS\setminus\calB},\bzero_{\calB},\bz_{-\calS})$, where the coordinates in $\calS\setminus\calB$ are set to $1$, those in $\calB$ to $0$, and the remaining coordinates to $\bz_{-\calS}$.} \revblue{For a singleton $\calS=\{k\}$, the contrast is $\tau_{i,\{k\}}(\bz_{-\{k\}})=Y_i(1,\bz_{-\{k\}})-Y_i(0,\bz_{-\{k\}})$, the effect of changing unit $k$'s assignment from control to treatment on unit $i$'s outcome. For a pair $\calS=\{k,\ell\}$, $\tau_{i,\{k,\ell\}}(\bz_{-\{k,\ell\}})=Y_i(1,1,\bz_{-\{k,\ell\}})-Y_i(1,0,\bz_{-\{k,\ell\}})-Y_i(0,1,\bz_{-\{k,\ell\}})+Y_i(0,0,\bz_{-\{k,\ell\}})$ measures how the effect of unit $k$'s assignment varies with unit $\ell$'s assignment (an interaction effect). More generally, when $|\calS|=q$, the contrast isolates the $q$-way interaction among the source units' assignments, conditional on the background assignment outside the source set.}

\revblue{To average over the background assignments, we further define $\theta_{i,\calS}(\bpi)=E_{\bpi_{-\calS}}\{\tau_{i,\calS}(\bZ_{-\calS})\}$. Partitioning the affected outcomes according to whether the receiving unit belongs to the source set gives $\Delta_{\calS}^{\mathrm D}(\bpi)=N^{-1}\sum_{i\in\calS}\theta_{i,\calS}(\bpi)$ and $\Delta_{\calS}^{\mathrm I}(\bpi)=N^{-1}\sum_{i\notin\calS}\theta_{i,\calS}(\bpi)$. Their sum is the overall effect (superscript `O'), $\Delta_{\calS}^{\mathrm O}(\bpi)=\Delta_{\calS}^{\mathrm D}(\bpi)+\Delta_{\calS}^{\mathrm I}(\bpi)=E_{\bpi_{-\calS}}\{\bar\tau_{\calS}(\bZ_{-\calS})\}$, where $\bar\tau_{\calS}(\bZ_{-\calS})=N^{-1}\sum_{i=1}^N \tau_{i,\calS}(\bZ_{-\calS})$. At higher orders, direct (superscript `D') and indirect (superscript `I') refer to outcomes inside and outside $\calS$, respectively. Supplementary Appendix~\ref{subsec:graph-distance-shells} further partitions the indirect effect by graph distance.}

\revblue{Finally, to express the overall effects in the form of Definition~\ref{def:pi-estimands-defi}, set $K=N$ and let $X_k$ represent unit $k$'s treatment. The contrast index $\calY=\{X_k:k\in\calS\}$ then corresponds to the source set $\calS$, while $\calT\subseteq\calY^c$ identifies the action variables outside the source set assigned control. Averaging over these background configurations gives $\Delta_{\calS}^{\mathrm O}(\bpi)
=
\sum_{\calT\subseteq\calY^c}
w_{\bpi}(\calT,\calY)
\left\{
\sum_{\calZ\subseteq\calY}
(-1)^{|\calZ|}f(\calZ\cup\calT)
\right\},$} \revblue{where $f(\calA)=W_N\{\ba(\calA)\}$ is the primitive response and $w_{\bpi}(\calT,\calY)$ is the assignment-induced weight in Equation~\eqref{eq:bernoulli-weights}. The inner sum contrasts assignments within the source set, and the outer sum averages over assignments outside it. Thus, the overall source set effects are instances of the general weighted construction in Definition~\ref{def:pi-estimands-defi}. Because treatment probabilities follow their associated units under relabeling, these weights are permutable. Theorem~\ref{thm:estimand-pi} therefore establishes that the collection of overall source set effects is permutation equivariant.}

\subsection{\revblue{Connection to \citet{hu2022average} and generalization}}
\label{subsec:network-cross-curvature}

\revblue{Summing over source sets of the same size gives an aggregate effect at each interaction order. For $q=1,\ldots,N$ and $r\in\{\mathrm D,\mathrm I,\mathrm O\}$, define $\Gamma_q^r(\bpi)=\sum_{|\calS|=q}\Delta_{\calS}^r(\bpi)$. At first order with $q=1$, each source unit's effect is partitioned into its effect on its own outcome and its effects on other units' outcomes. Aggregating the first-order effects recovers the estimands of \citet{hu2022average}, as shown below.}

\revblue{
\begin{proposition}[\emph{Hu--Li--Wager effects as the first-order case}]\label{prop:hu-first-order}
\revblue{Under independent Bernoulli assignment, $\Gamma_1^{\mathrm D}(\bpi)=\tau_{\mathrm{ADE}}(\bpi)$, $\Gamma_1^{\mathrm I}(\bpi)=\tau_{\mathrm{AIE}}(\bpi)$, and $\Gamma_1^{\mathrm O}(\bpi)=\tau_{\mathrm{AOE}}(\bpi)$, where the right-hand sides denote the average direct, indirect, and overall effects defined in \citet{hu2022average}.}
\end{proposition}
}

\revblue{The higher-order overall effects describe how expected welfare changes locally when treatment probabilities vary jointly. Define expected welfare under policy $\bpi$ by $G_N(\bpi)=E_{\bpi}\{W_N(\bZ)\}$. Under independent Bernoulli assignment, $G_N$ is multilinear in the treatment probabilities. The following result identifies the overall source set effects as its mixed derivatives.}

\revblue{
\begin{proposition}[\emph{Geometric derivative representation}]\label{prop:network-cross-curvature}
\revblue{For every nonempty $\calS\subseteq[N]$,}
\[
\Delta_{\calS}^{\mathrm O}(\bpi)
=
\frac{\partial^{|\calS|}G_N(\bpi)}
{\prod_{k\in\calS}\partial\pi_k}.
\]
Consequently, for each $q=1,\ldots,N$,
\begin{equation}\label{eq:directional-order-q}
\Gamma_q^{\mathrm O}(\bpi)
=
\frac{1}{q!}
\left.
\frac{d^q}{dt^q}G_N(\bpi+t\bone_N)
\right|_{t=0}.
\end{equation}
\end{proposition}
}

\revblue{For a singleton $\calS=\{k\}$, $\Delta_{\{k\}}^{\mathrm O}(\bpi)$ is the slope of expected welfare with respect to $\pi_k$. For a pair $\calS=\{k,\ell\}$, $\Delta_{\{k,\ell\}}^{\mathrm O}(\bpi)=\partial^2G_N(\bpi)/(\partial\pi_k\partial\pi_\ell)$ describes the cross-curvature of the welfare surface. Its sign indicates whether increasing $\pi_\ell$ raises or lowers the marginal welfare effect of increasing $\pi_k$. Equation~\eqref{eq:directional-order-q} gives the corresponding interpretation for a common change in all treatment probabilities. At first order, $\Gamma_1^{\mathrm O}(\bpi)=\tau_{\mathrm{AOE}}(\bpi)$ is the infinitesimal welfare effect in Definition~4 of \citet{hu2022average}; for $q\geq2$, $\Gamma_q^{\mathrm O}(\bpi)$ determines the order-$q$ contribution to the change in expected welfare along the same policy direction.}

\subsection{\revblue{Decomposition of finite policy changes}}
\label{subsec:network-finite-policy}

\revblue{The derivative representation in Proposition \ref{prop:network-cross-curvature} describes how expected welfare changes locally with treatment probabilities. Because expected welfare is multilinear, the full hierarchy yields an exact decomposition of welfare differences between Bernoulli policies, as shown in the following theorem.}

\revblue{
\begin{theorem}[\emph{Decomposition of finite policy changes}]\label{thm:network-policy-expansion}
\revblue{For any $\bpi,\bpi'\in[0,1]^N$,}
\begin{equation}\label{eq:finite-policy-expansion}
G_N(\bpi')-G_N(\bpi)
=
\sum_{\emptyset\neq\calS\subseteq[N]}
\left\{\prod_{k\in\calS}(\pi_k'-\pi_k)\right\}
\Delta_{\calS}^{\mathrm O}(\bpi).
\end{equation}
\end{theorem}}

\revblue{Equation~\eqref{eq:finite-policy-expansion} attributes the welfare difference to changes in individual treatment probabilities and their interactions, with no remainder term. For a common change $\bpi'=\bpi+t\bone_N\in[0,1]^N$, the expansion groups naturally by interaction order:}
\begin{revisionblue}
\begin{equation}\label{eq:common-finite-policy-expansion}
G_N(\bpi+t\bone_N)-G_N(\bpi)
=
\sum_{q=1}^N t^q\Gamma_q^{\mathrm O}(\bpi).
\end{equation}
\end{revisionblue}
\label{subsec:network-endpoint-exposure}\revblue{Taking $\bpi=\bone_N$ and $t=-1$ in Equation~\eqref{eq:common-finite-policy-expansion} yields}
\begin{revisionblue}
$W_N(\bone_N)-W_N(\bzero_N)
=
\sum_{q=1}^N(-1)^{q+1}\Gamma_q^{\mathrm O}(\bone_N)$.
\end{revisionblue}\revblue{Upon close investigation, this is precisely the residual-free decomposition of the endpoint contrast given in Equation~\eqref{eq:residual-free-identity} of Definition~\ref{def:residual-free}. It therefore appears as the all-treated boundary case of the general expansion for finite policy changes.} \revblue{These identities evaluate welfare differences between specified policies and complement the policy learning framework of \citet{viviano2025policy}, which optimizes welfare over feasible policy classes.}

\subsection{\revblue{Design-based estimation}}
\label{subsec:network-design-estimation}

\revblue{The source set effects can be estimated from observed outcomes using the known assignment probabilities. Under independent Bernoulli assignment with $0<\pi_k<1$ for every $k\in[N]$, let $Y_i^{\mathrm{obs}}=Y_i(\bZ)$. For each unit $i\in[N]$ and nonempty source set $\calS\subseteq[N]$, define}
\begin{revisionblue}
\[
\widehat\theta_{i,\calS}
=
Y_i^{\mathrm{obs}}
\prod_{k\in\calS}
\frac{Z_k-\pi_k}{\pi_k(1-\pi_k)}.
\]
\end{revisionblue}\revblue{With the potential outcomes fixed, expectation over the assignment distribution gives $E_{\bpi}(\widehat\theta_{i,\calS})=\theta_{i,\calS}(\bpi)$, as shown in Supplementary Appendix~\ref{app:design-identification}. Applying the same outcome partition as in the estimand construction gives}
\begin{revisionblue}
\[
\widehat\Delta_{\calS}^{\mathrm O}
=\widehat\Delta_{\calS}^{\mathrm D}+\widehat\Delta_{\calS}^{\mathrm I}
=\frac{1}{N}\sum_{i=1}^N\widehat\theta_{i,\calS},
\]
\end{revisionblue}\revblue{where $\widehat\Delta_{\calS}^{\mathrm D}=N^{-1}\sum_{i\in\calS}\widehat\theta_{i,\calS}$ and $\widehat\Delta_{\calS}^{\mathrm I}=N^{-1}\sum_{i\notin\calS}\widehat\theta_{i,\calS}$. These estimators are unbiased for the corresponding direct, indirect, and overall effects. Summing over all source sets of size $q$ likewise gives unbiased estimators of $\Gamma_q^r(\bpi)$.}

\revblue{It is worth mentioning that the interaction order and treatment probabilities can affect estimation precision. We consider the weight multiplying $Y_i^{\mathrm{obs}}$ in $\widehat\theta_{i,\calS}$. For $q=|\calS|$, its second moment is}
\begin{revisionblue}
\[
E_{\bpi}\left[
\left\{
\prod_{k\in\calS}
\frac{Z_k-\pi_k}{\pi_k(1-\pi_k)}
\right\}^{2}
\right]
=
\prod_{k\in\calS}
\frac{1}{\pi_k(1-\pi_k)}
\geq 4^q.
\]
\end{revisionblue}\revblue{Hence, each additional source unit contributes a factor of at least four to this second moment, with a larger factor when its treatment probability is close to zero or one. The resulting large weights can reduce precision, particularly for higher-order effects. Supplementary Appendix~\ref{sec:identification} provides further discussion.}

\section{Concluding remark}\label{sec:conclusion}

\revblue{Permutation equivariance offers a practical criterion for assessing whether a causal estimand remains attached to the scientific quantity it is intended to represent when action variables are relabeled. This matters when labels are arbitrary, as with treatment factors or unordered mediators, because an estimand could otherwise change its attribution solely through a change in notation. Our framework makes this concern explicit and shows how the choice of weights shapes the resulting estimands. For example, in factorial experiments, weights derived from a common assignment distribution can move with the factor labels and are therefore permutable. In causal mediation settings, the connection between weights and the action variables is generally less clear, making invariant weights a more transparent choice. These alternatives do not identify a universally preferred estimand. Rather, they clarify what one must consider when selecting and interpreting a contrast, namely how it averages over configurations of the remaining variables and how it allocates main and interaction effects when applicable. This distinction can help assess whether an existing estimand has a stable interpretation and explain the scientific rationale for their choice. Extending the framework to actions with multiple levels is an important direction for future work.}

\section*{Supplementary Material}
Additional technical details, derivations, and proofs are provided in the Online Supplementary Material.

\section*{Acknowledgement}
This work is supported by the Patient-Centered Outcomes Research Institute\textsuperscript{\textregistered} (PCORI\textsuperscript{\textregistered} Award ME-2023C1-31350). The statements are solely the responsibility of the authors and do not necessarily represent the views of PCORI\textsuperscript{\textregistered}, its Board of Governors or Methodology Committee.
	\par

\begingroup
\spacingset{1}

\endgroup

\clearpage
\pdfbookmark[0]{Supplementary Material}{supplementary-material}
\setcounter{page}{1}
\renewcommand{\thepage}{S\arabic{page}}
\setcounter{section}{0}
\setcounter{subsection}{0}
\setcounter{subsubsection}{0}
\setcounter{equation}{0}
\renewcommand{\thesection}{\Alph{section}}
\renewcommand{\thesubsection}{\thesection.\arabic{subsection}}
\renewcommand{\thesubsubsection}{\thesubsection.\arabic{subsubsection}}
\numberwithin{equation}{section}
\renewcommand{\theHsection}{supp.\arabic{section}}
\renewcommand{\theHsubsection}{\theHsection.\arabic{subsection}}
\renewcommand{\theHsubsubsection}{\theHsubsection.\arabic{subsubsection}}
\renewcommand{\theHequation}{\theHsection.\arabic{equation}}
\setcounter{table}{0}
\setcounter{figure}{0}
\renewcommand{\theHtable}{supp.\arabic{table}}
\renewcommand{\theHfigure}{supp.\arabic{figure}}
\captionsetup[table]{labelformat=addS}
\captionsetup[figure]{labelformat=addS}
\setcounter{theorem}{0}
\renewcommand{\thetheorem}{S\arabic{theorem}}
\providecommand{\theHtheorem}{}
\renewcommand{\theHtheorem}{supp.\arabic{theorem}}
\setcounter{assumption}{0}
\renewcommand{\theassumption}{S\arabic{assumption}}
\providecommand{\theHassumption}{}
\renewcommand{\theHassumption}{supp.\arabic{assumption}}
\setcounter{lemma}{0}
\renewcommand{\thelemma}{S\arabic{lemma}}
\providecommand{\theHlemma}{}
\renewcommand{\theHlemma}{supp.\arabic{lemma}}
\setcounter{corollary}{0}
\renewcommand{\thecorollary}{S\arabic{corollary}}
\providecommand{\theHcorollary}{}
\renewcommand{\theHcorollary}{supp.\arabic{corollary}}
\setcounter{condition}{0}
\renewcommand{\thecondition}{S\arabic{condition}}
\providecommand{\theHcondition}{}
\renewcommand{\theHcondition}{supp.\arabic{condition}}
\setcounter{proposition}{0}
\renewcommand{\theproposition}{S\arabic{proposition}}
\providecommand{\theHproposition}{}
\renewcommand{\theHproposition}{supp.\arabic{proposition}}
\setcounter{definition}{0}
\renewcommand{\thedefinition}{S\arabic{definition}}
\providecommand{\theHdefinition}{}
\renewcommand{\theHdefinition}{supp.\arabic{definition}}
\setcounter{example}{0}
\renewcommand{\theexample}{S\arabic{example}}
\providecommand{\theHexample}{}
\renewcommand{\theHexample}{supp.\arabic{example}}
\setcounter{remark}{0}
\renewcommand{\theremark}{S\arabic{remark}}
\providecommand{\theHremark}{}
\renewcommand{\theHremark}{supp.\arabic{remark}}
\setcounter{prop}{0}
\renewcommand{\theprop}{S\arabic{prop}}
\providecommand{\theHprop}{}
\renewcommand{\theHprop}{supp.\arabic{prop}}
\setcounter{inequality}{0}
\renewcommand{\theinequality}{S\arabic{inequality}}
\providecommand{\theHinequality}{}
\renewcommand{\theHinequality}{supp.\arabic{inequality}}
\setcounter{regularitycondition}{0}
\renewcommand{\theregularitycondition}{S\arabic{regularitycondition}}
\providecommand{\theHregularitycondition}{}
\renewcommand{\theHregularitycondition}{supp.\arabic{regularitycondition}}

% Restore the original supplement theorem styles.
\theoremstyle{definition}
\newtheorem{suppdefinition}[definition]{Definition}
\let\definition\suppdefinition
\expandafter\let\csname enddefinition\expandafter\endcsname\csname endsuppdefinition\endcsname
\newtheorem{suppassumption}[assumption]{Assumption}
\let\assumption\suppassumption
\expandafter\let\csname endassumption\expandafter\endcsname\csname endsuppassumption\endcsname
\newtheorem{suppexample}[example]{Example}
\let\example\suppexample
\expandafter\let\csname endexample\expandafter\endcsname\csname endsuppexample\endcsname
\newtheorem{suppremark}[remark]{Remark}
\let\remark\suppremark
\expandafter\let\csname endremark\expandafter\endcsname\csname endsuppremark\endcsname
\newtheorem{suppcondition}[condition]{Condition}
\let\condition\suppcondition
\expandafter\let\csname endcondition\expandafter\endcsname\csname endsuppcondition\endcsname
\newtheorem{suppinequality}[inequality]{Inequality}
\let\inequality\suppinequality
\expandafter\let\csname endinequality\expandafter\endcsname\csname endsuppinequality\endcsname
\newtheorem{suppregularitycondition}[regularitycondition]{Regularity condition}
\let\regularitycondition\suppregularitycondition
\expandafter\let\csname endregularitycondition\expandafter\endcsname\csname endsuppregularitycondition\endcsname

\theoremstyle{plain}
\makeatletter
\newcommand*{\section@cntformat}{Appendix~\thesection\quad}
\renewcommand*{\@seccntformat}[1]{\@ifundefined{#1@cntformat}{\csname the#1\endcsname\quad}{\csname #1@cntformat\endcsname}}
\makeatother
\allowdisplaybreaks
\spacingset{1.5}
\begin{center}
{\large\bfseries Supplementary Material\par}
\vspace{0.55em}
{\large\bfseries On the permutation equivariance principle for causal estimands\par}
\vspace{0.55em}
{\normalsize Jiaqi Tong$^{1}$ and Fan Li$^{1,\ast}$\par}
\vspace{0.15em}
{\small $^{1}$Yale School of Public Health\par}
\end{center}
\smallskip
The Supplementary Material is organized as follows. Appendix~\ref{app:algebraic-proofs} contains the proofs of the algebraic results. Appendix~\ref{app:small-k} presents explicit representations for $K=2$ and $K=3$, with numerical calculations illustrating how weight schemes determine the estimands and how both behave under relabeling. Appendix~\ref{app:group-primer} explains group actions, orbits, quotient sets, and row and column permutations, with worked row-multiset calculations for the introductory examples.

Appendix~\ref{app:policy-effects} develops the higher-order policy hierarchy under network interference, including its first-order special case, mixed-derivative interpretation, exact expansion for finite policy changes, and endpoint decomposition. Appendix~\ref{app:design-identification} derives unbiased estimators under independent Bernoulli assignment and decomposes indirect effects by graph distance.

Appendix~\ref{app:controlled-histories} treats partially ordered action variables. Appendix~\ref{app:dimension-reduction} examines exposure mappings and marginal structural models, and Appendix~\ref{sec:identification} discusses identification and estimation. Appendix~\ref{sec:application} presents the factorial and mediation applications, while Appendix~\ref{app:algorithm} gives the estimand permutation algorithm. Appendix~\ref{app:ratio-extensions} develops the ratio-scale extensions, and Appendix~\ref{app:application-comparison} compares the three applications.

\clearpage
\begingroup
\spacingset{1}
\hypersetup{linkcolor=black,linktoc=all}
\setcounter{tocdepth}{2}
\etocsettocstyle{\section*{\contentsname}}{}
\etocsetlocaltop{part}
\makeatletter
\renewcommand{\@pnumwidth}{2.5em}
\renewcommand{\@tocrmarg}{3.5em}
\renewcommand*{\l@subsection}{\@dottedtocline{2}{1.5em}{3.5em}}
\makeatother
\footnotesize
\localtableofcontents
\endgroup
\clearpage
% Supplementary appendices
\section{Proofs of Theorems, Propositions, and Corollaries}\label{app:algebraic-proofs}

We begin with three lemmas. The first expresses each weighted estimand and the residual as linear combinations of the primitive responses.

\begin{lemma}\label{lemma:alt-repre-cY-supp}
For every nonempty $\calY\subseteq\calX$, the weighted contrast admits the coordinate representation
\begin{equation}\label{eq:cy-alternative-supp}
\Delta_{\calY}
=
\sum_{\calZ\subseteq\calX}
(-1)^{|\calZ\cap\calY|}
w(\calZ\setminus\calY,\calY)f(\calZ).
\end{equation}
Writing $\text{res}=\operatorname{res}(f)$ for the residual defined in Section~\ref{ss:w-IEP} of the main manuscript, we obtain
\[
\text{res}
=
\sum_{\calZ\subseteq\calX}\text{coef}(\calZ)f(\calZ),
\]
where
\begin{equation}\label{eq:residual-effect-weights-supp}
\text{coef}(\calZ)
=
\begin{cases}
\displaystyle
\sum_{\emptyset\neq\calY\subseteq\calX}
(-1)^{|\calY|+|\calZ\cap\calY|+1}
w(\calZ\setminus\calY,\calY),
& \emptyset\neq\calZ\subsetneq\calX,\\[6pt]
\displaystyle
\sum_{\emptyset\neq\calY\subseteq\calX}
(-1)^{|\calY|+1}w(\emptyset,\calY)-1,
& \calZ=\emptyset,\\[6pt]
\displaystyle
1-\sum_{\emptyset\neq\calY\subseteq\calX}w(\calY^c,\calY),
& \calZ=\calX.
\end{cases}
\end{equation}
\end{lemma}

\begin{proof}
Every subset $\calZ\subseteq\calX$ has a unique decomposition $\calZ=\calT\cup\calZ'$, where $\calT=\calZ\setminus\calY\subseteq\calY^c$ and $\calZ'=\calZ\cap\calY\subseteq\calY$. Reindexing the weighted contrast in Equation~\eqref{eq:estimandsclass-unweighted} of the main manuscript therefore gives
\begin{align*}
\Delta_{\calY}
&=
\sum_{\calT\subseteq\calY^c}w(\calT,\calY)
\sum_{\calZ'\subseteq\calY}(-1)^{|\calZ'|}f(\calT\cup\calZ')\\
&=
\sum_{\calZ\subseteq\calX}
(-1)^{|\calZ\cap\calY|}
w(\calZ\setminus\calY,\calY)f(\calZ).
\end{align*}

Substituting Equation~\eqref{eq:cy-alternative-supp} into the definition of the residual and interchanging the finite sums yields
\begin{align*}
\text{res}
&=
\sum_{\calZ\subseteq\calX}
\left\{
\sum_{\emptyset\neq\calY\subseteq\calX}
(-1)^{|\calY|+|\calZ\cap\calY|+1}
w(\calZ\setminus\calY,\calY)
\right\}f(\calZ)-\{f(\emptyset)-f(\calX)\}.
\end{align*}
Subtracting the endpoint contrast reduces the coefficient of $f(\emptyset)$ by one and increases the coefficient of $f(\calX)$ by one. For $\calZ=\calX$, the sign in the inner sum is always $-1$. These observations give the three cases in Equation~\eqref{eq:residual-effect-weights-supp}.
\end{proof}

The second lemma is the weighted inclusion--exclusion principle \citep[Lemma~1,  p.~550]{bjorklund2009set}. Its element weights $\omega_X$ are distinct from the estimand weights $w$.

\begin{lemma}[\emph{Weighted Inclusion--Exclusion Principle}]\label{lemma:IEP-weighted}
Assign a real weight $\omega_X$ to each $X\in\calX$ and define the total weight of a subset $\calA\subseteq\calX$ by $\omega(\calA)=\sum_{X\in\calA}\omega_X$. For any subsets $\calG_1,\ldots,\calG_S\subseteq\calX$,
\[
\omega\left(\bigcap_{s=1}^{S}\calG_s\right)
=
\omega(\calX)
+
\sum_{s=1}^{S}(-1)^s
\sum_{1\leq i_1<\cdots<i_s\leq S}
\omega\left(\calG_{i_1}^{c}\cap\cdots\cap\calG_{i_s}^{c}\right),
\]
where $\calG_s^c=\calX\setminus\calG_s$.
\end{lemma}

The final lemma shows that a permutation of the action variables induces a bijection of their power set.

\begin{lemma}\label{lemma:surjective-power-set}
For $\sigma\in S_K$, define $\sigma^\ast(\calA)=\{\sigma(X):X\in\calA\}$ for each $\calA\subseteq\calX$. Then $\sigma^\ast:2^\calX\to2^\calX$ is a bijection.
\end{lemma}

\begin{proof}
Applying $\sigma^{-1}$ to each element of $\sigma^\ast(\calA)$ recovers $\calA$, so $\sigma^\ast$ is injective. To establish surjectivity, take any $\calB\subseteq\calX$ and set $\calA=\{\sigma^{-1}(X):X\in\calB\}$. Then $\calA\subseteq\calX$ and $\sigma^\ast(\calA)=\calB$.
\end{proof}

\subsection{Proof of Theorem~\ref{thm:characterization}}
\begin{proof}

Fix $\sigma\in\frakG$ and write $\mH'=\mH\mathbf{P}_c(\sigma)$. For necessity, Definition~\ref{def:permutation-invariance} provides a row permutation matrix $\mathbf{P}_r(\sigma)$ satisfying $\mathbf{P}_r(\sigma)\mH=\mH'$. Since permuting rows preserves their values and multiplicities,
\[
\mathcal R(\mH')
=
\mathcal R\{\mathbf{P}_r(\sigma)\mH\}
=
\mathcal R(\mH).
\]

For sufficiency, suppose $\mathcal R(\mH')=\mathcal R(\mH)$. Equality of these multisets ensures that each row of $\mH'$ can be matched to an equal row of $\mH$, with each row occurrence used exactly once. This bijection defines a row permutation matrix $\mathbf{P}_r(\sigma)$ satisfying $\mathbf{P}_r(\sigma)\mH=\mH'=\mH\mathbf{P}_c(\sigma)$. Since $\sigma$ was arbitrary, the contrast vector is equivariant under $\frakG$.

\end{proof}
\subsection{Proof of Theorem~\ref{thm:estimand-pi}}
\begin{proof}

Fix $\sigma\in\frakG$ and a nonempty subset $\calY\subseteq\calX$. Let $\sigma^\ast(\calA)=\{\sigma(X):X\in\calA\}$ denote the induced relabeling of subsets. This map is a bijection on $2^{\calX}$ by Lemma~\ref{lemma:surjective-power-set} and preserves cardinalities, intersections, and set differences.

We show that $\Delta_{\calY}^{\sigma^{-1}}=\Delta_{\sigma^\ast(\calY)}$. Under the relabeling convention of the main manuscript, applying $\sigma^{-1}$ replaces $f(\calA)$ with $f(\sigma^\ast(\calA))$. For permutable weights, it also replaces $w(\calT,\calY)$ with $w(\sigma^\ast(\calT),\sigma^\ast(\calY))$.

For permutable weights, Lemma~\ref{lemma:alt-repre-cY-supp} and the change of variable $\calB=\sigma^\ast(\calZ)$ give
\begin{align*}
\Delta_{\calY}^{\sigma^{-1}}
&=\sum_{\calZ\subseteq\calX}
(-1)^{|\calZ\cap\calY|}
w(\sigma^\ast(\calZ\setminus\calY),\sigma^\ast(\calY))
f(\sigma^\ast(\calZ))\\
&=\sum_{\calB\subseteq\calX}
(-1)^{|\calB\cap\sigma^\ast(\calY)|}
w(\calB\setminus\sigma^\ast(\calY),\sigma^\ast(\calY))
f(\calB)\\
&=\Delta_{\sigma^\ast(\calY)}.
\end{align*}
The second equality follows because $\sigma^\ast(\calZ\setminus\calY)=\calB\setminus\sigma^\ast(\calY)$ and $|\calZ\cap\calY|=|\calB\cap\sigma^\ast(\calY)|$.

For invariant weights, relabeling acts only on the response surface, so
\[
\Delta_{\calY}^{\sigma^{-1}}
=
\sum_{\calZ\subseteq\calX}
(-1)^{|\calZ\cap\calY|}
w(\calZ\setminus\calY,\calY)
f(\sigma^\ast(\calZ)).
\]
The orbit condition in Theorem~\ref{thm:estimand-pi}(i) gives $w(\sigma^\ast(\calT),\sigma^\ast(\calY))=w(\calT,\calY)$ for every admissible pair $(\calT,\calY)$. Taking $\calT=\calZ\setminus\calY$ therefore converts this expression into the first line of the preceding calculation. Hence, $\Delta_{\calY}^{\sigma^{-1}}=\Delta_{\sigma^\ast(\calY)}$ also holds for invariant weights satisfying the orbit condition.

Thus, under either weight specification, relabeling by $\sigma$ carries the component indexed by $\calY$ to the component indexed by $\sigma^\ast(\calY)$. Since relabeling preserves subset size, each order-$q$ subvector is equivariant for $q=1,\ldots,K$. The full vector is equivariant and complete because it contains a component for every nonempty subset of $\calX$.

\end{proof}

\begin{remark}\label{remark:df-pi-inva}
Under the full symmetric group $S_K$, any two disjoint pairs $(\calT,\calY)$ with the same cardinalities $|\calT|$ and $|\calY|$ belong to the same orbit. Invariant weights satisfying the orbit condition therefore depend only on these cardinalities. For each contrast order $q=1,\ldots,K$, the reference set can have size $0,\ldots,K-q$, giving $K-q+1$ weight parameters subject to one normalization constraint. The total degrees of freedom are thus
\[
\mathrm{df}
=
\sum_{q=1}^{K}\{(K-q+1)-1\}
=
\frac{K(K-1)}{2}.
\]
The number of free weight parameters therefore grows quadratically in $K$, whereas the complete estimand vector has $2^K-1$ components.
\end{remark}
\subsection{Proof of Theorem~\ref{prop:nie}}
\begin{proof}

For sufficiency, set $w(\calT,\calY)=\mathbf{1}(\calT=\emptyset)$. For each $\calZ\subseteq\calX$, the coefficient of $f(\calZ)$ in the inclusion--exclusion sum is
\[
\sum_{\substack{\emptyset\neq\calY\subseteq\calX\\\calZ\subseteq\calY}}
(-1)^{|\calY|+|\calZ|+1}.
\]
By the binomial identity $\sum_{r=0}^{m}\binom{m}{r}(-1)^r=0$ for $m\geq1$, this coefficient equals $1$ for $\calZ=\emptyset$, $-1$ for $\calZ=\calX$, and $0$ for every nonempty proper subset. The sum therefore equals $f(\emptyset)-f(\calX)$ for every response surface.

For necessity, hold the weights fixed and suppose that the residual-free identity holds for every primitive response surface $f$. The coordinate representation in Lemma~\ref{lemma:alt-repre-cY-supp} then has zero coefficients for every primitive response value. In particular, matching the coefficient of $f(\calX)$ yields
\[
\sum_{\emptyset\neq\calY\subseteq\calX}w(\calY^c,\calY)=1.
\]
Normalization gives $w(\emptyset,\calX)=1$, so nonnegativity forces $w(\calY^c,\calY)=0$ for every nonempty proper subset $\calY$. Thus, all weights with nonempty $\calT$ and $\calY$ satisfying $\calT\cap\calY=\emptyset$ and $|\calT\cup\calY|=K$ vanish. This establishes the base case for downward induction on $|\calT\cup\calY|$.

For the induction step, let $2\leq m<K$ and suppose that all such weights vanish when $|\calT\cup\calY|>m$. Fix $\calZ\subseteq\calX$ with $|\calZ|=m$. For every $\calY\supseteq\calZ$ and every nonempty $\calT\subseteq\calY^c$, we have $|\calT\cup\calY|>m$. The induction hypothesis and normalization therefore imply $w(\emptyset,\calY)=1$, including $w(\emptyset,\calZ)=1$.

The coefficient identity for $f(\calZ)$ is
\[
0=\sum_{\emptyset\neq\calY\subseteq\calX}
(-1)^{|\calY|+|\calZ\cap\calY|+1}
w(\calZ\setminus\calY,\calY).
\]
If $\calY$ neither contains nor is contained in $\calZ$, then $\calZ\setminus\calY$ is nonempty and $|\calZ\cup\calY|>m$, so the corresponding term vanishes by the induction hypothesis. For strict supersets $\calY\supsetneq\calZ$, the weights equal one, and their total contribution is
\[
\sum_{\calZ\subsetneq\calY\subseteq\calX}(-1)^{|\calY|+m+1}
=
\sum_{r=1}^{K-m}\binom{K-m}{r}(-1)^{r+1}
=1.
\]
For the remaining terms, indexed by $\emptyset\neq\calY\subseteq\calZ$, the sign is $-1$. Hence,
\begin{align*}
0
&=1-\sum_{\emptyset\neq\calY\subseteq\calZ}w(\calZ\setminus\calY,\calY)\\
&=-\sum_{\emptyset\neq\calY\subsetneq\calZ}w(\calZ\setminus\calY,\calY),
\end{align*}
where the second equality uses $w(\emptyset,\calZ)=1$. Nonnegativity forces every remaining summand to vanish. Taking $\calZ=\calT\cup\calY$ covers every pair of nonempty disjoint sets with $|\calT\cup\calY|=m$, completing the induction step.

Since two nonempty disjoint sets have union size at least two, the induction shows that $w(\calT,\calY)=0$ whenever $\calT\neq\emptyset$. Normalization then gives $w(\emptyset,\calY)=1$ for every nonempty $\calY$. This establishes uniqueness for the nonnegative, normalized invariant weights in Theorem~\ref{prop:nie}.

\end{proof}
\subsection{Proof of Proposition \ref{prop:alternative-repre}}
\begin{proof}
Fix a nonempty $\calY\subseteq\calX$ and $X_k\in\calY$. We first establish the identity
\begin{align*}
    \sum_{\calZ\subseteq \calY\backslash\{X_k\}}(-1)^{|\calZ|}(f(\calZ)-f(\calZ\cup\{X_k\}))=\sum_{\calZ\subseteq \calY}(-1)^{|\calZ|}f(\calZ).
\end{align*}
The left-hand side can be written as
\begin{align*}
    \text{LHS}=&\sum_{\calZ\subseteq \calY\backslash\{X_k\}}(-1)^{|\calZ|}f(\calZ)-\sum_{\calZ\subseteq \calY\backslash\{X_k\}}(-1)^{|\calZ|}f(\calZ\cup\{X_k\})\\
    =&\sum_{\calZ\subseteq \calY,X_k\notin \calZ}(-1)^{|\calZ|}f(\calZ)+\sum_{\calZ'\subseteq \calY,X_k\in \calZ'}(-1)^{|\calZ'|}f(\calZ')\\
    =&\sum_{\calZ\subseteq \calY}(-1)^{|\calZ|}f(\calZ),
\end{align*}
where the second equality follows from the change of variable $\calZ'=\calZ\cup\{X_k\}$.

For each $\calT\subseteq\calY^c$, apply this identity to the shifted response surface $f_{\calT}(\calA)=f(\calA\cup\calT)$. Multiplying by $w(\calT,\calY)$ and summing over $\calT\subseteq\calY^c$ gives the stated representation.
\end{proof}
\subsection{Proof of Corollaries \ref{eg:rr} and \ref{eg:or}}\label{app:ratio-proofs}
\begin{proof}
Both corollaries follow by applying Theorem~\ref{prop:nie} to the transformed response surface $\log(h\circ f)$, wherever the transformation is defined. Under the residual-free point mass weights, the estimands on the log scale satisfy
\begin{align*}
    \log(\Delta_{\calY})=&\sum_{\calZ\subseteq \calY}(-1)^{|\calZ|}\log(h\circ f(\calZ))\\
    =&\log\left[ \prod_{\calZ\subseteq \calY}\left(h\circ f(\calZ)\right)^{(-1)^{|\calZ|}}\right].
\end{align*}
The inclusion--exclusion identity on the log scale gives
\begin{align*}
    \sum_{\emptyset\neq \calY \subseteq \calX}(-1)^{|\calY|+1}\log(\Delta_{\calY})=\log(h\circ f(\emptyset))-\log(h\circ f(\calX)).
\end{align*}
Equivalently,
\begin{align*}
   \log\left(\prod_{\emptyset\neq \calY \subseteq \calX}\Delta_{\calY}^{(-1)^{|\calY|+1}}\right)=\log\left(\frac{h\circ f(\emptyset)}{h\circ f(\calX)}\right).
\end{align*}
Exponentiating gives Corollary~\ref{eg:rr} with $h(x)=x$ and Corollary~\ref{eg:or} with $h(x)=x/(1-x)$.
\end{proof}

\section{Representations for $K=2$ and $K=3$}\label{app:small-k}

We first show how the matrix formulation in Section~\ref{subsec:setup} of the main manuscript accommodates permutable weights by expanding the response vector to include products of weights and primitive responses. We give the general construction below, followed by explicit representations for $K=2$ and $K=3$. Appendix~\ref{app:worked-example} evaluates the estimands under different weight choices using the numerical response surface from the main manuscript. Background on permutation matrices is provided in Appendix~\ref{app:group-primer}.

For each nonempty $\calY\subseteq\calX$, $\calT\subseteq\calY^c$, and $\calZ\subseteq\calY$, define
\[
v^{\mathrm{aug}}_{\calY,\calT,\calZ}
=
w(\calT,\calY)f(\calT\cup\calZ),
\]
where $\calY$ identifies the variables being contrasted, while $\calT$ and $\calZ$ identify the variables assigned state $0$ outside and inside $\calY$, respectively. Collecting these coordinates in a fixed order gives a vector $\mathbf v_{\mathrm{aug}}$ of length $m=2^K(2^K-1)$. The estimands in Definition~\ref{def:pi-estimands-defi} then satisfy
\[
\boldsymbol\Delta
=
\mathbf H_{\mathrm{aug}}\mathbf v_{\mathrm{aug}},
\]
where $\mathbf H_{\mathrm{aug}}$ is a fixed $(2^K-1)\times m$ matrix. Its row indexed by $\calY$ assigns coefficient $(-1)^{|\calZ|}$ to each coordinate indexed by $(\calY,\calT,\calZ)$ and zero to coordinates whose first index differs from $\calY$. Thus, the weights enter the expanded response vector, while the contrast matrix contains only the coefficients $0$, $1$, and $-1$.

Relabeling the weights and primitive responses together moves the coordinate indexed by $(\calY,\calT,\calZ)$ to the coordinate indexed by $(\sigma(\calY),\sigma(\calT),\sigma(\calZ))$. Let $\mathbf P_{c,\mathrm{aug}}(\sigma)$ denote this coordinate permutation, and let $\mathbf P_r(\sigma)$ move the estimand indexed by $\calY$ to the position indexed by $\sigma(\calY)$. Since relabeling preserves $|\calZ|$, it preserves the coefficient attached to each term and gives
\[
\mathbf H_{\mathrm{aug}}\mathbf P_{c,\mathrm{aug}}(\sigma)
=
\mathbf P_r(\sigma)\mathbf H_{\mathrm{aug}}.
\]
The expanded representation therefore satisfies the matrix criterion in Definition~\ref{def:permutation-invariance}, and the row-multiset characterization follows by the argument of Theorem~\ref{thm:characterization}. The following example illustrates this construction for $K=2$ and compares it with the representation for invariant weights.

\begin{example}\label{eg:weighted-estimands}
\revblue{For $K=2$, permutable weights can be incorporated into the response vector by collecting products of weights and primitive responses. Relabeling then acts on both quantities together:}
\begin{align*}
\bv=\bigl(&w(\emptyset,\{X_1\})f(\emptyset),w(\emptyset,\{X_1\})f(\{X_1\}),\\
&w(\{X_2\},\{X_1\})f(\{X_2\}),w(\{X_2\},\{X_1\})f(\{X_1,X_2\}),\\
&w(\emptyset,\{X_2\})f(\emptyset),w(\emptyset,\{X_2\})f(\{X_2\}),\\
&w(\{X_1\},\{X_2\})f(\{X_1\}),w(\{X_1\},\{X_2\})f(\{X_1,X_2\}),\\
&w(\emptyset,\{X_1,X_2\})f(\emptyset),w(\emptyset,\{X_1,X_2\})f(\{X_1\}),\\
&w(\emptyset,\{X_1,X_2\})f(\{X_2\}),w(\emptyset,\{X_1,X_2\})f(\{X_1,X_2\})\bigr)^\top.
\end{align*}
Let $\bm\Delta=(\Delta_{\{X_1\}},\Delta_{\{X_2\}},\Delta_{\{X_1,X_2\}})^\top$. The estimands in Definition~\ref{def:pi-estimands-defi} then admit the linear representation in Equation~\eqref{eq:def-contrast} of the main manuscript:
\[
\bm\Delta=\mathbf{H}\bv=
\begin{pmatrix}
\mathbf{e}_1^\top & \mathbf{0} & \mathbf{0} \\
\mathbf{0} & \mathbf{e}_1^\top & \mathbf{0} \\
\mathbf{0} & \mathbf{0} & \mathbf{e}_2^\top
\end{pmatrix}\bv,
\]
where $\mathbf{e}_1=(1,-1,1,-1)^\top$, $\mathbf{e}_2=(1,-1,-1,1)^\top$, and each $\mathbf{0}$ denotes a row of four zeros.

\revblue{For invariant weights, the weights enter $\mathbf{H}$, while $\bv$ contains the four primitive responses ordered as $\emptyset$, $\{X_1\}$, $\{X_2\}$, and $\calX$. With equal weights on the two states of the variable outside each first-order contrast,}
\[
\bm\Delta=\mathbf{H}\bv=
\begin{pmatrix}
\frac{1}{2} & -\frac{1}{2} & \frac{1}{2} & -\frac{1}{2} \\
\frac{1}{2} & \frac{1}{2} & -\frac{1}{2} & -\frac{1}{2} \\
1 & -1 & -1 & 1
\end{pmatrix}\bv.
\]
\revblue{Exchanging $X_1$ and $X_2$ exchanges columns two and three of $\mathbf{H}$. The same matrix is obtained by exchanging its first two rows. Thus, the main effects exchange positions and the interaction remains unchanged, satisfying the equivariance criterion in Theorem~\ref{thm:characterization}.}
\end{example}

We illustrate the framework for $K=2$ and $K=3$, using the notation from Example~\ref{eg2} of the main manuscript. We write $\{k_1,\ldots,k_r\}$ for the set of action variables $\{X_{k_1},\ldots,X_{k_r}\}$. In factorial experiments, a common joint distribution of the action variables can generate weights for all contrasts. The following remark formalizes this compatibility.

\begin{remark}\label{remark:compatibility}
The weights $w(\calT,\calY)$ are \emph{compatible across contrast subsets} if they arise from a common probability mass function $\widetilde{w}:2^\calX\to\mathbb{R}_{\geq0}$, with $\sum_{\calA\subseteq\calX}\widetilde{w}(\calA)=1$. Specifically, for every nonempty $\calY\subseteq\calX$ and every $\calT\subseteq\calY^c$, they satisfy
\[
w(\calT,\calY)
=
\sum_{\calZ\subseteq\calY}\widetilde{w}(\calZ\cup\calT).
\]
Thus, each weight is the marginal probability of a reference configuration for the variables outside the contrast.
\end{remark}

\begin{example}[$K=2$]\label{example:K=2-practice}
Holding the variable outside each contrast in state $1$ gives the unweighted estimands
\begin{align*}
\Delta_{\lb1\rb}&=\mu(1,1)-\mu(0,1),\\
\Delta_{\lb2\rb}&=\mu(1,1)-\mu(1,0),\\
\Delta_{\lb1,2\rb}&=\mu(1,1)-\mu(0,1)-\mu(1,0)+\mu(0,0).
\end{align*}
Their inclusion--exclusion sum recovers the endpoint contrast: $\Delta_{\lb1\rb}+\Delta_{\lb2\rb}-\Delta_{\lb1,2\rb}=\mu(1,1)-\mu(0,0)$.

More generally, averaging each first-order contrast over the other variable gives
\begin{align*}
\Delta_{\lb1\rb}
&=w(\emptyset,\lb1\rb)\{\mu(1,1)-\mu(0,1)\}\\
&\quad+w(\lb2\rb,\lb1\rb)\{\mu(1,0)-\mu(0,0)\},\\
\Delta_{\lb2\rb}
&=w(\emptyset,\lb2\rb)\{\mu(1,1)-\mu(1,0)\}\\
&\quad+w(\lb1\rb,\lb2\rb)\{\mu(0,1)-\mu(0,0)\},\\
\Delta_{\lb1,2\rb}
&=\mu(1,1)-\mu(0,1)-\mu(1,0)+\mu(0,0),
\end{align*}
where the weights are nonnegative and satisfy $w(\emptyset,\lb1\rb)+w(\lb2\rb,\lb1\rb)=1$ and $w(\emptyset,\lb2\rb)+w(\lb1\rb,\lb2\rb)=1$. The interaction remains unchanged because it involves both variables.

For compatible permutable weights, let $\widetilde{w}(a_1,a_2)$ denote the probability of the binary configuration $(a_1,a_2)$ under the common reference distribution. Marginalizing this distribution gives
\begin{align*}
w(\emptyset,\lb1\rb)&=\widetilde{w}(0,1)+\widetilde{w}(1,1),\\
w(\lb2\rb,\lb1\rb)&=\widetilde{w}(0,0)+\widetilde{w}(1,0),\\
w(\emptyset,\lb2\rb)&=\widetilde{w}(1,1)+\widetilde{w}(1,0),\\
w(\lb1\rb,\lb2\rb)&=\widetilde{w}(0,1)+\widetilde{w}(0,0),
\end{align*}
where $\widetilde{w}(0,0)+\widetilde{w}(0,1)+\widetilde{w}(1,0)+\widetilde{w}(1,1)=1$. This reference distribution is relabeled together with the action variables.

For invariant weights under the full permutation group $S_2$, permutation equivariance requires the additional equalities $w(\emptyset,\lb1\rb)=w(\emptyset,\lb2\rb)$ and $w(\lb2\rb,\lb1\rb)=w(\lb1\rb,\lb2\rb)$. In either construction, exchanging the labels of $X_1$ and $X_2$ exchanges the first-order effects and preserves the interaction:
\[
\Delta^\dagger_{\lb1\rb}=\Delta_{\lb2\rb},
\qquad
\Delta^\dagger_{\lb2\rb}=\Delta_{\lb1\rb},
\qquad
\Delta^\dagger_{\lb1,2\rb}=\Delta_{\lb1,2\rb}.
\]

For general weights, the residual $\text{res}=\Delta_{\lb1\rb}+\Delta_{\lb2\rb}-\Delta_{\lb1,2\rb}-\{\mu(1,1)-\mu(0,0)\}$ has the coordinate representation
\begin{align*}
\text{res}
={}&\text{coef}(\emptyset)\mu(1,1)
+\text{coef}(\lb1\rb)\mu(0,1)\\
&+\text{coef}(\lb2\rb)\mu(1,0)
+\text{coef}(\lb1,2\rb)\mu(0,0),
\end{align*}
where the coefficients are
\begin{align*}
\text{coef}(\emptyset)&=w(\emptyset,\lb1\rb)+w(\emptyset,\lb2\rb)-2,\\
\text{coef}(\lb1\rb)&=-w(\emptyset,\lb1\rb)+w(\lb1\rb,\lb2\rb)+1,\\
\text{coef}(\lb2\rb)&=w(\lb2\rb,\lb1\rb)-w(\emptyset,\lb2\rb)+1,\\
\text{coef}(\lb1,2\rb)&=-\{w(\lb2\rb,\lb1\rb)+w(\lb1\rb,\lb2\rb)\}.
\end{align*}
\end{example}

\begin{example}[$K=3$]\label{example:K=3-practice}
For three action variables, holding all variables outside each contrast in state $1$ gives seven unweighted estimands:
\begin{align*}
\Delta_{\lb1\rb}&=\mu(1,1,1)-\mu(0,1,1),\\
\Delta_{\lb2\rb}&=\mu(1,1,1)-\mu(1,0,1),\\
\Delta_{\lb3\rb}&=\mu(1,1,1)-\mu(1,1,0),\\
\Delta_{\lb1,2\rb}&=\mu(1,1,1)-\mu(0,1,1)-\mu(1,0,1)+\mu(0,0,1),\\
\Delta_{\lb1,3\rb}&=\mu(1,1,1)-\mu(0,1,1)-\mu(1,1,0)+\mu(0,1,0),\\
\Delta_{\lb2,3\rb}&=\mu(1,1,1)-\mu(1,0,1)-\mu(1,1,0)+\mu(1,0,0),\\
\Delta_{\lb1,2,3\rb}&=\mu(1,1,1)-\mu(0,1,1)-\mu(1,0,1)-\mu(1,1,0)\\
&\quad+\mu(0,0,1)+\mu(0,1,0)+\mu(1,0,0)-\mu(0,0,0).
\end{align*}
These components yield the inclusion--exclusion decomposition of the endpoint contrast:
\begin{align*}
\mu(1,1,1)-\mu(0,0,0)
&=\sum_{j=1}^3\Delta_{\lb j\rb}
-\Delta_{\lb1,2\rb}-\Delta_{\lb1,3\rb}-\Delta_{\lb2,3\rb}+\Delta_{\lb1,2,3\rb}.
\end{align*}

Averaging each contrast over the variables outside it gives the weighted estimands
\[
\adjustbox{max width=\linewidth}{$\displaystyle\begin{aligned}
    \Delta_{\lb1\rb}=&w(\emptyset,\lb1\rb)(\mu(1,1,1)-\mu(0,1,1))+w(\lb2\rb,\lb1\rb)(\mu(1,0,1)-\mu(0,0,1))+\\
    &w(\lb3\rb,\lb1\rb)(\mu(1,1,0)-\mu(0,1,0))+w(\lb2,3\rb,\lb1\rb)(\mu(1,0,0)-\mu(0,0,0)),\\
    \Delta_{\lb2\rb}=&w(\emptyset,\lb2\rb)(\mu(1,1,1)-\mu(1,0,1))+w(\lb1\rb,\lb2\rb)(\mu(0,1,1)-\mu(0,0,1))+\\
    &w(\lb3\rb,\lb2\rb)(\mu(1,1,0)-\mu(1,0,0))+w(\lb1,3\rb,\lb2\rb)(\mu(0,1,0)-\mu(0,0,0)),\\
    \Delta_{\lb3\rb}=&w(\emptyset,\lb3\rb)(\mu(1,1,1)-\mu(1,1,0))+w(\lb1\rb,\lb3\rb)(\mu(0,1,1)-\mu(0,1,0))+\\
    &w(\lb2\rb,\lb3\rb)(\mu(1,0,1)-\mu(1,0,0))+w(\lb1,2\rb,\lb3\rb)(\mu(0,0,1)-\mu(0,0,0)),\\
    \Delta_{\lb1,2\rb}=&w(\emptyset,\lb1,2\rb)(\mu(1,1,1)-\mu(0,1,1)-\mu(1,0,1)+\mu(0,0,1))+\\
    &w(\lb3\rb,\lb1,2\rb)(\mu(1,1,0)-\mu(0,1,0)-\mu(1,0,0)+\mu(0,0,0)),\\
    \Delta_{\lb1,3\rb}=&w(\emptyset,\lb1,3\rb)(\mu(1,1,1)-\mu(0,1,1)-\mu(1,1,0)+\mu(0,1,0))+\\
    &w(\lb2\rb,\lb1,3\rb)(\mu(1,0,1)-\mu(0,0,1)-\mu(1,0,0)+\mu(0,0,0)),\\
    \Delta_{\lb2,3\rb}=&w(\emptyset,\lb2,3\rb)(\mu(1,1,1)-\mu(1,0,1)-\mu(1,1,0)+\mu(1,0,0))+\\
    &w(\lb1\rb,\lb2,3\rb)(\mu(0,1,1)-\mu(0,0,1)-\mu(0,1,0)+\mu(0,0,0)),\\
    \Delta_{ \lb1,2,3\rb}=&\mu(1,1,1)-\mu(0,1,1)-\mu(1,0,1)-\mu(1,1,0)+\\
    &\mu(0,0,1)+\mu(0,1,0)+\mu(1,0,0)-\mu(0,0,0),
\end{aligned}$}
\]
where the weights are nonnegative and normalized such that
\begin{align*}
    &w(\emptyset,\lb1\rb)+w(\lb2\rb,\lb1\rb)+w(\lb3\rb,\lb1\rb)+w(\lb2,3\rb,\lb1\rb)=1,\\
    &w(\emptyset,\lb2\rb)+w(\lb1\rb,\lb2\rb)+w(\lb3\rb,\lb2\rb)+w(\lb1,3\rb,\lb2\rb)=1,\\
    &w(\emptyset,\lb3\rb)+w(\lb1\rb,\lb3\rb)+w(\lb2\rb,\lb3\rb)+w(\lb1,2\rb,\lb3\rb)=1,\\
    &w(\emptyset,\lb1,2\rb)+w(\lb3\rb,\lb1,2\rb)=1,~~w(\emptyset,\lb1,3\rb)+w(\lb2\rb,\lb1,3\rb)=1,\\
    &w(\emptyset,\lb2,3\rb)+w(\lb1\rb,\lb2,3\rb)=1.
\end{align*}
For compatible permutable weights, a common probability mass function $\widetilde{w}$ on the trivariate binary configurations generates all weights by marginalization:
\[
\adjustbox{max width=\linewidth}{$\displaystyle\begin{aligned}
    &w(\emptyset,\lb1\rb)=\widetilde{w}(0,1,1)+\widetilde{w}(1,1,1),~~~w(\lb2\rb,\lb1\rb)=\widetilde{w}(0,0,1)+\widetilde{w}(1,0,1),\\
    &w(\lb3\rb,\lb1\rb)=\widetilde{w}(0,1,0)+\widetilde{w}(1,1,0),~~~w(\lb2,3\rb,\lb1\rb)=\widetilde{w}(0,0,0)+\widetilde{w}(1,0,0),\\[0.5ex]
    &w(\emptyset,\lb2\rb)=\widetilde{w}(1,1,1)+\widetilde{w}(1,0,1),~~~w(\lb1\rb,\lb2\rb)=\widetilde{w}(0,1,1)+\widetilde{w}(0,0,1),\\
    &w(\lb3\rb,\lb2\rb)=\widetilde{w}(1,1,0)+\widetilde{w}(1,0,0),~~~w(\lb1,3\rb,\lb2\rb)=\widetilde{w}(0,1,0)+\widetilde{w}(0,0,0),\\[0.5ex]
    &w(\emptyset,\lb3\rb)=\widetilde{w}(1,1,0)+\widetilde{w}(1,1,1),~~~w(\lb1\rb,\lb3\rb)=\widetilde{w}(0,1,0)+\widetilde{w}(0,1,1),\\
    &w(\lb2\rb,\lb3\rb)=\widetilde{w}(1,0,0)+\widetilde{w}(1,0,1),~~~w(\lb1,2\rb,\lb3\rb)=\widetilde{w}(0,0,0)+\widetilde{w}(0,0,1),\\[0.5ex]
    &w(\emptyset,\lb1,2\rb)=\widetilde{w}(0,0,1)+\widetilde{w}(0,1,1)+\widetilde{w}(1,0,1)+\widetilde{w}(1,1,1),\\
    &w(\lb3\rb,\lb1,2\rb)=\widetilde{w}(0,0,0)+\widetilde{w}(0,1,0)+\widetilde{w}(1,0,0)+\widetilde{w}(1,1,0),\\[0.5ex]
    &w(\emptyset,\lb1,3\rb)=\widetilde{w}(0,1,0)+\widetilde{w}(0,1,1)+\widetilde{w}(1,1,0)+\widetilde{w}(1,1,1),\\
    &w(\lb2\rb,\lb1,3\rb)=\widetilde{w}(0,0,0)+\widetilde{w}(0,0,1)+\widetilde{w}(1,0,0)+\widetilde{w}(1,0,1),\\[0.5ex]
    &w(\emptyset,\lb2,3\rb)=\widetilde{w}(1,0,0)+\widetilde{w}(1,0,1)+\widetilde{w}(1,1,0)+\widetilde{w}(1,1,1),\\
    &w(\lb1\rb,\lb2,3\rb)=\widetilde{w}(0,0,0)+\widetilde{w}(0,0,1)+\widetilde{w}(0,1,0)+\widetilde{w}(0,1,1),
\end{aligned}$}
\]
where the probability mass function $\widetilde{w}$ satisfies
\begin{align*}
    &\widetilde{w}(0,0,0)+\widetilde{w}(1,0,0)+\widetilde{w}(0,1,0)+\widetilde{w}(0,0,1)+\widetilde{w}(1,1,0)+\\
    &\widetilde{w}(1,0,1)+\widetilde{w}(0,1,1)+\widetilde{w}(1,1,1)=1.
\end{align*}
For invariant weights under the full group $S_3$, the orbit condition in Theorem~\ref{thm:estimand-pi}(i) gives
\[
\adjustbox{max width=\linewidth}{$\displaystyle\begin{aligned}
    &w(\emptyset,\{1\})=w(\emptyset,\{2\})=w(\emptyset,\{3\}),\\
    &w(\{2\},\{1\})=w(\{1\},\{2\})=w(\{1\},\{3\})=w(\{3\},\{1\})=w(\{2\},\{3\})=w(\{3\},\{2\}),\\
    &w(\{2,3\},\{1\})=w(\{1,3\},\{2\})=w(\{1,2\},\{3\}),\\
    &w(\emptyset,\{1,2\})=w(\emptyset,\{1,3\})=w(\emptyset,\{2,3\}),\\
    &w(\{3\},\{1,2\})=w(\{2\},\{1,3\})=w(\{1\},\{2,3\}).
\end{aligned}$}
\]

To verify equivariance, consider the cyclic relabeling $\sigma(1)=2$, $\sigma(2)=3$, and $\sigma(3)=1$. Relabeling the response surface and the permutable weights, or keeping invariant weights that satisfy the preceding restrictions, gives
\begin{align*}
    &\Delta^\dagger_{\lb1\rb}=\Delta_{\lb3\rb},~~~\Delta^\dagger_{\lb2\rb}=\Delta_{\lb1\rb},~~~\Delta_{\lb3\rb}^\dagger=\Delta_{\lb2\rb},\\
    &\Delta^\dagger_{\lb1,2\rb}=\Delta_{\lb1,3\rb},~~~\Delta^\dagger_{\lb1,3\rb}=\Delta_{\lb2,3\rb},~~~\Delta^\dagger_{\lb2,3\rb}=\Delta_{\lb1,2\rb},~~~\Delta^\dagger_{\lb1,2,3\rb}=\Delta_{\lb1,2,3\rb}.
\end{align*}
The residual effect for the weighted estimands, $\text{res}=\sum_{j=1}^3\Delta_{\lb j\rb}-\Delta_{\lb1,2\rb}-\Delta_{\lb1,3\rb}-\Delta_{\lb2,3\rb}+\Delta_{ \lb1,2,3\rb}-(\mu(1,1,1)-\mu(0,0,0))$, can be written as
\[
\adjustbox{max width=\linewidth}{$\displaystyle\begin{aligned}
    \text{res}=&\text{coef}(\emptyset)\mu(1,1,1)+\text{coef}(\lb1\rb)\mu(0,1,1)+\text{coef}(\lb2\rb)\mu(1,0,1)+\text{coef}(\lb3\rb)\mu(1,1,0)+\\
    &\text{coef}(\lb1,2\rb)\mu(0,0,1)+\text{coef}(\lb1,3\rb)\mu(0,1,0)+\text{coef}(\lb2,3\rb)\mu(1,0,0)+\text{coef}(\lb1,2,3\rb)\mu(0,0,0),
\end{aligned}$}
\]
where the coefficients are
\[
\adjustbox{max width=\linewidth}{$\displaystyle\begin{aligned}
    \text{coef}(\emptyset)=&w(\emptyset,\lb1\rb)+w(\emptyset,\lb2\rb)+w(\emptyset,\lb3\rb)-w(\emptyset,\lb1,2\rb)-w(\emptyset,\lb1,3\rb)-w(\emptyset,\lb2,3\rb),\\
    \text{coef}(\lb1\rb)=&-w(\emptyset,\lb1\rb)+w(\lb1\rb,\lb2\rb)+w(\lb1\rb,\lb3\rb)+w(\emptyset,\lb1,2\rb)+\\
                         &w(\emptyset,\lb1,3\rb)-w(\lb1\rb,\lb2,3\rb)-1,\\
    \text{coef}(\lb2\rb)=&-w(\emptyset,\lb2\rb)+w(\lb2\rb,\lb1\rb)+w(\lb2\rb,\lb3\rb)+w(\emptyset,\lb1,2\rb)+\\
                         &w(\emptyset,\lb2,3\rb)-w(\lb2\rb,\lb1,3\rb)-1,\\
    \text{coef}(\lb3\rb)=&-w(\emptyset,\lb3\rb)+w(\lb3\rb,\lb2\rb)+w(\lb3\rb,\lb1\rb)+w(\emptyset,\lb2,3\rb)+\\
                         &w(\emptyset,\lb1,3\rb)-w(\lb3\rb,\lb1,2\rb)-1,\\
    \text{coef}(\lb1,2\rb)=&-w(\lb2\rb,\lb1\rb)-w(\lb1\rb,\lb2\rb)+w(\lb1,2\rb,\lb3\rb)-w(\emptyset,\lb1,2\rb)+\\
                           &w(\lb2\rb,\lb1,3\rb)+w(\lb1\rb,\lb2,3\rb)+1,\\
    \text{coef}(\lb1,3\rb)=&-w(\lb3\rb,\lb1\rb)-w(\lb1\rb,\lb3\rb)+w(\lb1,3\rb,\lb2\rb)-w(\emptyset,\lb1,3\rb)+\\
                           &w(\lb3\rb,\lb1,2\rb)+w(\lb1\rb,\lb2,3\rb)+1,\\
    \text{coef}(\lb2,3\rb)=&-w(\lb2\rb,\lb3\rb)-w(\lb3\rb,\lb2\rb)+w(\lb2,3\rb,\lb1\rb)-w(\emptyset,\lb2,3\rb)+\\
                           &w(\lb2\rb,\lb1,3\rb)+w(\lb3\rb,\lb1,2\rb)+1,\\
    \text{coef}(\lb1,2,3\rb)=&-\left(w(\lb2,3\rb,\lb1\rb)+w(\lb1,3\rb,\lb2\rb)+w(\lb1,2\rb,\lb3\rb)+w(\lb3\rb,\lb1,2\rb)+\right.\\
                             &\left.w(\lb2\rb,\lb1,3\rb)+w(\lb1\rb,\lb2,3\rb)\right).
\end{aligned}$}
\]

\end{example}

\subsection{\revblue{A numerical illustration of weighted estimands}}
\label{app:worked-example}

\revblue{We use the response surface $\mu(0,0)=0$, $\mu(1,0)=2$, $\mu(0,1)=1$, and $\mu(1,1)=10$ from Section~\ref{sec:whatispi} to illustrate how weights determine the estimands and how these estimands behave under relabeling.}

\paragraph{\revblue{Invariant estimand weights.}}
\revblue{Let $\lambda\in[0,1]$ be the common weight assigned to state $1$ when averaging over the variable outside the contrast. Changing $X_1$ from state $0$ to state $1$ gives an effect of $10-1=9$ when $X_2$ is in state $1$ and $2-0=2$ when it is in state $0$. Averaging these differences, and proceeding similarly for $X_2$, gives}
\begin{revisionblue}
\begin{align*}
\Delta_{\{X_1\}}(\lambda)&=\lambda(10-1)+(1-\lambda)(2-0)=2+7\lambda,\\
\Delta_{\{X_2\}}(\lambda)&=\lambda(10-2)+(1-\lambda)(1-0)=1+7\lambda,\\
\Delta_{\{X_1,X_2\}}(\lambda)&=10-1-2+0=7.
\end{align*}
\end{revisionblue}\revblue{The interaction involves both variables, so there is no remaining variable to average over and it does not depend on $\lambda$. Table~S\ref{tab:reference-weight-example} compares three reference choices: (1) fixing the other variable at state $0$, (2) averaging equally over both states, and (3) fixing it at state $1$. Each choice yields a permutation equivariant system; that is, exchanging the variable labels while keeping $\lambda$ fixed preserves the effect attributed to each original variable.}

\begin{table}[htbp]

\centering
\setlength{\belowcaptionskip}{5pt}
\renewcommand{\baselinestretch}{1}\small
\renewcommand{\arraystretch}{1.15}
\caption{ Estimands under invariant weights. The reference choice specifies the state of the other action variable. The last column gives $\Delta_{\{X_1\}}+\Delta_{\{X_2\}}-\Delta_{\{X_1,X_2\}}$.}
\label{tab:reference-weight-example}
\begin{tabular}{@{}lcccc@{}}
\toprule
Reference choice & $\lambda$ & $\Delta_{\{X_1\}}$ & $\Delta_{\{X_2\}}$ & Inclusion--exclusion sum \\
\midrule
Fixed at state $0$ & $0$ & $2$ & $1$ & $-4$ \\
Equally weighted & $1/2$ & $5.5$ & $4.5$ & $3$ \\
Fixed at state $1$ & $1$ & $9$ & $8$ & $10$ \\
\bottomrule
\end{tabular}
\end{table}

\revblue{The inclusion--exclusion sum is $-4+14\lambda$, which recovers the endpoint contrast $\mu(1,1)-\mu(0,0)=10$ at $\lambda=1$. This choice holds the variable outside the contrast in state $1$. More generally, Theorem~\ref{prop:nie} establishes that, within the nonnegative, normalized invariant-weight class, point mass weights at $\calT=\emptyset$ uniquely yield the residual-free identity for every response surface.}

\paragraph{\revblue{Permutable estimand weights.}}
\revblue{Now take the weights from independent Bernoulli assignments with $(\pi_1,\pi_2)=(1/4,3/4)$. Each first-order effect averages over the other variable's assignment, such that the effect of $X_1$ uses $\pi_2$, and the effect of $X_2$ uses $\pi_1$. In other words,}
\begin{revisionblue}
\[
\Delta_{\{X_1\}}=2+7\pi_2=7.25,
\qquad
\Delta_{\{X_2\}}=1+7\pi_1=2.75.
\]
\end{revisionblue}\revblue{Exchanging the variable labels also exchanges their assignment probabilities, giving $(3/4,1/4)$ in the new labeling, and exchanges the response values $\mu(1,0)=2$ and $\mu(0,1)=1$. Applying the same averaging rule to the relabeled quantities gives}
\begin{revisionblue}
\[
\Delta_{\{X_1\}}^\dagger=\Delta_{\{X_2\}}=2.75,
\qquad
\Delta_{\{X_2\}}^\dagger=\Delta_{\{X_1\}}=7.25.
\]
\end{revisionblue}\revblue{The interaction remains $7$, so each effect follows the original variable or pair of variables it represents. If the probabilities are instead left at their original label positions, the effects become $\Delta_{\{X_1\}}^\dagger=1+7(3/4)=6.25$ and $\Delta_{\{X_2\}}^\dagger=2+7(1/4)=3.75$. These values are not a permutation of the original estimands. Thus, permutable weights must be relabeled together with the action variables to preserve permutation equivariance.}

\section{\revblue{A primer on group actions and permutation equivariance}}\label{app:group-primer}

\revblue{This appendix complements Sections~\ref{subsec:primitive-response} and~\ref{subsec:setup} of the main manuscript by providing some preliminary notation and concepts that are needed for understanding the abstract discussion of the permutation equivariance principle. We summarize the notation, distinguish response configurations from estimand indices, and explain the group action and matrix formulation through explicit examples.}

\subsection{\revblue{Notation and the indexing of response surfaces and estimands}}\label{app:group-primer-notation}

\revblue{The framework describes causal comparisons involving $K\geq2$ action variables, denoted by $\calX=\{X_1,\ldots,X_K\}$, each with two states labeled $0$ (control) and $1$ (treatment). The interpretation of an action variable depends on the application. In a factorial experiment, it represents a treatment factor. In causal mediation, its state selects whether a mediator takes its potential value under treatment $0$ or $1$. In a network experiment, it represents the treatment assigned to a unit, so $K$ equals the number of units. The binary states describe the conditions being compared, while outcomes and mediator values may be measured on other scales.}

\revblue{We index a response configuration by the subset $\calA\subseteq\calX$ of action variables in state $0$. The corresponding binary vector $\ba(\calA)$ has $a_k(\calA)=0$ when $X_k\in\calA$ and $a_k(\calA)=1$ otherwise. Thus, $\calA=\emptyset$ represents the configuration with all variables in state $1$, and $\calA=\calX$ represents the configuration with all variables in state $0$. The power set $2^{\calX}$ consists of all subsets of $\calX$ and indexes the $2^K$ possible configurations. The primitive response surface $f:2^{\calX}\to\mathbb R$ assigns a response value $f(\calA)$ to each configuration. In the causal applications considered here, this value is a mean potential outcome.}

\revblue{Subsets also index estimands. In $\Delta_{\calY}$, the nonempty subset $\calY$ identifies the variables whose states are jointly contrasted. The variables outside $\calY$ are held fixed within each contrast, and the resulting contrasts are averaged over their reference configurations. For each nonempty $\calY\subseteq\calX$, the weighted estimand is}
\begin{revisionblue}
\[
\Delta_{\calY}
=
\sum_{\calT\subseteq\calY^c}
w(\calT,\calY)
\left\{
\sum_{\calZ\subseteq\calY}
(-1)^{|\calZ|}f(\calZ\cup\calT)
\right\},
\]
\end{revisionblue}
\revblue{where $\calY^c=\calX\setminus\calY$, and the weights satisfy $w(\calT,\calY)\geq0$ and $\sum_{\calT\subseteq\calY^c}w(\calT,\calY)=1$. Within the inner sum, $\calZ$ identifies the variables in $\calY$ assigned state $0$, while $\calT$ identifies the variables outside $\calY$ held at state $0$. All variables outside $\calZ\cup\calT$ take state $1$, so $\calA=\calZ\cup\calT$ indexes the complete response configuration in that term. The inner sum forms a contrast of order $|\calY|$, and the outer sum averages it over the reference configurations specified by $\calT$. Once $f$ and the weights are specified, each nonempty $\calY$ therefore indexes one estimand. Table~S\ref{tab:group-primer-notation} summarizes these symbols and the notation for relabelings used below.}

\begin{table}[htbp]

\centering
\renewcommand{\baselinestretch}{1}\small
\renewcommand{\arraystretch}{1.15}
\caption{Summary of notation used in the main manuscript.}
\label{tab:group-primer-notation}
\begin{tabular}{@{}p{0.27\linewidth}p{0.68\linewidth}@{}}
\toprule
Symbol & Meaning \\
\midrule
$\calX=\{X_1,\ldots,X_K\}$ & Set of $K$ action variables. \\
$2^{\calX}$, $\calA$ & The power set of $\calX$ and a subset $\calA\subseteq\calX$. \\
$\ba(\calA)$ & Binary configuration with $a_k(\calA)=0$ if and only if $X_k\in\calA$. \\
$f(\calA)$ & Primitive response surface under configuration $\ba(\calA)$. \\
$\calY$, $\Delta_{\calY}$ & A nonempty subset of action variables and its associated contrast of order $|\calY|$. \\
$\calT$, $w(\calT,\calY)$ & Reference subset of action variables outside $\calY$, $\calT\subseteq\calY^c$, and its nonnegative standardization weight, where $\calY^c=\calX\setminus\calY$. For each $\calY$, the weights sum to one over $\calT$. \\
$S_K$, $\frakG$ & The group of all permutations of $\calX$ and its subgroup of scientifically admissible relabelings. \\
$\sigma(\calA)$ & Subset $\{\sigma(X):X\in\calA\}$ obtained by applying $\sigma\in\frakG$ to each member of $\calA$. \\
$2^{\calX}/\frakG$ & Quotient set whose elements are the distinct orbits of subsets under $\frakG$. \\
$[q]$, $\calQ$ & Under $\frakG=S_K$, the cardinality class $[q]=\{\calA\subseteq\calX:|\calA|=q\}$ and the quotient set $\calQ=\{[0],\ldots,[K]\}$. \\
$\bv$ & Vector of the $2^K$ primitive responses in a fixed ordering of subsets. \\
$\mathbf{H}$, $\bm\Delta$ & A $d\times2^K$ contrast matrix and the vector of $d$ estimands $\bm\Delta=\mathbf{H}\bv$. \\
$\mathbf{P}_c(\sigma)$ & A $2^K\times2^K$ permutation matrix acting on the primitive response coordinates, which correspond to the columns of $\mathbf{H}$. \\
$\mathbf{P}_r(\sigma)$ & A $d\times d$ permutation matrix acting on the estimand coordinates, which correspond to the rows of $\mathbf{H}$. \\
\bottomrule
\end{tabular}
\end{table}

\revblue{To illustrate these distinctions, consider a $2^3$ factorial experiment with $N$ units and three treatment factors $\calX=\{X_1,X_2,X_3\}$. For each factor combination $(a_1,a_2,a_3)\in\{0,1\}^3$, define the mean potential outcome}
\begin{revisionblue}
\[
\mu(a_1,a_2,a_3)
=
\frac{1}{N}\sum_{i=1}^N Y_i(a_1,a_2,a_3),
\]
\end{revisionblue}
\revblue{where $Y_i(a_1,a_2,a_3)$ is the potential outcome of unit $i$ under that combination. Under the design-based formulation, these potential outcomes are fixed, and randomness arises from treatment assignment. The primitive response is $f(\calA)=\mu\{a_1(\calA),a_2(\calA),a_3(\calA)\}$. For example, $f(\emptyset)=\mu(1,1,1)$, $f(\{X_1\})=\mu(0,1,1)$, and $f(\{X_1,X_3\})=\mu(0,1,0)$. Each of these quantities describes one factor combination.}

\revblue{The singleton $\calY=\{X_1\}$ indexes the main effect of $X_1$. Here, $\calY^c=\{X_2,X_3\}$, so the possible reference subsets are $\emptyset$, $\{X_2\}$, $\{X_3\}$, and $\{X_2,X_3\}$. These correspond to the outside factor states $(a_2,a_3)=(1,1)$, $(0,1)$, $(1,0)$, and $(0,0)$, respectively. Expanding the general definition gives}
\begin{revisionblue}
\[
\begin{aligned}
\Delta_{\{X_1\}}
={}&
w(\emptyset,\{X_1\})
\{\mu(1,1,1)-\mu(0,1,1)\}\\
&+
w(\{X_2\},\{X_1\})
\{\mu(1,0,1)-\mu(0,0,1)\}\\
&+
w(\{X_3\},\{X_1\})
\{\mu(1,1,0)-\mu(0,1,0)\}\\
&+
w(\{X_2,X_3\},\{X_1\})
\{\mu(1,0,0)-\mu(0,0,0)\}.
\end{aligned}
\]
\end{revisionblue}
\revblue{Each difference changes $X_1$ from state $0$ to state $1$ while holding $X_2$ and $X_3$ fixed. For example, $\calT=\{X_2\}$ fixes $X_2=0$ and $X_3=1$, giving the difference $f(\{X_2\})-f(\{X_1,X_2\})=\mu(1,0,1)-\mu(0,0,1)$. Thus, $\Delta_{\{X_1\}}$ averages four treatment contrasts, whereas $f(\{X_1\})=\mu(0,1,1)$ is a single response value. The same singleton identifies the factor being contrasted in the former ($\Delta_{\{X_1\}}$) and the factor assigned state $0$ in the latter ($f(\{X_1\})$).}

\revblue{The weights specify how these four comparisons contribute to the main effect. In a balanced full factorial experiment, equal weights $w(\calT,\{X_1\})=1/4$ average the four differences equally. Alternatively, suppose $X_2$ and $X_3$ are independently assigned state $1$ with probabilities $2/3$ and $3/4$. Averaging over their assignment distribution gives weights $1/2$, $1/4$, $1/6$, and $1/12$ for $\calT=\emptyset$, $\{X_2\}$, $\{X_3\}$, and $\{X_2,X_3\}$, respectively. For instance, $w(\{X_2\},\{X_1\})=\Pr(X_2=0,X_3=1)=(1/3)(3/4)=1/4$. A further choice, $w(\emptyset,\{X_1\})=1$ with all other weights zero, gives $\Delta_{\{X_1\}}=\mu(1,1,1)-\mu(0,1,1)$, the main effect with both outside factors held at state $1$.}

\revblue{For the pair $\calY=\{X_1,X_2\}$, the estimand is a second-order contrast that measures how the effect of $X_1$ changes between the two states of $X_2$. The only outside variable is $X_3$, so $\calT$ is either $\emptyset$ or $\{X_3\}$. The corresponding estimand is}
\begin{revisionblue}
\[
\begin{aligned}
\Delta_{\{X_1,X_2\}}
={}&
w(\emptyset,\{X_1,X_2\})
\bigl[
\{\mu(1,1,1)-\mu(0,1,1)\}
-
\{\mu(1,0,1)-\mu(0,0,1)\}
\bigr]\\
&+
w(\{X_3\},\{X_1,X_2\})
\bigl[
\{\mu(1,1,0)-\mu(0,1,0)\}
-
\{\mu(1,0,0)-\mu(0,0,0)\}
\bigr].
\end{aligned}
\]
\end{revisionblue}
\revblue{The first bracket is a difference in differences with $X_3=1$, and the second is the corresponding contrast with $X_3=0$. Equal weighting assigns weight $1/2$ to each bracket. Under the assignment distribution above, the weights are instead $\Pr(X_3=1)=3/4$ and $\Pr(X_3=0)=1/4$. With equal weights, this second-order contrast is twice the factorial interaction effect under the normalization of \citet{dasgupta2015causal}, consistent with the convention in the main manuscript.}

\revblue{The complete system for $K=3$ contains seven estimands, indexed by the three singleton subsets, the three pairs, and the full set $\calX$. For the third-order contrast indexed by $\calY=\calX$, there are no outside variables, so the only reference subset is $\calT=\emptyset$ and its weight is one. These seven contrasts are formed from the eight primitive response values. Thus, response configurations and estimands use subsets of the same action set, with $\calA$ specifying a complete configuration, $\calY$ specifying the variables jointly contrasted, and $\calT$ specifying the reference states of the remaining variables.}

\subsection{\revblue{Group actions, orbits, and quotient sets}}\label{app:group-primer-orbits}

\revblue{Let $\frakG\leq S_K$ denote the group of scientifically admissible relabelings. Each permutation $\sigma\in\frakG$ relabels a subset $\calA\subseteq\calX$ by replacing each variable with its new label:}
\begin{revisionblue}
\[
\sigma(\calA)=\{\sigma(X):X\in\calA\}.
\]
\end{revisionblue}\revblue{This defines a \emph{group action} because the identity leaves every subset unchanged, and successive relabelings agree with their composition. For example, exchanging $X_1$ and $X_2$ maps $\{X_1\}$ to $\{X_2\}$ while leaving $\emptyset$ and $\{X_1,X_2\}$ unchanged.}

\revblue{The \emph{orbit} of $\calA$ contains all subsets reachable from $\calA$ by admissible relabelings:}
\begin{revisionblue}
\[
\operatorname{Orb}_{\frakG}(\calA)
=
\{\sigma(\calA):\sigma\in\frakG\}.
\]
\end{revisionblue}\revblue{Every subset belongs to exactly one orbit, so the orbits partition $2^{\calX}$. The \emph{quotient set} treats each distinct orbit as a single element:}
\begin{revisionblue}
\[
2^{\calX}/\frakG
=
\{\operatorname{Orb}_{\frakG}(\calA):\calA\subseteq\calX\}.
\]
\end{revisionblue}

\revblue{Under the full group $S_K$, two subsets belong to the same orbit exactly when they have the same size, as permutations preserve size, and any two subsets of equal size can be relabeled into one another. The orbits are therefore the cardinality classes $[q]=\{\calA\subseteq\calX:|\calA|=q\}$ for $q=0,\ldots,K$, giving the quotient set $\calQ=\{[0],\ldots,[K]\}$. For $K=2$,}
\begin{revisionblue}
\[
[0]=\{\emptyset\},
\qquad
[1]=\bigl\{\{X_1\},\{X_2\}\bigr\},
\qquad
[2]=\bigl\{\{X_1,X_2\}\bigr\}.
\]
\end{revisionblue}\revblue{Thus, the power set has four elements, while its quotient under $S_2$ has three. The subsets $\{X_1\}$ and $\{X_2\}$ share an orbit because exchanging their labels maps one to the other; their response values $f(\{X_1\})$ and $f(\{X_2\})$ may differ.}

\revblue{Restricting the admissible relabelings can split a cardinality class into several orbits. For example, with $K=3$, suppose $X_1$ and $X_2$ may exchange labels while $X_3$ remains fixed. The singleton subsets then form two orbits, $\bigl\{\{X_1\},\{X_2\}\bigr\}$ and $\bigl\{\{X_3\}\bigr\}$, because no admissible relabeling maps $X_3$ to either of the other variables.}
\subsection{\revblue{Row and column permutation matrices}}\label{app:group-primer-matrices}

\revblue{A permutation matrix rearranges the coordinates of a vector. It is square, with exactly one entry equal to $1$ in each row and each column and all remaining entries equal to $0$. Fix an ordering of the subsets of $\calX$ for the coordinates of $\bv$. For $\calA,\calB\subseteq\calX$, define the matrix representing $\sigma$ by}
\begin{revisionblue}
\[
\{\mathbf{P}_c(\sigma)\}_{\calA,\calB}
=
\mathbf{1}\{\calA=\sigma(\calB)\}.
\]
\end{revisionblue}\revblue{This matrix moves the response at $\calB$ to the coordinate labeled $\sigma(\calB)$. The value at coordinate $\calA$ after relabeling is therefore the original value at $\sigma^{-1}(\calA)$:}
\begin{revisionblue}
\[
\{\mathbf{P}_c(\sigma)\bv\}_{\calA}
=
f\{\sigma^{-1}(\calA)\}.
\]
\end{revisionblue}\revblue{The correspondence $\sigma\mapsto\mathbf{P}_c(\sigma)$ is called a \emph{permutation representation}, and composing relabelings corresponds to composing their matrix actions.}

\revblue{The terms \emph{column permutation} and \emph{row permutation} describe how these matrices act on the contrast matrix $\mathbf{H}$. Its columns correspond to primitive responses, so right multiplication by $\mathbf{P}_c(\sigma)$ permutes its columns. Its rows correspond to estimands, so left multiplication by $\mathbf{P}_r(\sigma)$ permutes its rows. The matrices $\mathbf{P}_c(\sigma)$ and $\mathbf{P}_r(\sigma)$ have dimensions $2^K\times2^K$ and $d\times d$, respectively. Both act on their corresponding vectors by left multiplication, giving $\mathbf{P}_c(\sigma)\bv$ and $\mathbf{P}_r(\sigma)\bm\Delta$.}

\revblue{As explained after Definition 1 in Section 3.2 of the main manuscript, permutation equivariance means that relabeling the primitive responses before forming contrasts gives the same result as forming the contrasts first and then relabeling them. Thus, for each admissible $\sigma$, there must be a row permutation $\mathbf{P}_r(\sigma)$ such that, for every $\bv\in\mathbb{R}^{2^K}$,}
\begin{revisionblue}
\[
\mathbf{H}\{\mathbf{P}_c(\sigma)\bv\}
=
\mathbf{P}_r(\sigma)\{\mathbf{H}\bv\}.
\]
\end{revisionblue}\revblue{Because this equality holds for every response vector, it is equivalent to the \emph{intertwining relation} used to characterize Definition 1 in the main manuscript:}
\begin{revisionblue}
\[
\mathbf{P}_r(\sigma)\mathbf{H}
=
\mathbf{H}\mathbf{P}_c(\sigma).
\]
\end{revisionblue}
\subsection{\revblue{A worked example with two mediators}}\label{app:group-primer-example}

\revblue{Consider causal mediation with a binary treatment and two causally unordered mediators, $X_1$ and $X_2$. For $k=1,2$, let $X_k(a)$ denote the potential value of mediator $k$ under treatment $a\in\{0,1\}$. We hold treatment at $1$ in the outcome and compare configurations of the mediator potential values. For each $(a_1,a_2)\in\{0,1\}^2$, define}
\begin{revisionblue}
\[
\mu(a_1,a_2)
=
E\!\left[Y\{1,X_1(a_1),X_2(a_2)\}\right],
\]
\end{revisionblue}
\revblue{where $Y(1,x_1,x_2)$ is the potential outcome under treatment $1$ with the mediators set to values $x_1$ and $x_2$. Thus, each binary action state selects the treatment condition under which a mediator's potential value is taken.}

\revblue{Under the subset notation, $\calA$ identifies the mediators whose potential values are taken under treatment $0$, while the remaining mediators take their potential values under treatment $1$. Order the primitive responses as}
\begin{revisionblue}
\[
\begin{aligned}
\bv
&=
\{f(\emptyset),f(\{X_1\}),f(\{X_2\}),f(\{X_1,X_2\})\}^{\top}\\
&=
\{\mu(1,1),\mu(0,1),\mu(1,0),\mu(0,0)\}^{\top}.
\end{aligned}
\]
\end{revisionblue}

\revblue{Choose the point mass weights $w(\calT,\calY)=\mathbf{1}(\calT=\emptyset)$. These weights hold each mediator outside the contrasted subset $\calY$ at its potential value under treatment $1$. The resulting contrast vector is}
\begin{revisionblue}
\[
\bm\Delta
=
\begin{pmatrix}
\Delta_{\{X_1\}}\\
\Delta_{\{X_2\}}\\
\Delta_{\{X_1,X_2\}}
\end{pmatrix}
=
\underbrace{
\begin{pmatrix}
1&-1&0&0\\
1&0&-1&0\\
1&-1&-1&1
\end{pmatrix}}_{\mathbf{H}}
\bv.
\]
\end{revisionblue}
\revblue{The first row gives $\Delta_{\{X_1\}}=\mu(1,1)-\mu(0,1)$, the natural indirect main effect through $X_1$ with $X_2$ held at $X_2(1)$. The second row gives $\Delta_{\{X_2\}}=\mu(1,1)-\mu(1,0)$, the corresponding effect through $X_2$ with $X_1$ held at $X_1(1)$. The third row gives $\Delta_{\{X_1,X_2\}}=\mu(1,1)-\mu(0,1)-\mu(1,0)+\mu(0,0)$. This interaction measures how the indirect contrast through $X_1$ changes when the reference value of $X_2$ changes from $X_2(0)$ to $X_2(1)$.}

\revblue{The system is residual-free because its inclusion--exclusion sum recovers the natural indirect effect}
\begin{revisionblue}
\[
\Delta_{\{X_1\}}
+
\Delta_{\{X_2\}}
-
\Delta_{\{X_1,X_2\}}
=
\mu(1,1)-\mu(0,0),
\]
\end{revisionblue}
\revblue{where the right-hand side compares both mediators at their potential values under treatment $1$ with both at their potential values under treatment $0$, holding treatment at $1$ in the outcome.}

\revblue{Now let $\sigma$ exchange the mediator labels $X_1$ and $X_2$ together with their associated potential outcome coordinates. Under the ordering above, this relabeling exchanges the second and third primitive coordinates and the first and second estimand coordinates. It is represented by}
\begin{revisionblue}
\[
\mathbf{P}_c(\sigma)
=
\begin{pmatrix}
1&0&0&0\\
0&0&1&0\\
0&1&0&0\\
0&0&0&1
\end{pmatrix},
\qquad
\mathbf{P}_r(\sigma)
=
\begin{pmatrix}
0&1&0\\
1&0&0\\
0&0&1
\end{pmatrix}.
\]
\end{revisionblue}

\revblue{Direct multiplication gives}
\begin{revisionblue}
\[
\mathbf{P}_r(\sigma)\mathbf{H}
=
\begin{pmatrix}
1&0&-1&0\\
1&-1&0&0\\
1&-1&-1&1
\end{pmatrix}
=
\mathbf{H}\mathbf{P}_c(\sigma).
\]
\end{revisionblue}
\revblue{Thus, relabeling the mediators before forming the contrasts gives the same result as forming the contrasts first and then relabeling their coordinates. The two natural indirect main effects exchange positions, and the interaction involving both mediators remains unchanged. The main effects may have different values, but each continues to represent the same mediator after relabeling. In the row-multiset characterization of Theorem~\ref{thm:characterization}, the same calculation shows that permuting the columns preserves the collection of contrast rows, including their multiplicities.}
\subsection{The row-multiset criterion for the introductory examples}\label{app:row-multiset-examples}

\begin{example}\label{eg:test-thm1}
The three contrast matrices below correspond to Examples~\ref{eg1}, \ref{eg2}, and~\ref{eg3}, respectively, with equal weights for the balanced full factorial trial:
\[
\adjustbox{max width=\linewidth}{$\displaystyle\begin{aligned}
    &\mathbf{H}_1=\begin{pmatrix}
        0&0&1&-1\\
        1&0&-1&0
    \end{pmatrix},~~~\mathbf{H}_2=\begin{pmatrix}
        1&-1&0&0\\
        1&0&-1&0
    \end{pmatrix},~~~\mathbf{H}_3=\begin{pmatrix}
        \frac{1}{2}&-\frac{1}{2}&\frac{1}{2}&-\frac{1}{2}\\
        \frac{1}{2}&\frac{1}{2}&-\frac{1}{2}&-\frac{1}{2}
    \end{pmatrix},
\end{aligned}$}
\]
where $\bv=(\mu(1,1),\mu(0,1),\mu(1,0),\mu(0,0))^\top$ fixes the common ordering of primitive responses.

Exchanging the labels of $X_1$ and $X_2$ swaps the second and third columns of each matrix. This operation preserves the row multisets of $\mathbf{H}_2$ and $\mathbf{H}_3$, but gives a different row multiset for $\mathbf{H}_1$: $\{((0,1,0,-1),1),((1,-1,0,0),1)\}$, where the first entry of each pair $(\cdot,\cdot)$ denotes a row and the second denotes its multiplicity.
\end{example}

\section{\revblue{Proofs and additional results for causal estimands under network interference}}\label{app:policy-effects}
\revblue{We use the notation and definitions in Section~\ref{sec:network-policy-effects} of the main manuscript. Source sets $\calS\subseteq[N]$ index units, and the corresponding action-variable subsets are $\calY=\{X_k:k\in\calS\}$.}

\subsection{\revblue{Additional technical results and their proofs}}
\begin{prop}[\revblue{Assignment-induced weights}]\label{prop:assignment-induced-weights}
\revblue{For a nonempty source set $\calS\subseteq[N]$, let $\calY=\{X_k:k\in\calS\}$. For each $\calT\subseteq\calY^c$, independent Bernoulli assignment gives the reference weight}
\begin{revisionblue}\[
w_{\bpi}(\calT,\calY)
=
\prod_{k:X_k\in\calT}(1-\pi_k)
\prod_{k:X_k\in\calY^c\setminus\calT}\pi_k,
\]\end{revisionblue}\revblue{where $\calT$ indexes the action variables outside the contrast assigned state $0$. With $f(\calA)=W_N\{\ba(\calA)\}$, the weighted estimand $\Delta_{\calY}$ in Definition~\ref{def:pi-estimands-defi} equals the overall source set effect $\Delta_{\calS}^{\mathrm O}(\bpi)$.}
\end{prop}
\begin{proof}
\revblue{The event that exactly the action variables in $\calT\subseteq\calY^c$ are assigned state $0$ has probability $w_{\bpi}(\calT,\calY)$. On that event, the inner alternating sum in Definition~\ref{def:pi-estimands-defi} is $N^{-1}\sum_{i=1}^N\sum_{\calZ\subseteq\calY}(-1)^{|\calZ|}Y_i\{\ba(\calZ\cup\calT)\}$. This is the average mixed assignment difference $\bar\tau_{\calS}$ evaluated at the corresponding background assignment. Averaging over $\calT\subseteq\calY^c$ therefore gives $E_{\bpi_{-\calS}}\{\bar\tau_{\calS}(\bZ_{-\calS})\}=\Delta_{\calS}^{\mathrm O}(\bpi)$.}
\end{proof}

\subsection{\revblue{Proof of Proposition \ref{prop:hu-first-order}}}
\begin{proof}
\revblue{For each $k\in[N]$, the singleton source set $\calS=\{k\}$ gives}
\begin{revisionblue}\begin{align*}
\Delta_{\{k\}}^{\mathrm D}(\bpi)
&=N^{-1}\theta_{k,\{k\}}(\bpi),\\
\Delta_{\{k\}}^{\mathrm I}(\bpi)
&=N^{-1}\sum_{i\neq k}\theta_{i,\{k\}}(\bpi),
\end{align*}\end{revisionblue}\revblue{where $\theta_{i,\{k\}}(\bpi)=E_{\bpi_{-\{k\}}}\{Y_i(1,\bZ_{-\{k\}})-Y_i(0,\bZ_{-\{k\}})\}$ is the assignment-averaged effect of unit $k$'s treatment on unit $i$'s outcome. Summing $\Delta_{\{k\}}^{\mathrm D}(\bpi)$ and $\Delta_{\{k\}}^{\mathrm I}(\bpi)$ over $k\in[N]$ yields the average direct and indirect effects, respectively, and the sum of these two effects yields the average overall effect.}
\end{proof}

\subsection{\revblue{Proof of Proposition \ref{prop:network-cross-curvature}}}
\begin{proof}
\revblue{Write $\partial_{\calS}G_N(\bpi)=\partial^{|\calS|}G_N(\bpi)/\prod_{k\in\calS}\partial\pi_k$ for the mixed derivative indexed by a nonempty source set $\calS$. Under independent Bernoulli assignment, expected welfare is}
\begin{revisionblue}\begin{align*}
G_N(\bpi)
=&
\sum_{\bz\in\{0,1\}^N}
W_N(\bz)
\prod_{j=1}^N
\pi_j^{z_j}(1-\pi_j)^{1-z_j}.
\end{align*}\end{revisionblue}\revblue{Thus, \(G_N(\bpi)\) is affine in each coordinate of \(\bpi\) and hence is a multilinear polynomial. Fix \(k\in[N]\). Separating the terms with \(z_k=1\) and \(z_k=0\) gives}
\begin{revisionblue}\[
\adjustbox{max width=\linewidth}{$\displaystyle\begin{aligned}
G_N(\bpi)
=&
\sum_{\bz_{-\{k\}}\in\{0,1\}^{N-1}}
\bigl\{
\pi_k W_N(1,\bz_{-\{k\}})
+
(1-\pi_k)W_N(0,\bz_{-\{k\}})
\bigr\}
\prod_{j\neq k}
\pi_j^{z_j}(1-\pi_j)^{1-z_j}.
\end{aligned}$}
\]\end{revisionblue}\revblue{Because the product over \(j\neq k\) does not depend on \(\pi_k\), differentiating with respect to \(\pi_k\) yields}
\begin{revisionblue}\begin{align*}
\frac{\partial G_N(\bpi)}{\partial\pi_k}
=&
\sum_{\bz_{-\{k\}}\in\{0,1\}^{N-1}}
\bigl\{
W_N(1,\bz_{-\{k\}})-W_N(0,\bz_{-\{k\}})
\bigr\}
\prod_{j\neq k}
\pi_j^{z_j}(1-\pi_j)^{1-z_j}\\
=&
\sum_{\bz_{-\{k\}}\in\{0,1\}^{N-1}}
\bar\tau_{\{k\}}(\bz_{-\{k\}})
\prod_{j\neq k}
\pi_j^{z_j}(1-\pi_j)^{1-z_j}\\
=&
E_{\bpi_{-\{k\}}}
\left\{
\bar\tau_{\{k\}}(\bZ_{-\{k\}})
\right\}.
\end{align*}\end{revisionblue}

\revblue{For an arbitrary nonempty source set $\calS\subseteq[N]$, write $\calS=\{k_1,\ldots,k_q\}$, where $q=|\calS|$. For each $r=1,\ldots,q$, let $\calS_r=\{k_1,\ldots,k_r\}$. We show by induction that}
\begin{revisionblue}\begin{align}
\partial_{\calS_r}G_N(\bpi)
=&
\sum_{\bz_{-\calS_r}\in\{0,1\}^{N-r}}
\bar\tau_{\calS_r}(\bz_{-\calS_r})
\prod_{j\notin\calS_r}
\pi_j^{z_j}(1-\pi_j)^{1-z_j}.
\label{eq:mixed-derivative-induction}
\end{align}\end{revisionblue}\revblue{The preceding calculation establishes \eqref{eq:mixed-derivative-induction} for \(r=1\). Suppose that it holds for \(r-1\). Separating the sum in \eqref{eq:mixed-derivative-induction} according to the value of \(z_{k_r}\) gives}
\begin{revisionblue}\[
\adjustbox{max width=\linewidth}{$\displaystyle\begin{aligned}
&\partial_{\calS_{r-1}}G_N(\bpi)\\
=&
\sum_{\bz_{-\calS_r}\in\{0,1\}^{N-r}}
\Bigl[
\pi_{k_r}
\bar\tau_{\calS_{r-1}}(1,\bz_{-\calS_r})
+
(1-\pi_{k_r})
\bar\tau_{\calS_{r-1}}(0,\bz_{-\calS_r})
\Bigr]
\prod_{j\notin\calS_r}
\pi_j^{z_j}(1-\pi_j)^{1-z_j}.
\end{aligned}$}
\]\end{revisionblue}\revblue{Differentiating with respect to \(\pi_{k_r}\) therefore yields}
\begin{revisionblue}\[
\adjustbox{max width=\linewidth}{$\displaystyle\begin{aligned}
\partial_{\calS_r}G_N(\bpi)
=&
\sum_{\bz_{-\calS_r}\in\{0,1\}^{N-r}}
\Bigl\{
\bar\tau_{\calS_{r-1}}(1,\bz_{-\calS_r})
-
\bar\tau_{\calS_{r-1}}(0,\bz_{-\calS_r})
\Bigr\}\times
\prod_{j\notin\calS_r}
\pi_j^{z_j}(1-\pi_j)^{1-z_j}\\
=&
\sum_{\bz_{-\calS_r}\in\{0,1\}^{N-r}}
\bar\tau_{\calS_r}(\bz_{-\calS_r})
\prod_{j\notin\calS_r}
\pi_j^{z_j}(1-\pi_j)^{1-z_j},
\end{aligned}$}
\]\end{revisionblue}\revblue{where the final equality follows by separating the mixed assignment difference according to the state of $k_r$:}
\begin{revisionblue}\begin{align*}
 \bar\tau_{\calS_r}(\bz_{-\calS_r})=&N^{-1}\sum_{i=1}^N\sum_{\calB\subseteq\calS_r}(-1)^{|\calB|}
Y_i(\bone_{\calS_r\setminus\calB},\bzero_{\calB},\bz_{-\calS_r})\\
=&N^{-1}\sum_{i=1}^N\sum_{\calB\subseteq\calS_{r-1}}(-1)^{|\calB|}
Y_i(\bone_{(\calS_{r-1}\setminus\calB)\cup\{k_r\}},\bzero_{\calB},\bz_{-\calS_r})+\\
&N^{-1}\sum_{i=1}^N\sum_{\calB\subseteq\calS_{r-1}}(-1)^{|\calB|+1}
Y_i(\bone_{\calS_{r-1}\setminus\calB},\bzero_{\calB\cup \{k_r\}},\bz_{-\calS_r})\\
=&
\bar\tau_{\calS_{r-1}}(1,\bz_{-\calS_r})
-
\bar\tau_{\calS_{r-1}}(0,\bz_{-\calS_r}).
\end{align*}\end{revisionblue}\revblue{This completes the induction. Taking \(r=q\) gives}
\begin{revisionblue}\begin{align*}
\partial_{\calS}G_N(\bpi)
=&
\sum_{\bz_{-\calS}\in\{0,1\}^{N-|\calS|}}
\bar\tau_{\calS}(\bz_{-\calS})
\prod_{j\notin\calS}
\pi_j^{z_j}(1-\pi_j)^{1-z_j}\\
=&
E_{\bpi_{-\calS}}
\left\{
\bar\tau_{\calS}(\bZ_{-\calS})
\right\}\\
=&
\Delta_{\calS}^{\mathrm O}(\bpi),
\end{align*}\end{revisionblue}\revblue{where the last equality follows from the definition of $\Delta_{\calS}^{\mathrm O}(\bpi)$.}

\revblue{For a singleton source set, the derivative is the slope of expected welfare with respect to that unit's treatment probability. For a pair $\calS=\{k,\ell\}$, $\partial_{\{k,\ell\}}G_N(\bpi)=\partial^2G_N(\bpi)/(\partial\pi_k\partial\pi_\ell)$ is the cross-curvature of the welfare surface. For each $q=1,\ldots,N$, the chain rule and multilinearity give}
\begin{revisionblue}\[
\left.\frac{d^q}{dt^q}G_N(\bpi+t\bone_N)\right|_{t=0}
=q!\sum_{|\calS|=q}\partial_{\calS}G_N(\bpi)
=q!\Gamma_q^{\mathrm O}(\bpi).
\]\end{revisionblue}\revblue{Thus, the average overall effect of \citet{hu2022average} is the first derivative along a common change in treatment probabilities, and the order-$q$ aggregate is the corresponding $q$th derivative divided by $q!$.}
\end{proof}

\subsection{\revblue{Proof of Theorem \ref{thm:network-policy-expansion}}}
\begin{proof}
\revblue{We apply the multivariate Taylor expansion. The coefficient of $\prod_{k\in\calS}(\pi_k'-\pi_k)$ is $\partial_{\calS}G_N(\bpi)$, which equals $\Delta_{\calS}^{\mathrm O}(\bpi)$ by Proposition \ref{prop:network-cross-curvature}. Multilinearity rules out repeated-coordinate derivatives, so the expansion has no remainder. For a common change with $\bpi+t\bone_N\in[0,1]^N$, the decomposition reduces to}
\begin{revisionblue}\[
G_N(\bpi+t\bone_N)-G_N(\bpi)
=
\sum_{q=1}^N t^q\Gamma_q^{\mathrm O}(\bpi).
\]\end{revisionblue}\revblue{Setting $\bpi=\bone_N$ and $t=-1$ gives}
\begin{revisionblue}\[
W_N(\bone_N)-W_N(\bzero_N)
=
\sum_{q=1}^N(-1)^{q+1}\Gamma_q^{\mathrm O}(\bone_N),
\]\end{revisionblue}\revblue{which is the residual-free inclusion--exclusion identity in Equation~\eqref{eq:residual-free-identity} of the main manuscript.}
\end{proof}

\section{\revblue{Design-based estimation and graph distance decomposition}}
\label{app:design-identification}

\subsection{\revblue{Unbiased estimation under Bernoulli assignment}}\label{subsec:bernoulli-unbiased}
\begin{prop}[\revblue{Unbiasedness}]\label{prop:bernoulli-unbiased}
\revblue{Under independent Bernoulli assignment with $0<\pi_k<1$ for every $k\in[N]$, let $Y_i^{\mathrm{obs}}=Y_i(\bZ)$. For each unit $i\in[N]$ and nonempty source set $\calS\subseteq[N]$, define}
\begin{revisionblue}\[
\widehat\theta_{i,\calS}
=
Y_i^{\mathrm{obs}}
\prod_{k\in\calS}
\frac{Z_k-\pi_k}{\pi_k(1-\pi_k)}.
\]\end{revisionblue}\revblue{With the potential outcomes fixed, expectation over the assignment distribution gives $E_{\bpi}(\widehat\theta_{i,\calS})=\theta_{i,\calS}(\bpi)$.}
\end{prop}
\begin{proof}
\revblue{Condition on $\bZ_{-\calS}$. We first establish the identity for a single coordinate $k$. Conditional on $\bZ_{-k}$, only $Z_k$ is random, so}
\begin{revisionblue}\[
\adjustbox{max width=\linewidth}{$\displaystyle\begin{aligned}
E\left\{
Y_i(\bZ)\frac{Z_k-\pi_k}{\pi_k(1-\pi_k)}
\,\middle|\,
\bZ_{-k}
\right\}=
\sum_{z_k\in\{0,1\}}
Y_i(z_k,\bZ_{-k})
\frac{z_k-\pi_k}{\pi_k(1-\pi_k)}
\Pr(Z_k=z_k\mid\bZ_{-k}).
\end{aligned}$}
\]\end{revisionblue}\revblue{By independence, $\Pr(Z_k=1\mid\bZ_{-k})=\pi_k$ and $\Pr(Z_k=0\mid\bZ_{-k})=1-\pi_k$. Hence,}
\begin{revisionblue}\[
\adjustbox{max width=\linewidth}{$\displaystyle\begin{aligned}
E\left\{
Y_i(\bZ)\frac{Z_k-\pi_k}{\pi_k(1-\pi_k)}
\,\middle|\,
\bZ_{-k}
\right\}=&
Y_i(1,\bZ_{-k})
\frac{1-\pi_k}{\pi_k(1-\pi_k)}\pi_k
+
Y_i(0,\bZ_{-k})
\frac{-\pi_k}{\pi_k(1-\pi_k)}(1-\pi_k) \\
=&
Y_i(1,\bZ_{-k})-Y_i(0,\bZ_{-k})
=
\tau_{i,\{k\}}(\bZ_{-k}).
\end{aligned}$}
\]\end{revisionblue}\revblue{More generally, independence gives}
\begin{revisionblue}\[
\adjustbox{max width=\linewidth}{$\displaystyle\begin{aligned}
E\left\{
Y_i(\bZ)
\prod_{k\in\calS}
\frac{Z_k-\pi_k}{\pi_k(1-\pi_k)}
\,\middle|\,
\bZ_{-\calS}
\right\}=&
\sum_{\bz_{\calS}\in\{0,1\}^{|\calS|}}
Y_i(\bz_{\calS},\bZ_{-\calS})
\prod_{k\in\calS}
\left\{
\frac{z_k-\pi_k}{\pi_k(1-\pi_k)}
\Pr(Z_k=z_k)
\right\}\\
=&
\sum_{\bz_{\calS}\in\{0,1\}^{|\calS|}}
(-1)^{|\calS|-\sum_{k\in\calS}z_k}
Y_i(\bz_{\calS},\bZ_{-\calS}) \\
=&
\tau_{i,\calS}(\bZ_{-\calS}),
\end{aligned}$}
\]\end{revisionblue}\revblue{where, for each $k\in\calS$, the factor inside braces equals $1$ when $z_k=1$ and $-1$ when $z_k=0$. Taking expectation over $\bZ_{-\calS}$ gives $E_{\bpi}(\widehat\theta_{i,\calS})=E_{\bpi_{-\calS}}\{\tau_{i,\calS}(\bZ_{-\calS})\}=\theta_{i,\calS}(\bpi)$.}
\end{proof}
\revblue{Applying the same partition of outcome units as in the estimand construction gives}
\begin{revisionblue}\[
\widehat\Delta_{\calS}^{\mathrm O}
=\widehat\Delta_{\calS}^{\mathrm D}+\widehat\Delta_{\calS}^{\mathrm I}
=\frac1N\sum_{i=1}^N\widehat\theta_{i,\calS},
\]\end{revisionblue}\revblue{where $\widehat\Delta_{\calS}^{\mathrm D}=N^{-1}\sum_{i\in\calS}\widehat\theta_{i,\calS}$ and $\widehat\Delta_{\calS}^{\mathrm I}=N^{-1}\sum_{i\notin\calS}\widehat\theta_{i,\calS}$. Linearity of expectation establishes unbiasedness for the corresponding direct, indirect, and overall effects.}

\subsection{\revblue{Distance-specific decomposition of the indirect effect}}
\label{subsec:graph-distance-shells}

\revblue{The indirect effect admits an additive refinement according to the network distance between the source units and the affected outcome units. This construction is related to the distance-specific average marginalized response addressed by \citet{wang2025design}, but uses a different normalization to preserve an exact decomposition of the indirect effect.}

\revblue{Let $\mathbf{G}$ be a fixed adjacency matrix, and let $\operatorname{dist}_{\mathbf{G}}(i,k)$ denote the shortest-path distance between units $i$ and $k$ in the corresponding network, with $\operatorname{dist}_{\mathbf{G}}(i,k)=\infty$ if they are disconnected. For a nonempty source set $\calS$, define $\operatorname{dist}_{\mathbf{G}}(i,\calS)=\min_{k\in\calS}\operatorname{dist}_{\mathbf{G}}(i,k)$. For $r\geq1$, let}
\begin{revisionblue}\[
\calV_r(\calS)
=
\left\{
i\in[N]\setminus\calS:
\operatorname{dist}_{\mathbf{G}}(i,\calS)=r
\right\}
\]\end{revisionblue}\revblue{denote the distance-$r$ shell around $\calS$. For a disconnected network, $\calV_\infty(\calS)$ contains the units that are not connected to $\calS$.}

\revblue{Define the distance-specific contribution to the indirect effect by}
\begin{revisionblue}\[
\Delta_{\calS,r}^{\mathrm I}(\bpi)
=
\frac1N
\sum_{i\in\calV_r(\calS)}
\theta_{i,\calS}(\bpi).
\]\end{revisionblue}\revblue{Because the distance shells partition $[N]\setminus\calS$,}
\begin{revisionblue}\begin{align}\label{eq:distance-shell-decomposition}
\Delta_{\calS}^{\mathrm I}(\bpi)
=
\sum_{r\in\{1,2,\ldots,\infty\}}
\Delta_{\calS,r}^{\mathrm I}(\bpi).
\end{align}\end{revisionblue}\revblue{Thus, $\Delta_{\calS,1}^{\mathrm I}(\bpi)$ captures effects on immediate neighbors of the source set, $\Delta_{\calS,2}^{\mathrm I}(\bpi)$ captures effects on units two steps away, and so forth. The interaction order $|\calS|$ and the network distance $r$ characterize distinct features: the former is the number of source assignments jointly involved in the interaction, whereas the latter describes the locations of the affected outcome units.}

\revblue{For effects of order $q$, define}
\begin{revisionblue}\[
\Gamma_{q,r}^{\mathrm I}(\bpi)
=
\sum_{\substack{\calS\subseteq[N]\\|\calS|=q}}
\Delta_{\calS,r}^{\mathrm I}(\bpi).
\]\end{revisionblue}\revblue{Equation~\eqref{eq:distance-shell-decomposition} implies $\Gamma_q^{\mathrm I}(\bpi) = \sum_{r\in\{1,2,\ldots,\infty\}} \Gamma_{q,r}^{\mathrm I}(\bpi)$. In particular, Proposition~\ref{prop:hu-first-order} gives $\tau_{\mathrm{AIE}}(\bpi)=\sum_{r\in\{1,2,\ldots,\infty\}} \Gamma_{1,r}^{\mathrm I}(\bpi)$ and $\tau_{\mathrm{AOE}}(\bpi)=\tau_{\mathrm{ADE}}(\bpi)+\sum_{r\in\{1,2,\ldots,\infty\}} \Gamma_{1,r}^{\mathrm I}(\bpi)$. Hence, the average indirect effect of \citet{hu2022average} is decomposed additively into contributions from different network distances.}

\revblue{The average marginalized response in \citet{wang2025design} uses a different normalization. For a singleton source set $\{k\}$ with a nonempty distance-$r$ shell, the latter replaces $\sum_{i\in\calV_r(\{k\})} \theta_{i,\{k\}}(\bpi)$ by the within-shell average}
\begin{revisionblue}\[
\frac{1}{|\calV_r(\{k\})|}
\sum_{i\in\calV_r(\{k\})}
\theta_{i,\{k\}}(\bpi).
\]\end{revisionblue}\revblue{As emphasized by \citet{hu2022average}, this shell-size normalization changes the weighting of the target. The average marginalized response first averages within each source-specific shell and then weights source units equally. In contrast, $\Gamma_{1,r}^{\mathrm I}(\bpi)$ retains the original $1/N$ normalization, weights each ordered source--outcome pair equally, and measures the total contribution from distance $r$. This common normalization ensures that the distance-specific contributions sum to $\tau_{\mathrm{AIE}}(\bpi)$.}

\revblue{The distance-specific contributions also admit direct design-based estimators. Under the conditions of Proposition~\ref{prop:bernoulli-unbiased}, linearity of expectation gives the following unbiased estimators of $\Delta_{\calS,r}^{\mathrm I}(\bpi)$ and $\Gamma_{q,r}^{\mathrm I}(\bpi)$:}
\begin{revisionblue}\begin{align*}
&\widehat\Delta_{\calS,r}^{\mathrm I}=\frac1N
\sum_{i\in\calV_r(\calS)}
\widehat\theta_{i,\calS},\\
&\widehat\Gamma_{q,r}^{\mathrm I}=\sum_{\substack{\calS\subseteq[N]\\|\calS|=q}}
\widehat\Delta_{\calS,r}^{\mathrm I}.
\end{align*}\end{revisionblue}

\section{\revblue{Partially ordered action variables}}
\label{app:controlled-histories}

\revblue{Consider a longitudinal study with several mediators measured at each time point or multiple intervention components delivered at each visit. Although the time points or visits are ordered, the variables within a given time point or visit may be scientifically unordered. In such settings, the framework permits relabelings within time blocks while preserving the ordering between blocks.}

\revblue{Partition the action variables into disjoint, temporally ordered blocks $\calX=\calX_1\cup\cdots\cup\calX_T$, where $\calX_t=\{X_{t1},\ldots,X_{tK_t}\}$ contains the $K_t$ variables in block $t$ and $\calX_t\cap\calX_s=\emptyset$ for $t\neq s$. Relabelings that preserve block membership act within each block, so the scientifically admissible group satisfies $\frakG\leq S_{K_1}\times\cdots\times S_{K_T}$. The full product group allows every permutation within each block and is appropriate when all such permutations preserve the scientific structure. Otherwise, the admissible group must be restricted further.}

\revblue{In causal mediation, the blocks may correspond to collections of mediators measured at successive time points. Each action variable specifies the intervention state used to define the primitive response surface; the mediator itself need not be binary. An admissible permutation relabels the action variables and their associated mediator labels within a block, without exchanging time points or changing the specified causal structure.}

\revblue{To define effects within block $t$, first fix the action configuration in all other blocks. Let $\calB\subseteq\calX\setminus\calX_t$ index the outside variables assigned state $0$, with the remaining outside variables assigned state $1$. For each $\calA_t\subseteq\calX_t$, the response surface restricted to block $t$ is}
\begin{revisionblue}\[
f_t(\calA_t)
=
f(\calB\cup\calA_t).
\]\end{revisionblue}\revblue{Here, $\calA_t$ indexes the variables assigned state $0$ within the selected block. Varying $\calA_t$ changes only the configuration in block $t$, while $\calB$ remains fixed as part of the estimand definition.}

\revblue{For each nonempty $\calS\subseteq\calX_t$, the residual-free construction applied to $f_t$ gives}
\begin{revisionblue}\[
\Delta_{\calS}^{(t)}
=
\sum_{\calU\subseteq\calS}
(-1)^{|\calU|}f_t(\calU).
\]\end{revisionblue}\revblue{These effects form a complete collection for the action variables in block $t$ and satisfy}
\begin{revisionblue}\[
f_t(\emptyset)-f_t(\calX_t)
=
\sum_{\emptyset\neq\calS\subseteq\calX_t}
(-1)^{|\calS|+1}\Delta_{\calS}^{(t)}.
\]\end{revisionblue}\revblue{The left-hand side compares assigning every action in block $t$ to state $1$ with assigning every action in that block to state $0$. The identity decomposes this contrast exactly while holding the configurations of all other blocks fixed. Under an admissible relabeling within block $t$, the primitive response coordinates and associated contrasts are relabeled together, as in the general construction.}

\revblue{The same formulation applies to longitudinal interventions; that is, block $t$ contains the intervention components delivered at visit $t$, and the intervention configurations at other visits are fixed as part of the target estimand. Equivariance concerns admissible relabelings of the components within the selected visit, not permutations of the visits themselves. When each visit contains only one intervention, the only within-visit permutation is the identity. More generally, for fully ordered action variables, temporal or causal positions distinguish their scientific roles. The algebraic results continue to apply, with admissible relabelings determined by these scientific roles.}

\section{\revblue{Permutation equivariance for exposure mappings and marginal structural models}}
\label{app:dimension-reduction}

\revblue{Consider causal comparisons involving $K$ binary action variables $\calX=\{X_1,\ldots,X_K\}$, with joint assignment $\ba=(a_1,\ldots,a_K)^\top\in\{0,1\}^K$. In factorial studies, the action variables represent treatment factors, and $\ba$ specifies a combination of factor levels. Under network interference, the action variables represent treatments assigned to individual units, and a unit's outcome may depend on the full assignment. In both settings, the primitive response surface records the causal response of interest under each of the $2^K$ possible assignments.}

\revblue{Exposure mappings and marginal structural models provide ways to summarize or restrict this dependence on the assignment. We consider exposure mappings in a general formulation and then illustrate their specialization to network interference, where the mapping depends on both the treatment assignment and an adjacency matrix describing the network. For example, a network exposure mapping may record a unit's own treatment and the number of treated neighbors. We illustrate marginal structural models in factorial studies. Marginal structural models also apply in other causal settings, where the same analysis proceeds after specifying the response model and admissible relabelings. For exposure mappings, we examine compatibility with relabeling, meaning that relabeling an assignment induces a well-defined relabeling of its exposure. For marginal structural models, we examine closure of the model class under relabeling, meaning that every relabeled response surface remains within the specified class. In both settings, permutation equivariance of the reported estimand vector further requires that relabeling preserve the selected contrasts and permute their coordinates.}

\revblue{Throughout, $\frakG\leq S_K$ denotes the scientifically admissible relabeling group. Let $\mathbf{P}_\sigma$ be the $K\times K$ permutation matrix satisfying $(\mathbf{P}_\sigma\ba)_{\sigma(k)}=a_k$, where subscripts index vector coordinates. Here, $\mathbf{P}_\sigma$ acts on the assignment coordinates rather than on the $2^K$ coordinates of the primitive response vector. Write $m(\ba)=f\bigl(\{X_k:a_k=0\}\bigr)$ for the response surface expressed as a function of the assignment vector. Its relabeled version is $m^\sigma(\ba)=m(\mathbf{P}_\sigma^{-1}\ba)$.}

\subsection{\revblue{Exposure mappings}}

\revblue{Exposure mappings summarize joint treatment assignments into exposure conditions for individual units. This formulation connects to the literature on causal inference under interference. \citet{hudgens2008toward} developed estimands for direct, indirect, total, and overall effects in settings with interference within groups. \citet{manski2013identification} studied restrictions on how potential outcomes depend on the joint treatment assignment, and \citet{aronow2017estimating} developed a framework for randomization-based inference that combines assignment mechanisms, exposure mappings, and causal estimands.}

\revblue{For a fixed exposure mapping, compatibility with relabeling requires a well-defined action on the exposure space. That is, the exposure after relabeling must be determined by the original exposure alone. The reported estimand vector is equivariant when this action also preserves the selected contrasts and permutes their coordinates.}

\revblue{Let $h$ be an exposure mapping that assigns an exposure value $h(\bm a)$ to each assignment vector $\bm a\in\{0,1\}^K$. Its exposure space is $\calE=\{h(\bm a):\bm a\in\{0,1\}^K\}$, the set of all values attained by $h$. The mapping is compatible with $\frakG$ if each admissible relabeling induces a permutation $\rho_\sigma:\calE\to\calE$ of the exposure values satisfying, for every $\ba\in\{0,1\}^K$,}
\begin{revisionblue}\begin{align*}
h(\mathbf{P}_\sigma\ba)
=
\rho_\sigma\{h(\ba)\}.
\end{align*}\end{revisionblue}\revblue{Thus, relabeling the assignment and then computing its exposure must agree with computing the exposure first and then relabeling it.}

\revblue{Equivalently, for every $\sigma\in\frakG$, assignments with the same exposure must have equal exposures after relabeling:}
\begin{revisionblue}\[
h(\ba)=h(\ba')
\quad\Longrightarrow\quad
h(\mathbf{P}_\sigma\ba)=h(\mathbf{P}_\sigma\ba').
\]\end{revisionblue}
\revblue{This condition makes $\rho_\sigma\{h(\ba)\}=h(\mathbf{P}_\sigma\ba)$ well defined.}

\revblue{Compatibility determines how exposure values transform, but does not by itself ensure permutation equivariance of the reported contrasts. When the primitive response depends on the assignment only through its exposure $h(\ba)$, we can write}
\begin{revisionblue}\[
m(\ba)=f\bigl(\{X_k:a_k=0\}\bigr)=g\{h(\ba)\},
\]\end{revisionblue}\revblue{where $g:\calE\to\mathbb R$ is an exposure response surface. Thus, $g$ plays the same role as $f$, but is indexed by exposure values rather than full action configurations: $g(e)$ is the common primitive response for all assignments with exposure $e$.}

\revblue{For a compatible mapping, the relabeled exposure response surface is}
\begin{revisionblue}\[
g^\sigma(e)=g\{\rho_{\sigma^{-1}}(e)\},
\]\end{revisionblue}\revblue{so that $m^\sigma(\ba)=g^\sigma\{h(\ba)\}$. The inverse identifies the original exposure whose response is assigned to $e$ after relabeling. Let $\bm\delta(g)$ be a vector of selected contrasts of the exposure response values, such as differences $g(e)-g(e')$. This vector is permutation equivariant if, for every $\sigma\in\frakG$, there exists a permutation matrix $\mathbf{P}_r(\sigma)$ such that, for every $g:\calE\to\mathbb R$,}
\begin{revisionblue}\[
\bm\delta(g^\sigma)=\mathbf{P}_r(\sigma)\bm\delta(g).
\]\end{revisionblue}\revblue{Thus, relabeling must preserve the selected contrasts as functionals of $g$, including their multiplicities. Compatibility determines how exposure values are relabeled; closure of the contrast collection ensures that this relabeling merely permutes the reported estimands.}

\revblue{For example, let $K=4$ and consider the exposure mapping $h(\ba)=(a_1+a_2,a_3+a_4)^\top$ with $\calE=\{0,1,2\}^2$. Each exposure coordinate counts the actions assigned level $1$ within the corresponding pair. Let $\frakG$ be generated by $\sigma=(1\,3)(2\,4)$, which exchanges the two action pairs. The mapping is compatible with $\frakG$ because relabeling exchanges the exposure coordinates: $\rho_\sigma(e_1,e_2)=(e_2,e_1)$. For an exposure response surface $g:\calE\to\mathbb R$, consider the contrasts $\delta_1(g)=g(1,0)-g(0,0)$ and $\delta_2(g)=g(0,1)-g(0,0)$. They satisfy $\delta_1(g^\sigma)=\delta_2(g)$ and $\delta_2(g^\sigma)=\delta_1(g)$. Thus, reporting only $\delta_1$ fails permutation equivariance in general, whereas the vector $\bm\delta(g)=(\delta_1(g),\delta_2(g))^\top$ is equivariant under $\frakG$, with relabeling exchanging its two components. This example shows that compatibility of the exposure mapping and closure of the selected contrast collection are distinct requirements.}

\revblue{For a simple counterexample, consider two scientifically interchangeable action variables and the fixed-coordinate summary $h(\ba)=a_1$. Let $\sigma$ exchange coordinates $1$ and $2$. The assignments $(0,0)^\top$ and $(0,1)^\top$ both have exposure $0$, but their relabeled exposures are $h\{\mathbf{P}_\sigma(0,0)^\top\}=0$ and $h\{\mathbf{P}_\sigma(0,1)^\top\}=1$. No function of the original exposure alone can produce both values. Consequently, this mapping is incompatible with the exchange.}

\revblue{Consider network interference among $N$ units, where a unit's outcome may depend on its own treatment and the treatments assigned to other units. Here, $K=N$, the action variable $X_i$ represents the treatment assigned to unit $i$, and $\ba=(a_1,\ldots,a_N)^\top\in\{0,1\}^N$ records the joint assignment, with $0$ denoting control and $1$ denoting treatment. Let $\mathbf{G}=(G_{ij})$ be the fixed adjacency matrix, where $G_{ij}=1$ if unit $j$ is a neighbor of unit $i$ and $G_{ij}=0$ otherwise. Unit indices and sums over units range from $1$ to $N$. A network exposure mapping $h_i(\ba,\mathbf{G})$ assigns an exposure value to unit $i$ based on the treatment assignment and the network.}

\revblue{An admissible relabeling $\sigma\in\frakG$ acts on both the treatment vector and the adjacency matrix. Unit $i$ receives label $\sigma(i)$, the assignment becomes $\mathbf{P}_\sigma\ba$, and the adjacency matrix becomes $\mathbf{P}_\sigma\mathbf{G}\mathbf{P}_\sigma^\top$. For every unit $i$, assignment $\ba\in\{0,1\}^N$, and admissible relabeling $\sigma$, the compatibility condition takes the form}
\begin{revisionblue}
\begin{align}\label{eq:app-network-exposure-compatibility}
h_{\sigma(i)}
(\mathbf{P}_\sigma\ba,\mathbf{P}_\sigma \mathbf{G} \mathbf{P}_\sigma^\top)
=
\rho_\sigma\{h_i(\ba,\mathbf{G})\},
\end{align}
\end{revisionblue}\revblue{where $\rho_\sigma$ is the permutation of exposure values induced by $\sigma$. Thus, computing the exposure of the corresponding unit after relabeling the assignment and network must agree with applying $\rho_\sigma$ to its original exposure.}

\revblue{For example, define}
\begin{revisionblue}
\[
h_i(\ba,\mathbf{G})
=
\left(
a_i,\,
\sum_jG_{ij},\,
\sum_jG_{ij}a_j
\right),
\]
\end{revisionblue}\revblue{which records own treatment, node degree, and treated-neighbor count. Writing $\ba'=\mathbf{P}_\sigma\ba$ and $\mathbf{G}'=\mathbf{P}_\sigma \mathbf{G} \mathbf{P}_\sigma^\top$ gives}
\begin{revisionblue}
\[
a'_{\sigma(i)}=a_i,\qquad
\sum_jG'_{\sigma(i),j}=\sum_jG_{ij},\qquad
\sum_jG'_{\sigma(i),j}a'_j=\sum_jG_{ij}a_j.
\]
\end{revisionblue}\revblue{Hence, Equation~\eqref{eq:app-network-exposure-compatibility} holds with the identity action $\rho_\sigma(e)=e$ on the exposure space. Relabeling changes the unit labels but leaves the exposure recorded for the corresponding unit unchanged.}

\subsection{\revblue{Marginal structural models}}

\revblue{Marginal structural mean models describe how the population mean potential outcome varies across treatment interventions \citep{robins2000marginal}. For such a model, closure under relabeling means that each relabeled response surface remains in the specified model class. Equivariance of the reported estimand vector also requires that relabeling preserve the selected contrasts and permute their coordinates. These are separate conditions on the model and its reported contrasts. In a factorial setting with $K$ binary treatment factors, let $Y(\ba)$ denote the potential outcome under the joint treatment combination $\ba\in\{0,1\}^K$. Consider the marginal structural mean model}
\begin{revisionblue}\[
m_{\bm\beta}(\ba)
=
\mathbb{E}\{Y(\ba)\}
=
\bm\beta^\top\boldsymbol b(\ba),
\]\end{revisionblue}\revblue{where $\bm\beta\in\mathbb R^p$, the expectation is taken over the target population, and $\boldsymbol b(\ba)$ is a vector of prespecified basis functions. If, for every $\sigma\in\frakG$, a $p\times p$ matrix $\mathbf{R}_\sigma$ satisfies}
\begin{revisionblue}\[
\boldsymbol b(\mathbf{P}_\sigma\ba)
=
\mathbf{R}_\sigma\boldsymbol b(\ba),
\]
then $m_{\bm\beta}^{\sigma}(\ba)=m_{\bm\beta}(\mathbf P_\sigma^{-1}\ba)=m_{\mathbf R_{\sigma^{-1}}^\top\bm\beta}(\ba)$, so relabeling maps each surface to another member of the same model class. Equivariance of the reported estimand vector further requires that relabeling permute its coordinates rather than map a reported contrast to one omitted from the vector or introduce a new contrast.\end{revisionblue}

\revblue{For example, the additive model}
\begin{revisionblue}\[
m_{\bm\beta}(\ba)
=
\beta_0+\sum_{k=1}^K\beta_k a_k
\]\end{revisionblue}\revblue{is closed under relabeling, and the full vector of treatment contrasts $(\beta_1,\ldots,\beta_K)^\top$ is permutation equivariant even when the slopes are unequal. Invariance would require the response surface itself to remain unchanged under relabeling; equivariance requires only that the reported contrasts follow the relabeling.}

\revblue{These properties can fail when model terms or reported contrasts are selected according to arbitrary coordinate labels. Consider three scientifically interchangeable action variables and the model}
\begin{revisionblue}\begin{align*}
m_{\bm\beta}(\ba)
=
\beta_0+\beta_1a_1+\beta_2a_2+\beta_3a_3
+\beta_{12}a_1a_2.
\end{align*}\end{revisionblue}\revblue{When $\beta_{12}\neq0$, exchanging coordinates $2$ and $3$ produces an $a_1a_3$ interaction, so the relabeled surface lies outside the original model class. To examine the reported contrasts, let $\mathbf e_j\in\{0,1\}^3$ be the assignment vector whose $j$th coordinate is $1$ and whose other coordinates are $0$, and let $\mathbf 0=(0,0,0)^\top$. For each $j=1,2,3$, define the main contrast $\delta_j$ and the selected interaction $\delta_{12}$ by}
\begin{revisionblue}\begin{align*}
\delta_j(m_{\bm\beta})
&=
m_{\bm\beta}(\mathbf e_j)-m_{\bm\beta}(\mathbf 0),\\
\delta_{12}(m_{\bm\beta})
&=
m_{\bm\beta}(1,1,0)-m_{\bm\beta}(1,0,0)-m_{\bm\beta}(0,1,0)+m_{\bm\beta}(0,0,0).
\end{align*}\end{revisionblue}\revblue{For the selected estimand vector}
\begin{revisionblue}\[
\bm\delta(m_{\bm\beta})=\{\delta_1(m_{\bm\beta}),\delta_2(m_{\bm\beta}),\delta_3(m_{\bm\beta}),\delta_{12}(m_{\bm\beta})\}^\top,
\]\end{revisionblue}\revblue{consider the surface $m_{\bm\beta}(\ba)=a_1a_2$. Exchanging coordinates $2$ and $3$ gives the relabeled surface $m_{\bm\beta}^\sigma(\ba)=a_1a_3$, and hence}
\begin{revisionblue}\[
\bm\delta(m_{\bm\beta})=(0,0,0,1)^\top,
\qquad
\bm\delta(m_{\bm\beta}^\sigma)=(0,0,0,0)^\top.
\]\end{revisionblue}\revblue{These vectors cannot be related by a permutation, because the interaction has moved to $\{X_1,X_3\}$, whose contrast is absent from the selected collection. This failure persists even if the model includes all pairwise interactions, showing that model-class closure alone does not ensure estimand equivariance.}

\section{\revblue{Identification and implications for estimation}}
\label{sec:identification}

\revblue{Permutation equivariance, completeness, and residual-free decomposition characterize the structure of causal estimands. Identification depends on the causal assumptions and study design, and its requirements can differ between an endpoint contrast and the individual components of its decomposition. We provide additional discussion about estimation issues in different scientific contexts.}

\subsection{\revblue{Factorial studies}}
\label{subsec:identification-factorial}

\revblue{In a randomized full factorial experiment, positive assignment probabilities for the required treatment combinations permit unbiased estimation of their finite population means and fixed linear contrasts \citep{zhao2023covariate}. A design supporting fewer combinations may identify the endpoint contrast while leaving some components unidentified. In observational studies, identification of joint potential outcome means requires appropriate consistency, exchangeability, and positivity assumptions \citep{imbens2000role,yu2026balancing}. Identification also depends on whether we average potential outcomes over a common distribution of the other treatments or keep those treatments at each person's observed values. For example, $\sum_{a_2=0}^1\Pr(X_2=a_2)E\{Y(a_1,a_2)\}$ generally differs from $E\{Y(a_1,X_2)\}$. The two agree when $E\{Y(a_1,a_2)\mid X_2=a_2\}=E\{Y(a_1,a_2)\}$ for each $a_2$, as under randomized experiments. The weights are either known by design or must be estimated in observational settings. In the latter case, uncertainty calculations must account for weight estimation \citep{murphy1985estimation,lunceford2004stratification}.}

\subsection{\revblue{Causal mediation with multiple mediators}}
\label{subsec:identification-mediation}

\revblue{For two mediators, the natural indirect effect is $\mu(1,1)-\mu(0,0)$, whereas the exit indirect effects, corresponding to the first-order main effects for the two mediators, and their interaction additionally involve $\mu(1,0)$ and $\mu(0,1)$ \citep{xia2022decomposition}. The natural indirect effect is identified under consistency, positivity, and sequential ignorability \citep{imai2010identification}. The exit indirect effects and their interaction additionally depend on the joint distribution of mediator potential values under different treatment conditions for the same individual. The observed mediator distributions in the two treatment arms do not identify this cross-world dependence. Consequently, these effects may remain unidentified under sequential ignorability alone. Their identification may require additional restrictions on cross-world dependence or a suitable structural model \citep{xia2022decomposition,ohnishi2024bayesian,cheng2024semiparametric,tong2026causal}, and sensitivity analysis can assess how conclusions depend on these additional cross-world assumptions \citep{daniel2015causal}.}
\subsection{\revblue{Design-based network interference}}
\label{subsec:identification-network}

\revblue{Under interference, identification depends on the assignments supported by the study design. Under an unrestricted fixed potential-outcome model, a linear welfare contrast is identified from the randomization distribution if and only if every assignment with a nonzero contrast coefficient has positive probability. A design that permits only the all-control and all-treated assignments therefore identifies their contrast but generally leaves its individual components unidentified. Under independent Bernoulli assignment with probabilities strictly between zero and one, every assignment has positive probability. In this case, Proposition~\ref{prop:bernoulli-unbiased} provides unbiased design-based estimators of the source set components, with unbiasedness taken over repeated assignments to fixed potential outcomes. Unbiasedness follows from the assignment mechanism with fixed potential outcomes; see Appendix~\ref{app:design-identification} for details and \citet{harshaw2025general} for a broader treatment of design-based estimation. Precision in a single study also depends on the estimand structure and can deteriorate rapidly with source set size, as illustrated by the second moment calculation in Section~\ref{subsec:network-design-estimation} of the main manuscript.}

\section{Additional, more concrete examples in observational factorial studies and causal mediation with $K=3$}
\label{sec:application}

\subsection{Factorial observational studies}\label{sec:application-eg-factorial-os}
Using the notation of Example~\ref{eg3}, consider a factorial observational study with three treatment factors ($K=3$). We define $Y(a_1,X_2,X_3)$ as the potential outcome when the first factor takes the value $X_1=a_1$, while the other two factors take their natural values according to the treatment assignment distribution. That is, $Y(a_1,X_2,X_3)=\sum_{a_2,a_3}\mathbf{1}(X_2=a_2,X_3=a_3)Y(a_1,a_2,a_3)$. This notation naturally extends to other combinations of treatment factors, such as $Y(X_1,a_2,X_3)$, $Y(X_1,X_2,a_3)$, $Y(a_1,a_2,X_3)$, $Y(a_1,X_2,a_3)$, and $Y(X_1,a_2,a_3)$. The main effects are
\begin{align*}
    \Delta_{\{X_1\}}=&E\{Y(1,X_2,X_3)\}-E\{Y(0,X_2,X_3)\},\\
    \Delta_{\{X_2\}}=&E\{Y(X_1,1,X_3)\}-E\{Y(X_1,0,X_3)\},\\
    \Delta_{\{X_3\}}=&E\{Y(X_1,X_2,1)\}-E\{Y(X_1,X_2,0)\}.
\end{align*}
The corresponding two-way interactions are
\[
\adjustbox{max width=\linewidth}{$\displaystyle\begin{aligned}
        \Delta_{\{X_1,X_2\}}=&E\{Y(1,1,X_3)\}-E\{Y(0,1,X_3)\}-E\{Y(1,0,X_3)\}+E\{Y(0,0,X_3)\},\nonumber\\
        \Delta_{\{X_1,X_3\}}=&E\{Y(1,X_2,1)\}-E\{Y(0,X_2,1)\}-E\{Y(1,X_2,0)\}+E\{Y(0,X_2,0)\},\nonumber\\
        \Delta_{\{X_2,X_3\}}=&E\{Y(X_1,1,1)\}-E\{Y(X_1,0,1)\}-E\{Y(X_1,1,0)\}+E\{Y(X_1,0,0)\}.
\end{aligned}$}
\]

The three-way interaction is
\[
\adjustbox{max width=\linewidth}{$\displaystyle\begin{aligned}
   \Delta_{\{X_1,X_2,X_3\}}= &E\{Y(1,1,1)\}-E\{Y(0,1,1)\}-E\{Y(1,0,1)\}-E\{Y(1,1,0)\}+\\
   &E\{Y(1,0,0)\}+E\{Y(0,1,0)\}+E\{Y(0,0,1)\}-E\{Y(0,0,0)\}.
\end{aligned}$}
\]

\revblue{These estimands cover all nonempty factor subsets and retain the remaining factors at their natural values. The corresponding factorial estimands in Definition~\ref{def:pi-estimands-defi} instead average contrasts of mean potential outcomes under fixed treatment combinations using assignment probabilities. The two agree when the potential outcomes are mean independent of the remaining factors for each contrast. Jointly relabeling the factors and potential outcomes preserves the interpretation of these effects. See Appendix~\ref{subsec:identification-factorial} for further discussion.} For example, consider the cyclic relabeling $\sigma(X_j)=X_{j+1}$ for $j=1,2$ and $\sigma(X_3)=X_1$. The original main effect $\Delta_{\{X_1\}}$ then becomes $\Delta_{\{X_2\}}^\dagger$ in the new labeling. Both represent the effect of the same factor, with the other factors retaining their natural values under the jointly relabeled distribution. The two-way and three-way interactions retain their interpretations in the same way.

\subsection{Causal mediation with multiple unordered mediators}\label{sec:application-mediation}
Using the notation of Example~\ref{eg2}, consider three causally unordered mediators and define $\mu(a_1,a_2,a_3)=E\{Y(1,X_1(a_1),X_2(a_2),X_3(a_3))\}$, where, for each $j=1,2,3$, $X_j(a_j)$ is mediator $j$'s potential value under treatment $a_j$. The first-order natural indirect effects, termed exit indirect effects by \citet{xia2022decomposition}, are
\begin{align*}
    \Delta_{\{X_1\}}=\mu(1,1,1)-\mu(0,1,1),\\
    \Delta_{\{X_2\}}=\mu(1,1,1)-\mu(1,0,1),\\
    \Delta_{\{X_3\}}=\mu(1,1,1)-\mu(1,1,0).
\end{align*}
Each first-order effect compares the outcome when one mediator takes its potential value under treatment $1$ rather than treatment $0$, holding the other mediators at their potential values under treatment $1$. The second-order natural indirect interactions are
\begin{align*}
    \Delta_{\{X_1,X_2\}}=\mu(1,1,1)-\mu(0,1,1)-\mu(1,0,1)+\mu(0,0,1),\\
    \Delta_{\{X_1,X_3\}}=\mu(1,1,1)-\mu(0,1,1)-\mu(1,1,0)+\mu(0,1,0),\\
    \Delta_{\{X_2,X_3\}}=\mu(1,1,1)-\mu(1,0,1)-\mu(1,1,0)+\mu(1,0,0).
\end{align*}

By Proposition~\ref{prop:alternative-repre}, each second-order effect measures how the contrast for one mediator changes when the other mediator moves from its potential value under treatment $0$ to its potential value under treatment $1$, holding the third mediator at its potential value under treatment $1$. The third-order natural indirect interaction is
\begin{align*}
       \Delta_{\{X_1,X_2,X_3\}}= &\mu(1,1,1)-\mu(0,1,1)-\mu(1,0,1)-\mu(1,1,0)+\\
   &\mu(1,0,0)+\mu(0,1,0)+\mu(0,0,1)-\mu(0,0,0).
\end{align*}

This contrast measures how a first-order natural indirect effect is modified by the interaction between the remaining two mediators. Permutation equivariance guarantees that these interpretations are preserved under mediator relabeling. Theorem \ref{prop:nie} further gives the endpoint decomposition
\begin{align*}
\mu(1,1,1)-\mu(0,0,0)
=
\sum_{j=1}^3\Delta_{\{X_j\}}
-
\sum_{j<j'}\Delta_{\{X_j,X_{j'}\}}
+
\revblue{\Delta_{\{X_1,X_2,X_3\}}}.
\end{align*}
\revblue{Identification of the individual components may require stronger assumptions on cross-world dependence than identification of the endpoint contrast under classical sequential ignorability \citep{imai2010identification}. Appendix~\ref{subsec:identification-mediation} provides further discussion.}

\section{Algorithm for identifying the row permutation matrix for a given relabeling}\label{app:algorithm}

\revblue{The following procedure implements the row matching in Theorem~\ref{thm:characterization} for one specified admissible relabeling. Matching only unused rows ensures a bijection even when the contrast matrix contains repeated rows.}

\begin{algorithm}[htbp]

\caption{Estimand permutation for a specified relabeling}
\label{algo:compute}
\DontPrintSemicolon
\SetKwInput{KwAssume}{Assume}
\begin{revisionblue}\KwIn{A contrast matrix $\mathbf H\in\mathbb R^{d\times2^K}$ and an admissible column permutation matrix $\mathbf P_c(\sigma)$.}
\KwAssume{The row multisets of $\mathbf H$ and $\mathbf H\mathbf P_c(\sigma)$ agree.}
\KwOut{A permutation matrix $\mathbf P_r$ satisfying $\mathbf P_r\mathbf H=\mathbf H\mathbf P_c(\sigma)$.}
Set $\mathbf H'=\mathbf H\mathbf P_c(\sigma)$, $J=\{1,\ldots,d\}$, and $\mathbf P_r=\mathbf0_{d\times d}$\;
Let $r_j$ and $r'_i$ denote rows of $\mathbf H$ and $\mathbf H'$, respectively\;
\For{$i=1$ \KwTo $d$}{
Choose the smallest $j\in J$ such that $r_j=r'_i$\;
Set $(\mathbf P_r)_{ij}=1$ and remove $j$ from $J$\;
}
Return $\mathbf P_r$\;
\end{revisionblue}\end{algorithm}

\section{Details on ratio scale extensions}\label{app:ratio-extensions}
This appendix develops the ratio estimands summarized in Section~\ref{sec:change-scale} of the main manuscript. Proofs of the corollaries appear in Appendix~\ref{app:ratio-proofs}. For binary outcomes, applying the additive construction to the log response or log odds and then exponentiating yields risk ratio or odds ratio estimands. For each nonempty $\calY\subseteq\calX$, the weighted construction gives
\begin{align}\label{eq:ratio-product-pe-estimands-supp}
\Delta_{\calY}=&\prod_{\calT\subseteq \calY^c}\prod_{\calZ\subseteq \calY}\{h\circ f(\calZ\cup \calT)\}^{(-1)^{|\calZ|}w(\calT,\calY)},
\end{align}
where $h(x)=x$ for risk ratios and $h(x)=x/(1-x)$ for odds ratios, and the transformation $\log(h\circ f)$ is assumed to be defined at every primitive response value. The weights satisfy the same nonnegativity and normalization conditions as in Definition~\ref{def:pi-estimands-defi}. The two choices of $h$ give
  \begin{align*}
\Delta_{\calY}^{\text{RR}}=&\prod_{\calT\subseteq \calY^c}\prod_{\calZ\subseteq \calY}f(\calZ\cup \calT)^{(-1)^{|\calZ|}w(\calT,\calY)},\\
\Delta_{\calY}^{\text{OR}}=&\prod_{\calT\subseteq \calY^c}\prod_{\calZ\subseteq \calY} \left\{\frac{f(\calZ\cup \calT)}{1-f(\calZ\cup \calT)}\right\}^{(-1)^{|\calZ|}w(\calT,\calY)}.
\end{align*}
In Equation~\eqref{eq:ratio-product-pe-estimands-supp}, the weights average contrasts on the log scale and appear as exponents on the ratio scale. The resulting estimands are geometric averages, and the inclusion--exclusion sum becomes a product. Permutation equivariance and completeness follow from Section~\ref{ss:class-estimands} under the same weight conditions, applied to the transformed response surface. The following corollaries give the residual-free construction with point mass weights at $\calT=\emptyset$. They apply, for example, to the binary-outcome versions of the mediation and network settings in Examples~\ref{eg:meidation-multiple} and~\ref{eg:network-interference}.

\begin{corollary}
\label{eg:rr}
Assume that $\log f$ is defined at every primitive response value. Applying the additive construction with point mass weights at $\calT=\emptyset$ and exponentiating gives a permutation equivariant, complete, and residual-free system. For each nonempty $\calY\subseteq\calX$, its component is
\begin{align*}
    \Delta_{\calY}^{\text{RR}}=\prod_{\calZ\subseteq \calY}f(\calZ)^{(-1)^{|\calZ|}}
\end{align*}
and the inclusion--exclusion product satisfies
\begin{align*}
    \prod_{\emptyset\neq \calY\subseteq \calX}(\Delta_{\calY}^{\text{RR}})^{(-1)^{|\calY|+1}}={f(\emptyset)}/{f(\calX)}.
\end{align*}
\end{corollary}

\begin{corollary}
\label{eg:or}
Assume that $\log\{f/(1-f)\}$ is defined at every primitive response value. Applying the additive construction with point mass weights at $\calT=\emptyset$ and exponentiating gives a permutation equivariant, complete, and residual-free system. For each nonempty $\calY\subseteq\calX$, its component is
\begin{align*}
    \Delta_{\calY}^{\text{OR}}=&\prod_{\calZ\subseteq \calY}\left\{\frac{f(\calZ)}{1-f(\calZ)}\right\}^{(-1)^{|\calZ|}},
\end{align*}
and the inclusion--exclusion product satisfies
\begin{align*}
    \prod_{\emptyset\neq \calY\subseteq \calX}(\Delta_{\calY}^{\text{OR}})^{(-1)^{|\calY|+1}}=\frac{f(\emptyset)\{1-f(\calX)\}}{f(\calX)\{1-f(\emptyset)\}}.
\end{align*}
\end{corollary}
We illustrate these formulas for the mediation setting in Example~\ref{eg2} and the balanced factorial design in Example~\ref{eg3}.

\begin{example}
For two mediators, the exit indirect main effects and their interaction on the risk ratio scale are
\begin{align}
    &\Delta_{\{X_1\}}=\frac{\mu(1,1)}{\mu(0,1)},~~\Delta_{\{X_2\}}=\frac{\mu(1,1)}{\mu(1,0)},~~\Delta_{\{X_1,X_2\}}=\frac{\mu(0,0)\mu(1,1)}{\mu(0,1)\mu(1,0)}.\label{eq:rr-interaction-two}
\end{align}
Their inclusion--exclusion product recovers the endpoint risk ratio:
\begin{align*}
    \frac{\Delta_{\{X_1\}} \Delta_{\{X_2\}}} { \Delta_{\{X_1,X_2\}}}=\frac{\mu(1,1)}{\mu(0,0)}.
\end{align*}
On the odds ratio scale, the corresponding effects and interaction are
\begin{align}
    &\Delta_{\{X_1\}}=\frac{\mu(1,1)\{1-\mu(0,1)\}}{\mu(0,1)\{1-\mu(1,1)\}},~~~\Delta_{\{X_2\}}=\frac{\mu(1,1)\{1-\mu(1,0)\}}{\mu(1,0)\{1-\mu(1,1)\}},\nonumber\\
    &\Delta_{\{X_1,X_2\}}=\frac{\mu(0,0)\mu(1,1)\{1-\mu(0,1)\}\{1-\mu(1,0)\}}{\mu(0,1)\mu(1,0)\{1-\mu(0,0)\}\{1-\mu(1,1)\}}.\label{eq:or-interaction-two}
\end{align}
Their inclusion--exclusion product recovers the endpoint odds ratio:
\begin{align*}
    \frac{\Delta_{\{X_1\}} \Delta_{\{X_2\}} }{\Delta_{\{X_1,X_2\}}}=\frac{\mu(1,1)\{1-\mu(0,0)\}}{\mu(0,0)\{1-\mu(1,1)\}}.
\end{align*}

\end{example}

\begin{example}
In a balanced full factorial trial with two factors, equal weights give the main effects on the risk ratio scale:
\begin{align*}
    &\Delta_{\{X_1\}}^{\text{RR}}=\sqrt{\frac{\mu(1,1)}{\mu(0,1)}}\times\sqrt{\frac{\mu(1,0)}{\mu(0,0)}},~~~\Delta_{\{X_2\}}^{\text{RR}}=\sqrt{\frac{\mu(1,1)}{\mu(1,0)}}\times\sqrt{\frac{\mu(0,1)}{\mu(0,0)}}.
\end{align*}
Each main effect is the geometric mean of the two risk ratios obtained by fixing the other factor at state $0$ or $1$. The corresponding main effects on the odds ratio scale are
\begin{align*}
    &\Delta_{\{X_1\}}^{\text{OR}}=\sqrt{\frac{\mu(1,1)\{1-\mu(0,1)\}}{\{1-\mu(1,1)\}\mu(0,1)}}\times\sqrt{\frac{\mu(1,0)\{1-\mu(0,0)\}}{\{1-\mu(1,0)\}\mu(0,0)}},\\
    &\Delta_{\{X_2\}}^{\text{OR}}=\sqrt{\frac{\mu(1,1)\{1-\mu(1,0)\}}{\{1-\mu(1,1)\}\mu(1,0)}}\times\sqrt{\frac{\mu(0,1)\{1-\mu(0,0)\}}{\{1-\mu(0,1)\}\mu(0,0)}}.
\end{align*}
These are the geometric means of the two odds ratios obtained by fixing the other factor at state $0$ or $1$. The two-way interactions on the risk ratio and odds ratio scales are given by Equations~\eqref{eq:rr-interaction-two} and~\eqref{eq:or-interaction-two}, respectively.
\end{example}

\clearpage
\section{A conceptual summary of comparison across the three application contexts}\label{app:application-comparison}
\thispagestyle{plain}

\begingroup
\renewcommand{\baselinestretch}{1}\fontsize{10}{12}\selectfont

\noindent Table~S\ref{tab:application-comparison} summarizes the action variables, mean responses, illustrative weights, admissible relabelings, and identification requirements in the three applications. In each setting, the analyst specifies which variables are contrasted and how the remaining variables are set or averaged over.

\smallskip
\setlength{\tabcolsep}{3pt}
\renewcommand{\arraystretch}{1.08}
\setlength{\belowcaptionskip}{5pt}
\captionof{table}{Comparison of the three applications.}
\label{tab:application-comparison}
\noindent\begin{tabularx}{\linewidth}{@{}>{\raggedright\arraybackslash}p{0.16\linewidth}*{3}{>{\raggedright\arraybackslash}X}@{}}
\toprule
& Factorial studies & Multiple mediators & Network interference \\
\midrule
Action variables
& $K$ treatment factors
& Treatment conditions selecting potential values of $K$ mediators
& Treatments of $N$ units, with $K=N$ \\
\addlinespace[3pt]
Mean response
& Mean potential outcome under a factor combination
& Mean potential outcome under the selected mediator potential values
& Mean potential outcome across units under the full assignment \\
\addlinespace[3pt]
Illustrative weights and effects
& Average over other factors using equal weights or assignment probabilities to define main effects and interactions
& Hold other mediators at their potential values under treatment~$1$ to define effects and interactions
& Average over other units' assignments to define direct, indirect, and overall effects and interactions \\
\addlinespace[3pt]
What is relabeled
& Factors and their assignment distributions
& Scientifically unordered mediators
& Units and their outcomes, treatment probabilities, and network connections \\
\addlinespace[3pt]
Identification
& Randomization or confounding adjustment, with positive probabilities for required factor combinations
& Combining mediator potential values across treatment conditions may require assumptions on their joint distribution
& Positive probabilities for required assignments; assess statistical precision separately \\
\bottomrule
\end{tabularx}

\endgroup

\clearpage


\begin{thebibliography}{}

\bibitem[\protect\citeauthoryear{Aronow and Samii}{Aronow and Samii}{2017}]{aronow2017estimating}
Aronow, P.~M. and C.~Samii (2017).
\newblock {Estimating average causal effects under general interference, with application to a social network experiment}.
\newblock {\em The Annals of Applied Statistics\/}~{\em 11\/}(4), 1912 -- 1947.

\bibitem[\protect\citeauthoryear{Bj{\"o}rklund, Husfeldt, and Koivisto}{Bj{\"o}rklund et~al.}{2009}]{bjorklund2009set}
Bj{\"o}rklund, A., T.~Husfeldt, and M.~Koivisto (2009).
\newblock Set partitioning via inclusion-exclusion.
\newblock {\em SIAM Journal on Computing\/}~{\em 39\/}(2), 546--563.

\bibitem[\protect\citeauthoryear{Cheng and Li}{Cheng and Li}{2024}]{cheng2024semiparametric}
Cheng, C. and F.~Li (2024).
\newblock Semiparametric causal mediation analysis in cluster-randomized experiments.
\newblock {\em arXiv preprint arXiv:2404.18256\/}.

\bibitem[\protect\citeauthoryear{Cheng and Ye}{Cheng and Ye}{2004}]{cheng2004geometric}
Cheng, S.-W. and K.~Q. Ye (2004).
\newblock {Geometric isomorphism and minimum aberration for factorial designs with quantitative factors}.
\newblock {\em The Annals of Statistics\/}~{\em 32\/}(5), 2168 -- 2185.

\bibitem[\protect\citeauthoryear{Daniel, De~Stavola, Cousens, and Vansteelandt}{Daniel et~al.}{2015}]{daniel2015causal}
Daniel, R.~M., B.~L. De~Stavola, S.~N. Cousens, and S.~Vansteelandt (2015).
\newblock Causal mediation analysis with multiple mediators.
\newblock {\em Biometrics\/}~{\em 71\/}(1), 1--14.

\bibitem[\protect\citeauthoryear{Dasgupta, Pillai, and Rubin}{Dasgupta et~al.}{2015}]{dasgupta2015causal}
Dasgupta, T., N.~S. Pillai, and D.~B. Rubin (2015).
\newblock Causal inference from {$2^K$} factorial designs by using potential outcomes.
\newblock {\em Journal of the Royal Statistical Society Series B: Statistical Methodology\/}~{\em 77\/}(4), 727--753.

\bibitem[\protect\citeauthoryear{Harshaw, S\"avje, and Wang}{Harshaw et~al.}{2025}]{harshaw2025general}
Harshaw, C., F.~S\"avje, and Y.~Wang (2025).
\newblock A general design-based framework and estimator for randomized experiments.
\newblock arXiv:2210.08698, version 3.

\bibitem[\protect\citeauthoryear{Hu, Li, and Wager}{Hu et~al.}{2022}]{hu2022average}
Hu, Y., S.~Li, and S.~Wager (2022).
\newblock Average direct and indirect causal effects under interference.
\newblock {\em Biometrika\/}~{\em 109\/}(4), 1165--1172.

\bibitem[\protect\citeauthoryear{Hudgens and Halloran}{Hudgens and Halloran}{2008}]{hudgens2008toward}
Hudgens, M.~G. and M.~E. Halloran (2008).
\newblock Toward causal inference with interference.
\newblock {\em Journal of the American Statistical Association\/}~{\em 103\/}(482), 832--842.

\bibitem[\protect\citeauthoryear{Imai, Keele, and Yamamoto}{Imai et~al.}{2010}]{imai2010identification}
Imai, K., L.~Keele, and T.~Yamamoto (2010, February).
\newblock Identification, inference and sensitivity analysis for causal mediation effects.
\newblock {\em Statistical Science\/}~{\em 25\/}(1), 51--71.

\bibitem[\protect\citeauthoryear{Imbens}{Imbens}{2000}]{imbens2000role}
Imbens, G.~W. (2000).
\newblock The role of the propensity score in estimating dose-response functions.
\newblock {\em Biometrika\/}~{\em 87\/}(3), 706--710.

\bibitem[\protect\citeauthoryear{Lange, Rasmussen, and Thygesen}{Lange et~al.}{2014}]{lange2014assessing}
Lange, T., M.~Rasmussen, and L.~C. Thygesen (2014).
\newblock Assessing natural direct and indirect effects through multiple pathways.
\newblock {\em American Journal of Epidemiology\/}~{\em 179\/}(4), 513--518.

\bibitem[\protect\citeauthoryear{Lunceford and Davidian}{Lunceford and Davidian}{2004}]{lunceford2004stratification}
Lunceford, J.~K. and M.~Davidian (2004).
\newblock Stratification and weighting via the propensity score in estimation of causal treatment effects: a comparative study.
\newblock {\em Statistics in Medicine\/}~{\em 23\/}(19), 2937--2960.

\bibitem[\protect\citeauthoryear{Ma, Fang, and Lin}{Ma et~al.}{2001}]{ma2001isomorphism}
Ma, C.-X., K.-T. Fang, and D.~K. Lin (2001).
\newblock On the isomorphism of fractional factorial designs.
\newblock {\em Journal of Complexity\/}~{\em 17\/}(1), 86--97.

\bibitem[\protect\citeauthoryear{Manski}{Manski}{2013}]{manski2013identification}
Manski, C.~F. (2013).
\newblock Identification of treatment response with social interactions.
\newblock {\em The Econometrics Journal\/}~{\em 16\/}(1), S1--S23.

\bibitem[\protect\citeauthoryear{McCullagh}{McCullagh}{2000}]{mccullagh2000invariance}
McCullagh, P. (2000).
\newblock Invariance and factorial models.
\newblock {\em Journal of the Royal Statistical Society Series B: Statistical Methodology\/}~{\em 62\/}(2), 209--256.

\bibitem[\protect\citeauthoryear{Mukerjee and Wu}{Mukerjee and Wu}{2006}]{mukerjee2006modern}
Mukerjee, R. and C.~F.~J. Wu (2006).
\newblock {\em A modern theory of factorial design}.
\newblock Springer.

\bibitem[\protect\citeauthoryear{Murphy and Topel}{Murphy and Topel}{1985}]{murphy1985estimation}
Murphy, K.~M. and R.~H. Topel (1985).
\newblock Estimation and inference in two-step econometric models.
\newblock {\em Journal of Business \& Economic Statistics\/}~{\em 3\/}(4), 370--379.

\bibitem[\protect\citeauthoryear{O'Donnell}{O'Donnell}{2014}]{o2014analysis}
O'Donnell, R. (2014).
\newblock {\em Analysis of {Boolean} functions}.
\newblock Cambridge University Press.

\bibitem[\protect\citeauthoryear{Ohnishi and Li}{Ohnishi and Li}{2026}]{ohnishi2024bayesian}
Ohnishi, Y. and F.~Li (2026).
\newblock A {Bayesian} nonparametric approach to mediation and spillover effects with multiple mediators in cluster-randomized trials.
\newblock {\em Journal of the American Statistical Association\/}~{\em 121\/}(553), 716--728.

\bibitem[\protect\citeauthoryear{Robins, Hern{\'a}n, and Brumback}{Robins et~al.}{2000}]{robins2000marginal}
Robins, J.~M., M.~A. Hern{\'a}n, and B.~Brumback (2000).
\newblock Marginal structural models and causal inference in epidemiology.
\newblock {\em Epidemiology\/}~{\em 11\/}(5), 550--560.

\bibitem[\protect\citeauthoryear{Rota}{Rota}{1964}]{rota1964foundations}
Rota, G.-C. (1964).
\newblock On the foundations of combinatorial theory {I}. theory of {M{\"o}bius} functions.
\newblock {\em Zeitschrift f{\"u}r Wahrscheinlichkeitstheorie und Verwandte Gebiete\/}~{\em 2}, 340--368.

\bibitem[\protect\citeauthoryear{Tong, Cheng, and Li}{Tong et~al.}{2026}]{tong2026causal}
Tong, J., C.~Cheng, and F.~Li (2026).
\newblock Causal mediation in cluster-randomized trials with multiple mediators: spillover-aware decomposition, identification, and semiparametric efficient inference.
\newblock {\em arXiv preprint arXiv:2604.10710\/}.

\bibitem[\protect\citeauthoryear{Viviano}{Viviano}{2025}]{viviano2025policy}
Viviano, D. (2025).
\newblock Policy targeting under network interference.
\newblock {\em Review of Economic Studies\/}~{\em 92\/}(2), 1257--1292.

\bibitem[\protect\citeauthoryear{Wang, Samii, Chang, and Aronow}{Wang et~al.}{2025}]{wang2025design}
Wang, Y., C.~Samii, H.~Chang, and P.~M. Aronow (2025).
\newblock Design-based inference for spatial experiments under unknown interference.
\newblock {\em The Annals of Applied Statistics\/}~{\em 19\/}(1), 744--768.

\bibitem[\protect\citeauthoryear{Xia and Chan}{Xia and Chan}{2022}]{xia2022decomposition}
Xia, F. and K.~C.~G. Chan (2022).
\newblock Decomposition, identification and multiply robust estimation of natural mediation effects with multiple mediators.
\newblock {\em Biometrika\/}~{\em 109\/}(4), 1085--1100.

\bibitem[\protect\citeauthoryear{Yu and Ding}{Yu and Ding}{2026}]{yu2026balancing}
Yu, R. and P.~Ding (2026).
\newblock Balancing weights for causal inference in observational factorial studies.
\newblock {\em Journal of the American Statistical Association\/}, 1--12.

\bibitem[\protect\citeauthoryear{Zhao and Ding}{Zhao and Ding}{2022}]{zhao2022regression}
Zhao, A. and P.~Ding (2022).
\newblock Regression-based causal inference with factorial experiments: estimands, model specifications and design-based properties.
\newblock {\em Biometrika\/}~{\em 109\/}(3), 799--815.

\bibitem[\protect\citeauthoryear{Zhao and Ding}{Zhao and Ding}{2023}]{zhao2023covariate}
Zhao, A. and P.~Ding (2023).
\newblock Covariate adjustment in multiarmed, possibly factorial experiments.
\newblock {\em Journal of the Royal Statistical Society Series B: Statistical Methodology\/}~{\em 85\/}(1), 1--23.

\end{thebibliography}
\end{document}